\documentclass[aip,cha,preprint]{revtex4-1}

\usepackage{color}
\usepackage{graphicx} 
\usepackage{subfig}
\usepackage{amssymb, amsmath, amsthm, amsfonts}
\usepackage{url}

\newcommand{\BE}{\begin{equation}}
\newcommand{\EE}{\end{equation}}
\newcommand{\BA}{\begin{align}}
\newcommand{\EA}{\end{align}}

\newcommand{\bfP}{\mathbf{P}}
\newcommand{\bfx}{\mathbf{x}}
\newcommand{\bfX}{\mathbf{X}}
\newcommand{\bfu}{\mathbf{u}}
\newcommand{\bfs}{\mathbf{s}}
\newcommand{\bff}{\mathbf{f}}
\newcommand{\bffp}{\mathbf{f}_\parallel}
\newcommand{\bfn}{\mathbf{n}}
\newcommand{\bfv}{\mathbf{v}}
\newcommand{\rmd}{\mathrm{d}}
\newcommand{\zacc}{a}
\newcommand{\sigh}{\sigma_\parallel}
\newcommand{\tv}{\tilde{v}}
\newcommand{\htv}{\hat{\tilde{v}}}
\newcommand{\tbfv}{\tilde{\mathbf{v}}}
\newcommand{\htbfv}{\hat{\tilde{\mathbf{v}}}}

\begin{document}


\title{Inhomogeneities and caustics in the sedimentation of noninertial particles in incompressible flows} 



\author{G\'{a}bor Dr\'{o}tos}
\email[]{gabor@ifisc.uib-csic.es}
\affiliation{Instituto de F\'{\i}sica Interdisciplinar y Sistemas Complejos (IFISC, CSIC-UIB), Campus Universitat de les Illes Balears, E-07122 Palma de Mallorca, Spain}
\affiliation{MTA-ELTE Theoretical Physics Research Group, P\'azmany P\'eter s\'etany 1/A, H-1117 Budapest, Hungary}
\author{Pedro Monroy}
\affiliation{Instituto de F\'{\i}sica Interdisciplinar y Sistemas Complejos (IFISC, CSIC-UIB), Campus Universitat de les Illes Balears, E-07122 Palma de Mallorca, Spain}
\author{Emilio Hern\'{a}ndez-Garc\'{\i}a}
\affiliation{Instituto de F\'{\i}sica Interdisciplinar y Sistemas Complejos (IFISC, CSIC-UIB), Campus Universitat de les Illes Balears, E-07122 Palma de Mallorca, Spain}
\author{Crist\'obal L\'opez}
\affiliation{Instituto de F\'{\i}sica Interdisciplinar y Sistemas Complejos (IFISC, CSIC-UIB), Campus Universitat de les Illes Balears, E-07122 Palma de Mallorca, Spain}


\date{\today}

\begin{abstract}
In an incompressible flow, fluid density remains invariant
along fluid element trajectories. This implies that the spatial
distribution of non-interacting noninertial particles in such
flows cannot develop density inhomogeneities beyond those that are
already introduced in the initial condition. However, in
certain practical situations, density is measured or
accumulated on (hyper-) surfaces of dimensionality lower than
the full dimensionality of the flow in which the particles
move. An example is the observation of particle distributions
sedimented on the floor of the ocean. In such cases, even if
the initial distribution of noninertial particles is uniform
within a finite support in an incompressible flow, advection in the
flow will give rise to inhomogeneities in the observed density.
In this paper we analytically derive, in the framework of an
initially homogeneous particle sheet sedimenting towards a
bottom surface, the relationship between the geometry of the
flow and the emerging distribution. From a physical point of
view, we identify the two processes that generate
inhomogeneities to be the stretching within the sheet, and the
projection of the deformed sheet onto the target surface. We
point out that an extreme form of inhomogeneity, caustics, can
develop for sheets. We exemplify our geometrical results with
simulations of particle advection in a simple kinematic flow,
study the dependence on various parameters involved, and
illustrate that the basic mechanisms work similarly if the
initial (homogeneous) distribution occupies a more general
region of finite extension rather than a sheet.
\end{abstract}


\maketitle 

\textbf{Sedimentation of small particles in complex flows is an
outstanding problem in science and technology. In particular,
the sinking of biogenic particles from the marine surface to
the bottom is a fundamental process of the biological carbon
pump, playing a key role in the global carbon cycle. A complete
understanding of this problem is still lacking. It has been
recently shown that despite fluid incompressibility, sedimenting
particles moving as noninertial particles in the ocean on large
scales show density inhomogeneities when accumulated on some
bottom surface. Here, we analytically derive the relation
between the geometry of the flow and the emerging distribution
for an initially homogeneous sheet of particles. From a
physical point of view, we identify the two processes that
generate inhomogeneities to be the stretching within the sheet,
and the projection of the deformed sheet onto the target
surface. We point out conditions under which an extreme form of
inhomogeneity, caustics, can develop.}


%


%
%

%

\section{Introduction}
\label{sec:intro}

The sinking of small particles immersed in fluids is a problem
of great importance both from theoretical and practical points
of view \cite{Michaelides1997,Falkovich2002}. In an
environmental context, the sinking of biogenic particles in the
ocean is a fundamental process. It plays a key role in the
Earth carbon cycle through the biological carbon pump, i.e.,
the sequestration of carbon from the atmosphere performed by
phytoplankton via photosynthesis in the surface waters, and
posterior sedimentation over the oceanic floor
\cite{Sabine2004}. This is a complex problem, which involves
the interplay of biogeochemical processes with oceanic
transport phenomena where many open questions remain. In
particular, some of these open questions concern the
identification of the catchment area (the place near the
surface where the particles come from) of a given oceanic floor
zone, and which the mechanisms are that lead to the observed
inhomogeneous distribution of particles in surface and
subsurface oceanic layers
\cite{Logan1990,Buesseler2007,Mitchell2008} or when collected
at a given depth by sediment traps
\cite{siegel1997,Schlitzer2003,Buesseler2007,Siegel2008,Qiu2014}).

In this paper, we shall describe basic ingredients for the
processes that lead to large-scale
inhomogeneities in the density of the particles after
sedimentation \cite{Monroy2017}. These inhomogeneities emerge
as a result of advection of the particles by flows in the
ocean. For the range of parameters that is relevant for marine
biogenic particles, a very good approximation for the equation
of motion of the particles
\cite{siegel1997,Siegel2008,Qiu2014,Roullier2014}, as it has
been explicitly shown in \cite{Monroy2017}, simply consists of
motion following the fluid velocity with an additional settling
term.

Such an equation of motion, if the fluid flow is
incompressible, preserves phase-space volume (note that the
phase space coincides here with the configuration space). Thus,
inertial effects, which have been typically identified as the
main causes for particle clustering (also called preferential
concentration) in other situations
\cite{Balkovsky2001,Bec2003,Vilela2007,Cartwright2010,Guseva2013,Guseva2016},
cannot explain inhomogeneities in mesoscale oceanic
sedimentation. Then the question is which the mechanisms are
that lead to sedimentation inhomogeneities in the absence of
particle inertia.

In incompressible flows, density is conserved along
trajectories, so that inhomogeneities can occur only if they
are already present in the initial distribution of the
particles, and these initial inhomogeneities are carried over
as intact during the entire time evolution, as long as
characterizing the concentration of particles by a density is
an appropriate framework. Note that this fact could already be
sufficient for explaining the presence of inhomogeneities: for
example, biogenic particles in the ocean are not produced in a
uniform distribution, of course.

At the same time, one can also argue that particles become
uniformly distributed for asymptotically long times in bounded
incompressible flows of chaotic nature \cite{Ott1993}, which
translates to a uniform particle density, at least when
coarse-grained.
For localized initial particle distributions in
unbounded chaotic systems, a (growing and flattening) Gaussian
is obtained instead of a uniform density, but such a shape can
also be regarded as trivial.

However, if the initial distribution is localized, even if being homogeneous within the localized support, it is well-known that complicated structures can be observed
before reaching the asymptotic state \cite{Ottino1989,Pierrehumbert1993}. In particular,
stretching and folding of the phase-space volume in which the
particles are located can, at least when looking at
coarse-grained scales, considerably rearrange the density. That is, (coarse-grained) inhomogeneities \emph{emerge} due to advection, which can be regarded as clustering or preferential concentration. In
fact, it is the same process that leads to the above-mentioned
asymptotic simplification, but the effect of this process is opposite on non-asymptotic time scales.

\begin{figure}
\includegraphics[width=0.45\textwidth]{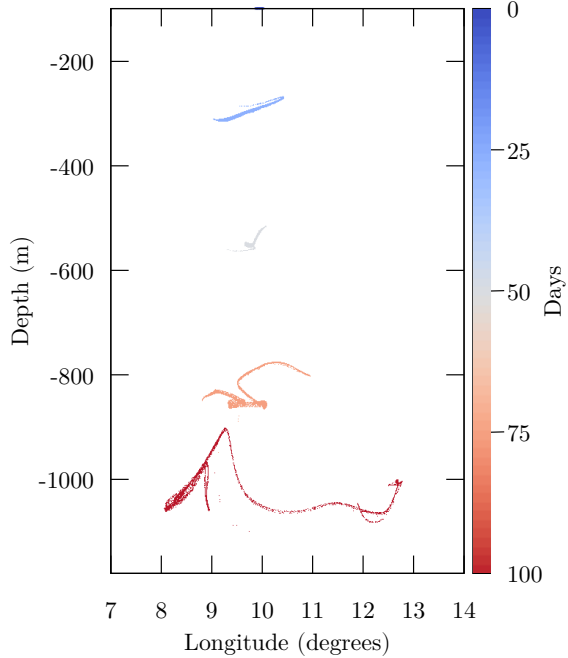}
\caption{\label{fig:benguela}The positions of particles (projected onto a vertical plane)
at different times in a realistic ROMS simulation \cite{Gutknecht2013} of the Benguela zone.
The numerical experiment consisted of releasing $6000$ particles from initial conditions
randomly chosen in a square with sides of $1/6^{\circ}$, centered
at $10.0^{\circ}$ E $29.12^{\circ}$ S and $100\mathrm{m}$ depth. The particles'
trajectories $\mathbf{X}(t)$ were calculated from
$\rmd\mathbf{X}/\rmd t = \bfv_\mathrm{ROMS} - W \hat{\mathbf{k}}$,
where $\bfv_\mathrm{ROMS}$ is the velocity from the ROMS model, and $W = 10\mathrm{m/day}$
corresponds to the sinking velocity \cite{Monroy2017}, pointing in the vertical direction
given by the unit vector $\hat{\mathbf{k}}$.}
\end{figure}
Preliminary numerical studies in a realistic oceanic setting
showed that a homogeneous layer of particles (with neglecting
the interaction between them) indeed evolves to complicated
shapes by stretching and folding while it is sinking.
As a motivation, Fig. \ref{fig:benguela} presents such a direct
numerical simulation. It is clear that homogeneization or
simplification is \emph{not} reached on the time scale of the
sinking process.
The example of oceanic sedimentation thus emphasizes the
practical importance of the investigation of non-asymptotic
time scales in general, which, from a practical point of view, has received relatively little attention
in the literature so far (an important exception is the
paradigmatic problem of weather forecasting).

Beyond stretching and folding during the sinking process, an
important additional ingredient for the emergence of observable
inhomogeneities in the density of sedimented particles is the
\emph{accumulation} at the bottom of the domain. This is a
time-integration of the density at a two-dimensional subset of
the full three-dimensional space, and this results in the
translation of the complicated shapes to actual inhomogeneities
without any coarse-graining: the conservation of density no
longer holds for such time-integrated projections.

In this paper, we shall describe in detail how inhomogeneities
in the accumulated density emerge in incompressible flows on
non-asymptotic time scales. We will derive the basic mechanisms
analytically, and we will investigate the properties of these
mechanisms in a simplified kinematic flow, in order to focus on
the particle dynamics.

The main results we achieve are: i) we identify and quantify
two geometrical mechanisms contributing to the creation of
inhomogeneities in the density: the stretching due to the flow
and the projection onto the constant depth where the particles
accumulate; ii) we obtain an explicit expression for the
density at an arbitrary position of the accumulation level in
terms of the trajectories arriving to that particular position;
and iii) in the context of a simplified kinematic flow we study
the dependence on parameters that are generic to the problem:
the settling velocity, the depth of the accumulation level, and
the amplitude of the fluctuating flow.

The paper is organized as follows. In Section \ref{sec:setup}
we establish the setup for our analysis. In Section
\ref{sec:formulae} we obtain the expression for the final
density, and quantify the above-mentioned two effects leading
to inhomogeneities. In Section \ref{sec:ex} we evaluate these
results in the kinematic flow model. This flow is defined in
two dimensions (one horizontal and one vertical), and it may
show chaotic behavior. The role of the chaoticity of the flow
will be explicitly addressed. In Section \ref{sec:pardep} we
study the parameter dependence. Finally, in Section
\ref{sec:concl} we summarize and comment on the results. A
number of appendices contain the more technical aspects of our
Paper.

\section{Formulation of the setup}
\label{sec:setup}

\subsection{Equations of motion}
\label{subsec:eqmotion}

In this work we will consider the motion of particles that
follow closely the velocity of the fluid in which they are
dispersed, except for the addition of a constant vertical
velocity arising from the particle weight. This description is
adequate in a variety of circumstances. In particular it was
shown by Monroy et al. \cite{Monroy2017} to be the adequate one
to describe a wide range of biogenic particles sedimenting in
ocean flows with turbulent intensity typical of the open ocean.
We revise in the following the arguments leading to that
conclusion.

The dynamics of spherical particles advected in fluids is
described, in the small-particle limit, by the
Maxey--Riley--Gatignol equation
\cite{Maxey1983,Haller2008,Monroy2017}. When writing it in a
nondimensionalized form that uses the characteristic length $L$
and magnitude $U$ of the fluid velocity field as units of space
and velocity, two relevant dimensionless parameters appear. The
first one is the Stokes number:
\BE
\mathrm{St} = \frac{a^2 U}{3\nu\beta L} , \label{eq:St}
\EE
where $a$ is the radius of the particle, $\nu$ is the kinematic
viscosity of the fluid, and $\beta=
3\rho_\mathrm{f}/(2\rho_\mathrm{p}+\rho_\mathrm{f})$, with
$\rho_\mathrm{p}$ and $\rho_\mathrm{f}$ being the densities of the
particle and of the fluid, respectively. This number quantifies
the importance of inertia with respect to viscous drag. The
second dimensionless quantity is the Froude number, quantifying
the importance of inertia with respect to gravity:
\BE
\mathrm{Fr} = \frac{U}{\sqrt{gL}} , \label{eq:Fr}
\EE
where $g$ is the gravitational acceleration. In terms of these
numbers, the dimensionless terminal settling velocity of a
particle in still fluid is
\BE
W = (1-\beta) \frac{\mathrm{St}}{\mathrm{Fr^2}} \ .
\label{eq:settlingvelocity}
\EE


In complex turbulent flows such as in the ocean, the values of
$\mathrm{St}$ and $\mathrm{Fr}$ vary with scale. Monroy et al.
\cite{Monroy2017} showed that for a relevant range of sizes and
densities of biogenic particles, $\mathrm{St}$ is very small.
For example, it takes values\cite{Monroy2017} in the range
$10^{-7}\ldots10^{-1}$ in wind-driven surface turbulence in the
open ocean at the Kolmogorov scale ($\sim 0.3 \ldots
2\mathrm{mm}$), where\cite{Jimenez1997} typical turbulent
velocities are in the range $0.5 \ldots
3\mathrm{mm}/\mathrm{s}$. At larger scales, $\mathrm{St}$ is
still smaller. For example, for mesoscale oceanic motions,
$L_\mathrm{h}= 100\mathrm{km}$ and $U_\mathrm{h}=
0.1\mathrm{m/s}$ for horizontal motion, and $L_\mathrm{v}=
100\mathrm{m}$ and $U_\mathrm{v}= 10\mathrm{m/day} \approx
10^{-4}\mathrm{m/s}$ for vertical motion. This gives $\mathrm{St}\approx
10^{-6}$ for both horizontal and vertical motion. In any case,
$\mathrm{St}$ is typically very small for the type of particles
we are interested in. Under these circumstances a standard
approach\cite{Haller2008} can be used to approximate the
equation of motion for the particle in the limit of vanishing
$\mathrm{St}$ (see Appendix \ref{sec:ad:ParticleEquations}, and Eq. \eqref{eq:MRG} in particular),
provided that the settling velocity $W$ is also small (Eq.
(\ref{eq:settlingvelocity})).

In our
ocean situation, the Froude number ranges from $10^{-5}$ at the
mesoscale to a maximum of $10^{-2}$ at the Kolmogorov scale.
Thus, the combination $\mathrm{St}/\mathrm{Fr}^2$, appearing in
the settling velocity $W$, Eq. (\ref{eq:settlingvelocity}), is within few orders of
magnitude from $1$, and is typically larger than $1$. This means, first, that $W$ is always orders of magnitude larger than $\mathrm{St}$, $W \gg \mathrm{St}$, and, second, that $W$
is typically not a small quantity.

If $W \ll 1$ does not hold, the standard approach\cite{Haller2008} for the small-$\mathrm{St}$ approximation is incorrect. In this case,
what is appropriate is to take the
limit defined by $\mathrm{St}\to 0$ and $\mathrm{Fr}\to 0$ with
the value of $W \sim \mathrm{St}/\mathrm{Fr}^2$ remaining constant.
Both in this new limit (see Appendix \ref{sec:ad:ParticleEquations})
and in the standard approach\cite{Haller2008} with $W \gg \mathrm{St}$, the \emph{leading
order} contribution in $\mathrm{St}$ to the equation of motion for the particle
is a
well-known\cite{siegel1997,Siegel2008,Qiu2014,Roullier2014,Monroy2017}
approximation:
\BE
\dot{\bfX} = \bfv(\bfX,t) \equiv \bfv_\mathrm{fluid}(\bfX,t) -
W \hat{\mathbf{k}} \ ,\label{eq:eqmotion_noninertial}
\EE
where we have introduced the notation $\bfv$ for the ``velocity
field of the particle''. An important feature of the
approximate Eq. \eqref{eq:eqmotion_noninertial} is the absence
of any inertial term.

The description \eqref{eq:eqmotion_noninertial} would be
applicable in other circumstances beyond the ones described
above, but, of course, there would be situations --- for
example, coastal wave-breaking turbulence environments,
industrial flows, or (other) cases in which $\mathrm{St}$ is not small
enough --- in which inertial terms will have central importance,
with effects that have been studied in recent works
\cite{Balkovsky2001,Bec2003,Vilela2007,Cartwright2010,Guseva2013,Guseva2016}.

In our paper, we shall restrict our investigations to dynamics
of the form of Eq. \eqref{eq:eqmotion_noninertial}.
Additionally, we shall assume $|v_{\mathrm{fluid},z}(\bfX,t)| <
W$ for the vertical component of the fluid velocity field,
which ensures $v_z < 0$ for the vertical component of the
``particle velocity field'' $\bfv$. This assumption excludes
the presence of particle trajectories that would be trapped
forever to the system, which simplifies the technical treatment
of the problem and the interpretation of the phenomenology in
that the accumulated density at the bottom of the domain is
obtained by integrating over finite times. This assumption is
reasonable in the above-discussed example of oceanic biogenic
particles serving as our motivation \cite{Monroy2017}.



\subsection{Definitions}
\label{subsec:def}

Let us consider a flow in a $d$-dimensional space in which we
distinguish a `vertical' direction, characterized by the
`vertical' coordinate $z$, and the remaining
$(d-1)$-dimensional subspace, which we call `horizontal', with
the position vector $\bfx = (x,y,\ldots) =
(x_1,x_2,\ldots,x_{d-1})$. We analyze the case $d=2$ in detail,
with mentioning $d=3$ at some points due to its practical
relevance, but all results can easily be generalized to higher
dimensions, which can be useful when analyzing problems with
phase spaces of higher dimensionality. The flow is defined by
the velocity field $\bfv(\bfX,t)$, $\bfX = (\bfx,z) =
(x,y,\ldots,z) = (x_1,x_2,\ldots,x_d)$ being the position
vector in the full space and $t$ being time. $v_z < 0$ is
assumed for all $\bfX$ and $t$.

We initialize noninertial particles at $t = t_0$ on a given
level $z=z_0$ whose density within the so-defined horizontal
subspace (a material line and surface for $d=2$ and $d=3$,
respectively) is described by a ``surface'' density $\sigma$.
We let the particles fall until all of them reach a depth
$z=-\zacc$ where they accumulate. We are interested in the
resulting horizontal ``surface'' density $\sigh$ of the
particles measured within the accumulation level.

In our notation, a vertical line with a variable in the lower
index, $\left.\right|_\alpha$, corresponds to keeping that
particular variable, $\alpha$, constant, while a vertical line
with the declaration of a value,
$\left.\right|_{\beta=\beta_0}$, denotes evaluating the
preceding expression at the indicated value, $\beta_0$. These
two notations can also occur together. As an implicit rule in
our notation, when taking derivatives with respect to a
horizontal coordinate, all other horizontal coordinates are
assumed to be kept constant.

\section{Relating the density to particle trajectories}
\label{sec:formulae}

The final density $\sigh$ forming at any position of the
accumulation level can be related to geometric properties of
the flow observable along the trajectory of a particle that was
initialized on the initial level $z_0$ at $t_0$ and that
arrives at the given position. If we have more particles, the
corresponding densities are to be added. In this Section we
first explain that the relation can be given in terms of a
special Jacobian, and analyze the formula from some practical
aspects. Then we present (for simplicity, taking $d=2$ in the
main text) an intuitive way of building up our formula, which
lets us distinguish between the contribution of two simple
effects: the stretching within the material line or material
surface in which the particles are distributed, and the
horizontal kinematic projection (i.e., a projection that takes
the horizontal component of the velocity into account) of the
density at the points of arrival at the accumulation level.
Each of these two effects is well-defined even in setups in
which the other is absent.

\subsection{General results}
\label{subsec:formulae_general}

Let the endpoint of a trajectory at time $t$ that was
initialized at $\bfx_0$ be denoted by $\bff(t;\bfx_0) =
(f_x(t;\bfx_0),f_y(t;\bfx_0),\ldots,f_z(t;\bfx_0)) =
(f_1(t;\bfx_0),f_2(t;\bfx_0),\ldots,f_d(t;\bfx_0))$. The
horizontal density at the point where a particular trajectory
crosses the accumulation plane $z=-\zacc$ is proportional to
the density at the initial position of the given trajectory:
\BE
\sigh(t(f_z=-\zacc,\bfx_0),\bfx_0) = \sigma(t=t_0,\bfx_0) \mathcal{F}(t(f_z=-\zacc,\bfx_0),\bfx_0) ,
\label{eq:sh_t_bfx0}
\EE
where $\bfx_0$ is the $d-1$ dimensional initial position at $t
= t_0$ of the particular trajectory within the initial level
$z=z_0$, $\sigma(t=t_0,\bfx_0)$ is the initial ``surface''
density at $\bfx_0$, and $\sigh(t,\bfx_0)$ is the horizontal
``surface'' density at the endpoint, at some time $t$, of the
trajectory that was initialized at $\bfx_0$. The time $t$ of
arrival at the accumulation level is unique, since $v_z < 0$ is
assumed, see Section \ref{sec:setup}. This time depends on the
vertical position of the accumulation level, where $f_z =
-\zacc$, and also on which trajectory is chosen, which is
defined by the initial position $\bfx_0$. (More generally, an
arbitrary time $t$ can be regarded as a function of any final
vertical position $f_z$ and of the initial position $\bfx_0$,
$t=t(f_z,\bfx_0)$. The relation $t(f_z,\bfx_0)$ is
single-valued because of the assumption $v_z < 0$.) In case
more than one trajectory arrives at the same position within the
accumulation level, the corresponding densities are summed up.

The total factor, $\mathcal{F}(t(f_z=-\zacc,\bfx_0),\bfx_0)$,
that multiplies the original density at the starting point of
the given trajectory, is the reciprocal of the determinant of a
Jacobian:
\BE
\mathcal{F}(t(f_z=-\zacc,\bfx_0),\bfx_0) = \det\left( J(t(f_z=-\zacc,\bfx_0),\bfx_0) \right)^{-1} ,
\label{eq:F_t_bfx0}
\EE
where $J$ is a $(d-1)\times(d-1)$ Jacobian:
\BE
J_{ij}(t(f_z,\bfx_0),\bfx_0) = \left.\frac{\partial f_j(t(f_z,\bfx_0),\bfx_0)}{\partial x_{0i}}\right|_{f_z}
\label{eq:J_t_bfx0}
\EE
for $i,j \in \{1,\ldots,d-1\}$. This Jacobian is \emph{not} a
usual one in two aspects. First, it is not a full-dimensional
Jacobian, but it is restricted to the horizontal coordinates.
In particular, for flows with $d=2$, it is a scalar. Second,
the derivatives with respect to the coordinates of $\bfx_0$ are
taken at a constant value of the vertical coordinate $f_z$, and
\emph{not at a constant time}. For this reason, the direct
numerical evaluation of Eq. \eqref{eq:J_t_bfx0} for a given
trajectory is not straightforward. Nevertheless, Eqs.
\eqref{eq:F_t_bfx0}-\eqref{eq:J_t_bfx0} are intuitive in the
sense that they give the ratio between the final and the
initial values of the ``area'' of an infinitesimal ``surface''
element neighboring the given trajectory within the material
``surface'' of particles. For a more rigorous derivation, see
Appendix \ref{sec:ad:F_t_bfx0}. Note that the determinant of a
full-dimensional Jacobian taken at a constant time is always
one for volume-preserving flows. In our setup, the reduced
dimensionality and the non-instantaneous accumulation process
lead to changes in the density, and thus the formation of
inhomogeneities becomes possible.

We show in Appendix \ref{sec:ad:constz_to_constt} that the
derivatives taken at a constant $f_z$ in the Jacobian
\eqref{eq:J_t_bfx0} can be replaced by derivatives taken at a
constant time $t$ in the following way:
\BE
\left.\frac{\partial f_i(t(f_z,\bfx_0),\bfx_0)}{\partial x_{0j}}\right|_{f_z} = \left.\frac{\partial f_i(t,\bfx_0)}{\partial x_{0j}}\right|_t - \frac{v_i(t,\bff(t;\bfx_0))}{v_z(t,\bff(t;\bfx_0))} \left.\frac{\partial f_z(t,\bfx_0)}{\partial x_{0j}}\right|_t ,
\label{eq:constz_to_constt}
\EE
for $i,j \in \{1,\ldots,d-1\}$. The difference between taking
derivatives at constant $f_z$ and constant $t$ stems from the
fact that different trajectories in the material ``surface''
reach a given level $f_z$ at different times $t$. From a
practical point of view, Eq. \eqref{eq:constz_to_constt} can
easily be evaluated numerically.

Transforming the right-hand side of Eq. \eqref{eq:F_t_bfx0} in
an alternative way, we learn that it can be obtained from an
integral along the given trajectory (as derived in Appendix
\ref{sec:ad:F_t_bfx0_alt}):
\begin{equation}
\mathcal{F}(t(f_z=-\zacc,\bfx_0),\bfx_0) = \exp \left( -\int_{z_0}^{-\zacc} \left. \sum_{i=1}^{d-1} \frac{\partial}{\partial f_i} \left( \frac{\hat{v}_i(f_z,\bffp)}{\hat{v}_z(f_z,\bffp)} \right) \right|_{f_z,\bffp = \bffp(f_z,\bfx_0)} \rmd f_z \right) ,
\label{eq:F_t_bfx0_alt}
\end{equation}
where $\bffp = (f_1,\ldots,f_{d-1})$ denotes the horizontal
coordinates of the trajectory, and $\hat{v}_i(f_z,\bffp) =
v_i(t(f_z,\bfx_0(f_z,\bffp)),\bff(t(f_z,\bfx_0(f_z,\bffp)),\bfx_0(f_z,\bffp)))$
for $i \in \{1,\ldots,d\}$, i.e., $\hat{\bfv}(f_z,\bffp)$ is
the velocity as regarded as a function of the endpoints of the
trajectories (instead of the time and the ``bare'' geometrical
coordinates of the domain of the fluid flow). When keeping
$f_z$ constant, the derivatives taken with respect to the
coordinates $f_i$, with $i \in \{1,\ldots,d-1\}$, correspond to
varying the selected trajectory and also the time $t$, so that
these derivatives are not the instantaneous geometrical
derivatives of the velocity field (see Appendix
\ref{sec:ad:F_t_bfx0_alt} for a more detailed explanation). By
replacing the derivatives taken at a constant $f_z$ with those
taken at a constant $t$, we can further transform our formula
such that it can be directly evaluated numerically, see
Appendix \ref{sec:further_transform}.

One important aspect of the results presented in this Section
is that the final density at a given point can be obtained in
terms of the initial density at one point (or, at least, a
countable number of them) and of the particle trajectory (or
trajectories) linking the points: these are all local
properties, and no spatially extended information (within the
material ``surface'') is needed to determine the final density
at the given point. In the next subsection we rewrite Eqs.
\eqref{eq:F_t_bfx0}-\eqref{eq:constz_to_constt} in alternative
ways which highlight the contributions from two different and
physically intuitive processes.

\subsection{Stretching and projection}
\label{subsec:formulae_st_and_pr}

In this Section we obtain Eq.
\eqref{eq:F_t_bfx0}-\eqref{eq:constz_to_constt} via two
physically intuitive steps which correspond to two individual
effects that modify the original density. For simplicity, we
restrict ourselves to $d=2$. In order to be able to precisely
formulate our considerations, we use a parametric notation for
the material line in this Section.

Let $\bff(t=t_0;u)$ describe a line segment of initial
conditions at time $t = t_0$ (a material line of particles)
embedded in 2 dimensions, parameterized by the arc length $u$,
and let $\sigma(t=t_0;u)$ be the initial density along the line
segment at $u$. Note that the initial line segment need not be
horizontal: the results of this Section apply for a
1-dimensional initial subset of arbitrary shape, which extends
the validity of these considerations to more general setups.

Let us denote the image of the initial line segment at time $t$
by $\bff(t;u)$. The density $\sigma(t;u)$ along this image at
$t$ in a point whose initial position was characterized by $u$
is given by
\BE
\sigma(t;u) = \sigma(t=t_0;u) \mathcal{S}(t;u) ,
\label{eq:s_t_u}
\EE
where
\BE
\mathcal{S}(t;u) = \left| \frac{\rmd \bff(t;u)}{\rmd u} \right|^{-1} .
\label{eq:S_t_u}
\EE
This simply follows from imposing the conservation of mass
(i.e., continuity) within the material line of the particles.
Note that the density (due to the incompressibility of the
fluid) is conserved only in the full space, but not along
subsets with lower dimensionality. For a precise derivation
based on the full-dimensional density, see Appendix
\ref{sec:ad:S_t_u}. The factor $\mathcal{S}(t;u)$, multiplying
the original density, describes the \emph{stretching} along the
material line up to time $t$ experienced near a particle
initialized at position $u$.

We can obtain the horizontal density $\sigh(t;u)$ by projecting
the instantaneous density $\sigma(t;u)$, which is measured
along the material line, to the horizontal direction taking
into account the kinematics of the problem. In particular, we
need to take into account the instantaneous orientation of the
material line at the position characterized by $u$, and also
the velocity at the same position:
\BE
\sigh(t;u) = \sigma(t;u) \mathcal{P}(t;u) ,
\label{eq:sh_t_u}
\EE
where, according to simple geometry relating the pre- and the
post-projection length of an infinitesimal segment of the
material line around the position characterized by $u$,
\BE
\mathcal{P}(t;u) = \left| \frac{\rmd f_x(t;u)}{\rmd s} - \frac{\rmd f_z(t;u)}{\rmd s} \frac{v_x(\bff(t;u),t)}{v_z(\bff(t;u),t)} \right|^{-1} .
\label{eq:P_t_u}
\EE
Here $s$ is the arc length along the \emph{image} of the line
segment at $t$, and $u$ can be regarded as a function of $s$.
The first term holds alone when there is no horizontal velocity
at the given time instant at the position of the given
particle, and the second term originates from an additional
change in the length, which is due to the presence of
horizontal motion. It is worth emphasizing that the presence of
the second term is due to projecting the material line to a
given depth, instead of taking the projection at a given time,
in agreement with Eq. \eqref{eq:J_t_bfx0}. For a more detailed
explanation of the formula, see Appendix \ref{sec:ad:P_t_u}.
This relation is valid for any $t$ and $u$, so that it also
applies to the time instant when a given particle arrives at
the accumulation level.

In total, there are two independent effects modifying the
initial density $\sigma(t=t_0;u)$: the stretching and the
projection, and both of them appear as a factor multiplying
$\sigma(t=t_0;u)$:
\BE
\sigh(t;u) = \sigma(t=t_0;u) \mathcal{F}(t;u) = \sigma(t=t_0;u) \mathcal{S}(t;u) \mathcal{P}(t;u) ,
\label{eq:fact_sh_t_u}
\EE
where $\mathcal{F}(t;u)$ is the total factor (the same as in
\eqref{eq:F_t_bfx0}, for $d=2$), and $\mathcal{S}(t;u)$ and
$\mathcal{P}(t;u)$ correspond to the stretching and the
projection as defined by Eqs. \eqref{eq:S_t_u} and Eq.
\eqref{eq:P_t_u}, respectively.

We can simplify the total factor to obtain \eqref{eq:F_t_bfx0}
with \eqref{eq:constz_to_constt} as follows. Applying the chain
rule for the partial derivatives in \eqref{eq:P_t_u} yields
\BE
\mathcal{P}(t;u) = \left| \frac{\rmd \bff(t;u)}{\rmd u} \right| \left| \frac{\rmd f_x(t;u)}{\rmd u} - \frac{\rmd f_z(t;u)}{\rmd u} \frac{v_x(\bff(t;u),t)}{v_z(\bff(t;u),t)} \right|^{-1} ,
\label{eq:P_t_u_by_u}
\EE
where
\BE
\left| \frac{\rmd u}{\rmd s} \right| = \left| \frac{\rmd \bff(t;u)}{\rmd u} \right|^{-1}
\EE
has been used (see Eq. \eqref{eq:duds} and the preceding
discussion in Appendix \ref{sec:ad:S_t_u}). Note that,
according to \eqref{eq:S_t_u},
\BE
\left| \frac{\rmd \bff(t;u)}{\rmd u} \right| = \mathcal{S}(t;u)^{-1} ,
\label{eq:S_t_u_inv}
\EE
the substitution of which into \eqref{eq:P_t_u_by_u} cancels
out $\mathcal{S}(t;u)$ in \eqref{eq:fact_sh_t_u}:
\BE
\mathcal{F}(t;u) = \left| \frac{\rmd f_x(t;u)}{\rmd u} - \frac{\rmd f_z(t;u)}{\rmd u} \frac{v_x(\bff(t;u),t)}{v_z(\bff(t;u),t)} \right|^{-1} ,
\label{eq:simple_F_t_u}
\EE
which is equivalent to \eqref{eq:F_t_bfx0}-\eqref{eq:constz_to_constt} for $d=2$.

The first term in Eq. \eqref{eq:simple_F_t_u},
\BE
\delta_x(t;u) = \frac{\rmd f_x(t;u)}{\rmd u} ,
\label{eq:delta_x_t_u}
\EE
is the parametric derivative, with respect to the position
along the initial line segment, of the horizontal component of
the current position vector, while the second term,
\BE
\tilde{\delta}_z(t;u) = - \delta_z(t;u) \frac{v_x(\bff(t;u),t)}{v_z(\bff(t;u),t)} = - \frac{\rmd f_z(t;u)}{\rmd u} \frac{v_x(\bff(t;u),t)}{v_z(\bff(t;u),t)} ,
\label{eq:delta_z_t_u}
\EE
is its vertical counterpart, but it is weighted by the ratio of
the two velocity components. As in Eq. \eqref{eq:P_t_u}, the
former one is due to a ``static'' change in length (i.e., not
influenced by any horizontal motion), and the latter one is the
``correction'' when horizontal motion is present. The
possibility of simplifying Eq. \eqref{eq:fact_sh_t_u} (with
Eqs. \eqref{eq:S_t_u} and Eq. \eqref{eq:P_t_u}) to Eq.
\eqref{eq:simple_F_t_u} is not a surprise: it is only the ratio
between the final and the initial length of an infinitesimal
line segment that is relevant, which we have already learnt in
Section \ref{subsec:formulae_general}.

Results for $d=3$ corresponding to those of this Section
discussed so far are given in Appendix \ref{sec:st_and_pr_3D},
and formulae for $d > 3$ can be constructed similarly.

For $d=2$, we can summarize our final expression as
\begin{align}
\sigh(t;u) &= \sigma(t=t_0;u) \mathcal{F}(t;u) \nonumber \\
&= \sigma(t=t_0;u) \mathcal{S}(t;u) \mathcal{P}(t;u) \nonumber \\
&= \sigma(t=t_0;u) \left| \delta_x(t;u) + \tilde{\delta}_z(t;u) \right|^{-1} ,
\label{eq:paracaustic}
\end{align}
with the particular quantities collected in Table
\ref{tab:quantities}. Note that a special situation may occur
for those trajectories for which
$|\delta_x+\tilde{\delta}_z|=0$ at the accumulation level. In
this case the final horizontal density is unbounded. The
corresponding positions within the accumulation level
characterize the so-called (density) caustics
\cite{Wilkinson2005}, and they refer to the maximum levels of
inhomogeneity in the accumulated density, so that their
identification and dependence on parameters is of great
relevance in our work. Of course, the integral of the density
(with respect to the final horizontal coordinate $x$) over such
caustics remains finite. In particular, the generic form of a
density caustic originating from a standard parabolic fold with its
vertex located at $x_\mathrm{c}$ is $\sim
1/\sqrt{x-x_\mathrm{c}}$.
\begin{table}
\begin{tabular}{| l | p{0.5\textwidth} | r |}
\hline
Notation & Name & Defining formula \\
\hline
$\mathcal{F}$ & Total factor & \eqref{eq:fact_sh_t_u} \\
\hline
$\mathcal{S}$ & Stretching factor & \eqref{eq:S_t_u} \\
\hline
$\mathcal{P}$ & Projection factor & \eqref{eq:P_t_u} \\
\hline
$\delta_x$ & Parametric derivative\newline of the horizontal position & \eqref{eq:delta_x_t_u} \\
\hline
$\delta_z$ & Parametric derivative\newline of the vertical position & \eqref{eq:delta_z_t_u} \\
\hline
$\tilde{\delta}_z$ & Weighted parametric derivative\newline of the vertical position & \eqref{eq:delta_z_t_u} \\
\hline
\end{tabular}
\caption{\label{tab:quantities}The main quantities relevant for changes in the density.}
\end{table}

We can give a more intuitive condition for the positions of the
caustics. We first recognize a simplification of
\eqref{eq:P_t_u}, which is useful in general, too, and reads as
\BE
\mathcal{P}(t;u) = \left| \frac{v_z(\bff(t;u),t)}{\bfn(t;u) \cdot \bfv(\bff(t;u),t)} \right| ,
\label{eq:P_t_u_withn}
\EE
where $\bfn(t;u)$ is the normal vector of the line $\bff$ at
time $t$ at a position that is characterized by $u$. Eq.
\eqref{eq:P_t_u_withn} is true, since $\bfn$ is obtained by
rotating the tangent vector $\rmd \bff / \rmd s$ of the line by
$\pi/2$:
\BE
\bfn(t;u) = \left( - \frac{\rmd f_z(t;u)}{\rmd s} , \frac{\rmd f_x(t;u)}{\rmd s} \right) .
\label{eq:n_t_u}
\EE
A remarkable property of \eqref{eq:P_t_u_withn} is that it
remains valid for $d = 3$, see Appendix \ref{sec:n_t_vecu} for
the derivation.

The presence of caustics actually originates from the
projection factor $\mathcal{P}$ alone, and Eq.
\eqref{eq:P_t_u_withn} gives a particularly intuitive
interpretation by identifying the positions of the caustics as
\BE
\bfn(t;u) \cdot \bfv(\bff(t;u),t) = 0 .
\label{eq:caustic_withn}
\EE
That is, caustics appear in the accumulation plane wherever the
local normal vector of the line is perpendicular to the local
velocity, or, equivalently, where the local tangent of the line
coincides with the direction of the local velocity.

\section{Numerical examples}
\label{sec:ex}

In this Section we present the basic phenomenology of our setup
via numerical examples in a 2D model flow.

\subsection{Model flow}
\label{subsec:ex_model}

The equation of motion for the particles, Eq.
\eqref{eq:eqmotion_noninertial}, relies on a fluid flow
$\bfv_\mathrm{fluid}(\bfX,t)$. For clarity, we choose this
velocity field to have zero mean integrated over space. Note,
however, that as long as the spatial distribution of the
particles is inhomogeneous, the vertical velocity averaged over
all particles will be different from $-W$ due to the
inhomogeneities of the velocity field \cite{Drotos2011}.

In order to present relevant phenomena in a clear way, we use a
$d=2$ model flow $\bfv_\mathrm{fluid}(\bfX,t)$ for our
numerical examples: we choose a modified version of the
paradigmatic double-shear flow \cite{Pierrehumbert1994}. In its
classical version, it is a periodic velocity field consisting
of a horizontal shear during the first half of the temporal
period and of a vertical shear during the other half. We modify
this in two aspects: First, we smooth the discontinuous
transition between the two orientations by introducing a
hyperbolic-tangent-type transition \cite{Vilela2007}. Second,
we rotate the shear directions by 45 degrees, to break the
coincidence of the two instantaneous velocity directions with
the horizontal and vertical axes, which in our sedimentation
setup have a very specific role. The resulting velocity field
is written as:
\begin{align}
v_{\mathrm{fluid},x}(\bfX,t) &= \frac{1}{\sqrt{2}} (v_{\mathrm{fluid},\xi}(\bfX,t)-v_{\mathrm{fluid},\eta}(\bfX,t)) , \\
v_{\mathrm{fluid},z}(\bfX,t) &= \frac{1}{\sqrt{2}} (v_{\mathrm{fluid},\xi}(\bfX,t)+v_{\mathrm{fluid},\eta}(\bfX,t)) ,
\label{eq:shearflow_main}
\end{align}
where
\begin{align}
v_{\mathrm{fluid},\xi}(\bfX,t) &= A (1+\tanh[\gamma\sin(2\pi t)]) \sin[\sqrt{2}\pi(z-x)] , \\
v_{\mathrm{fluid},\eta}(\bfX,t) &= A (1-\tanh[\gamma\sin(2\pi t)]) \sin[\sqrt{2}\pi(z+x)] .
\label{eq:shearflow_detail}
\end{align}
$\gamma = 20/\pi$ controls the temporal sharpness of the
shear-direction switching, it is fixed throughout the paper (as
well as the temporal period of the fluid, which is set to $1$).
$A$ is half of the amplitude of each elementary velocity
component (in what follows: the `amplitude'). By increasing $A$
we increase the strength of the flow and, as a consequence,
also its chaoticity, i.e., the (largest positive) Lyapunov
exponent, which is associated to the separation with time of
fluid particle trajectories. Note that the velocity field
\eqref{eq:shearflow_main}-\eqref{eq:shearflow_detail} is also
periodic in space, with a period of $\sqrt{2}$ in both $x$ and
$z$. For the trajectories, at variance with other
implementations of flows related to the double shear, we do not
impose any periodic boundary conditions, so that the particles'
positions evolve in the unbounded directions $x$ and $z$.

If we regard the accumulation level as the bottom of the domain
of a realistic fluid flow, the velocity field
$\bfv_\mathrm{fluid}(\bfX,t)$ would have to fulfill a no-flux
or even a no-slip boundary condition at $z = -\zacc$, which is
not satisfied by
\eqref{eq:shearflow_main}-\eqref{eq:shearflow_detail}.

As for the no-flux boundary condition, we
do not expect to introduce any qualitative difference compared to the results obtained in our example flow, since in all our theoretical
formulae the relevant quantity at the accumulation level
appears to be not the fluid velocity $\bfv_\mathrm{fluid}$, but
the particle velocity $\bfv$, which would not have a
vanishing vertical component. Indeed, we carried out our main analyses
in a different flow, namely a
spatially periodic sheared vortex flow with temporal modulation
\cite{Feudel2005,Lindner2017}, with accumulation levels fulfilling the no-flux boundary
condition, and obtained the very same qualitative results.

In principle, a viscous boundary layer with a so-slip boundary condition, or any kind of a separate flow regime at the bottom of the fluid with different characteristics compared to the bulk (e.g. length and time scales, magnitude of the velocity), cannot be excluded to leave an important, specific imprint on the qualitative properties of the accumulated particle density. However, with our assumptions and parameters, as well as in oceanic settings, the time that is spent by a particle in a given layer is mainly determined by the settling velocity $W$, independent of the flow, hence the effects of any boundary layer or separate flow regime are expected to be negligible if the boundary layer is thin compared to the bulk of the fluid (like in the ocean).

Beyond all of the above, in
experimental set-ups such as in sediment traps, the
accumulation points are not at the bottom of the sea, but at
some intermediate depth at which no boundary conditions
apply at all.



\subsection{Illustrative results}
\label{subsec:ex_ill}

We now present typical examples for the final density within
the accumulation level, and show how its form emerges from the
reshaping of the material line, which gives rise to the
different density-modifying contributions that have been
introduced in Section \ref{sec:formulae}. We always initialize,
at $t_0 = 0$, $10\,000$ particles at $z_0 = 0$ uniformly in a
line segment $x \in [0,1]$. (Note that any initial length of
the order of unity would suffice for our examples.) We follow
the particles' trajectories in the double-shear flow Eq.
\eqref{eq:shearflow_main}, and compute the relevant quantities
numerically (see Table \ref{tab:quantities}). When more than
one branch of the material line arrives at the same position (as
a result of folding), we additionally calculate the sum
$\sum\mathcal{F}$ of the total factors $\mathcal{F}$
corresponding to the individual branches. Furthermore, we
compare $\sum\mathcal{F}$ to a normalized histogram $h$
calculated directly from the arrival positions of the
individual particles.

We start with a parameter setting that does not produce
noticeable chaos, but leads to regular motion: the portrait of
the corresponding stroboscopic map consists of slightly
undulating quasi-vertical lines. However, the net horizontal
displacement of a trajectory after vertically traversing one
spatial period of the flow is not zero generally, it is just
very small.

\begin{figure}[h!]
	\subfloat{\label{fig:pos_ampl0.06_vsettl0.6_t4}\includegraphics{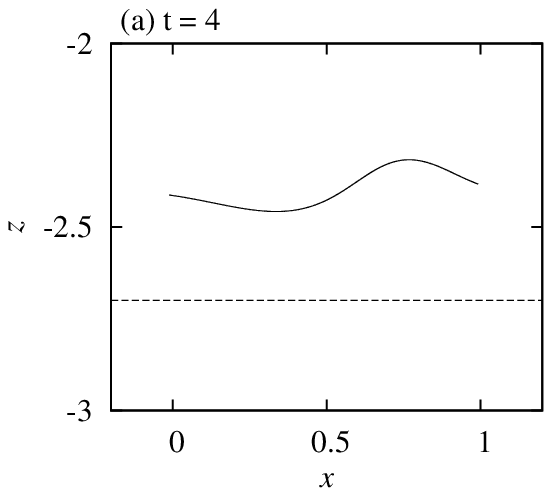}}
	\subfloat{\label{fig:pos_ampl0.06_vsettl0.6_t27}\includegraphics{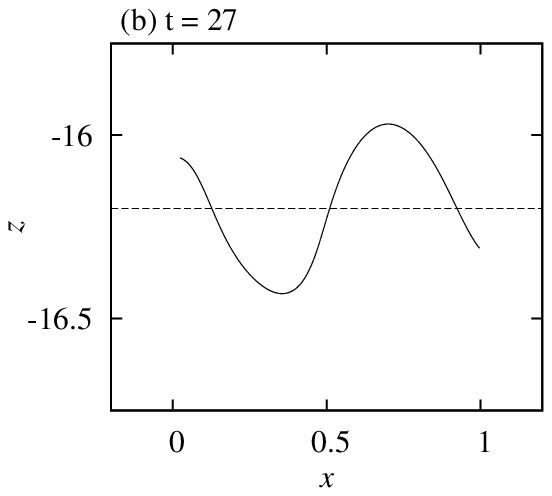}}
	\subfloat{\label{fig:pos_ampl0.06_vsettl0.6_t161}\includegraphics{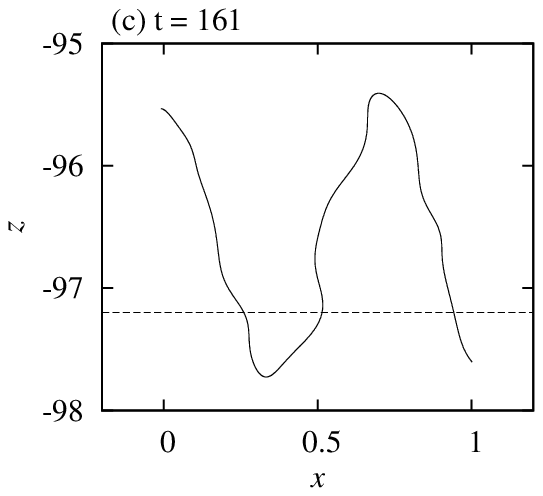}}
\caption{\label{fig:pos_ampl0.06_vsettl0.6}The positions of the particles of the initially horizontal
material line of unit length, at the indicated time instants. Dashed lines mark the accumulation levels
taken for Figs. \ref{fig:ampl0.06_vsettl0.6_disc-2.7_selisc0}, \ref{fig:ampl0.06_vsettl0.6_disc-16.2_selisc0},
and \ref{fig:ampl0.06_vsettl0.6_disc-16.2_selisc5}. $A = 0.06$, $W = 0.6$.}
\end{figure}
Snaphots from the time evolution of the line of particles are
shown in Fig. \ref{fig:pos_ampl0.06_vsettl0.6}. At the
beginning, both horizontal and relative vertical displacements
of neighboring particles remain small, and the line becomes
slightly undulated (Fig. \ref{fig:pos_ampl0.06_vsettl0.6_t4}).
Later on, relative vertical displacements become much larger,
see Fig. \ref{fig:pos_ampl0.06_vsettl0.6_t27}. When they become
large enough, it can happen that certain, more slowly falling,
parts of the line are folded above the faster parts, as can be
observed in Fig. \ref{fig:pos_ampl0.06_vsettl0.6_t161}. Such
folds, together with the nearly vertical velocity vector,
result in caustics after accumulation.

\begin{figure}[h!]
	\subfloat{\label{fig:densities_factor_total_foldsummed_with_histogram_ampl0.06_vsettl0.6_disc-2.7_selisc0}\includegraphics{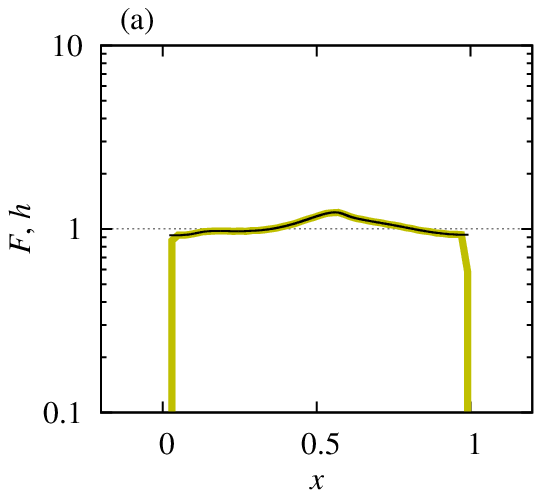}}
	\subfloat{\label{fig:densities_factors_ampl0.06_vsettl0.6_disc-2.7_selisc0}\includegraphics{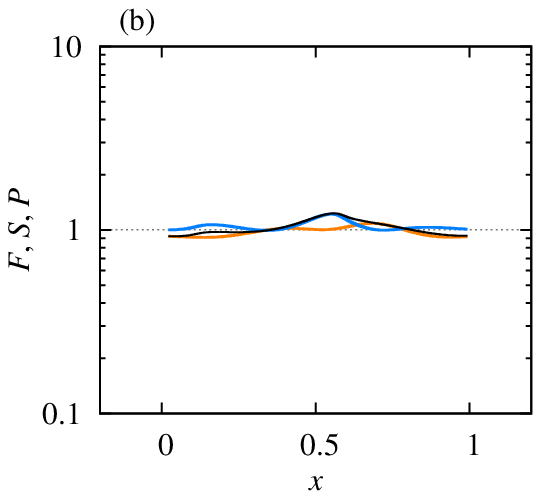}}
	\subfloat{\label{fig:densities_terms_ampl0.06_vsettl0.6_disc-2.7_selisc0}\includegraphics{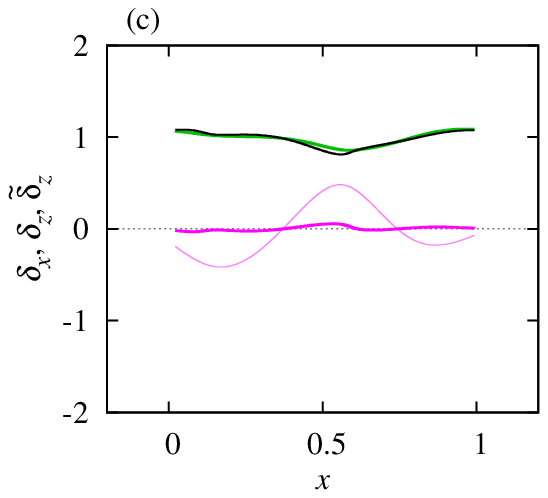}}
\caption{\label{fig:ampl0.06_vsettl0.6_disc-2.7_selisc0}(a) The total factor $\mathcal{F}$,
at the accumulation level $\zacc = 2.7$ (marked by a horizontal dashed line in Fig.
\ref{fig:pos_ampl0.06_vsettl0.6_t4}),
computed along the individual trajectories according to \eqref{eq:simple_F_t_u} (in black),
and the histogram $h$ (with bin size $0.02$) obtained from the positions of the trajectories
on the accumulation level (in dark yellow), both as a function of the position along the
accumulation level. (b) The total factor $\mathcal{F}$ (black) compared to the stretching factor
$\mathcal{S}$ (orange) and to the projection factor $\mathcal{P}$ (blue). (c) The reciprocal of
the total factor $\mathcal{F}$ (black) compared to the parametric derivative of the horizontal
position $\delta_x$ (green), to the parametric derivative of the vertical position $\delta_z$
(thin magenta) and to the weighted parametric derivative of the vertical position $\tilde{\delta}_z$
(thick magenta). See Table \ref{tab:quantities} to locate the corresponding formulae. $A = 0.06$,
$W = 0.6$.}
\end{figure}
Figure \ref{fig:ampl0.06_vsettl0.6_disc-2.7_selisc0} considers
the parameter setting of Fig. \ref{fig:pos_ampl0.06_vsettl0.6},
and shows the quantities of Table \ref{tab:quantities} for an
accumulation level placed at $z = -\zacc = -2.7$ (marked also
in Fig. \ref{fig:pos_ampl0.06_vsettl0.6_t4}). We can see in
Fig.
\ref{fig:densities_factor_total_foldsummed_with_histogram_ampl0.06_vsettl0.6_disc-2.7_selisc0}
that the total factor $\mathcal{F}$ computed along the
individual trajectories according to \eqref{eq:simple_F_t_u}
gives practically perfectly the same result as directly
calculating the histogram $h$ from the positions of the
trajectories on the accumulation level. The total factor does
not take a constant value of $1$, so that the density develops
\emph{inhomogeneities}, but only weak ones, which depend
smoothly on the position along the accumulation level.

In Fig.
\ref{fig:densities_factors_ampl0.06_vsettl0.6_disc-2.7_selisc0},
we can observe that the smooth ``undulation'' of the total
factor $\mathcal{F}$ originates from rather generic
``undulations'' of the stretching factor $\mathcal{S}$ and the
projection factor $\mathcal{P}$, the deviation of which from
$1$ is of similar magnitude as that of $\mathcal{F}$. A little
bit more interesting is Fig.
\ref{fig:densities_terms_ampl0.06_vsettl0.6_disc-2.7_selisc0},
which analyzes the contributions from the different terms in
the \emph{reciprocal} of the total factor $\mathcal{F}$. Due to
the weak horizontal displacements, a unit change along the
initial, \emph{horizontally oriented} material line segment
approximately results in a unit change along the accumulation
depth as well. As a consequence, the value of the parametric
derivative $\delta_x$ is near to $1$. The parametric derivative
$\delta_z$, however, deviates more from its initial value of
$0$, which indicates that relative vertical displacements are
stronger. Nevertheless, from the point of view of the density,
weighting this parametric derivative by $v_x/v_z$ to obtain
$\tilde{\delta}_z$ keeps the effect of stronger relative
vertical displacements small: the amplitude $A$ of the
fluctuating part of the velocity field, contributing alone to
$v_x$, is much smaller than $W$, which dominates in $v_z$.

\begin{figure}[h!]
	\subfloat{\label{fig:densities_factor_total_foldsummed_with_histogram_ampl0.06_vsettl0.6_disc-16.2_selisc0}\includegraphics{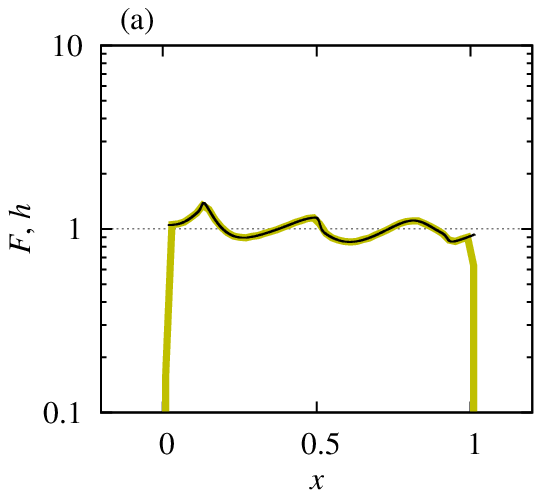}}
	\subfloat{\label{fig:densities_factors_ampl0.06_vsettl0.6_disc-16.2_selisc0}\includegraphics{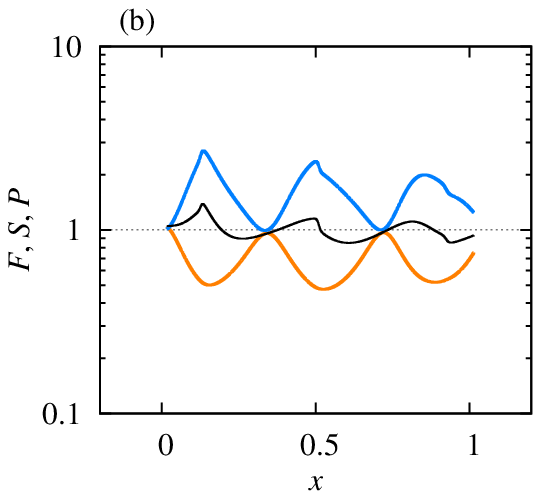}}
	\subfloat{\label{fig:densities_terms_ampl0.06_vsettl0.6_disc-16.2_selisc0}\includegraphics{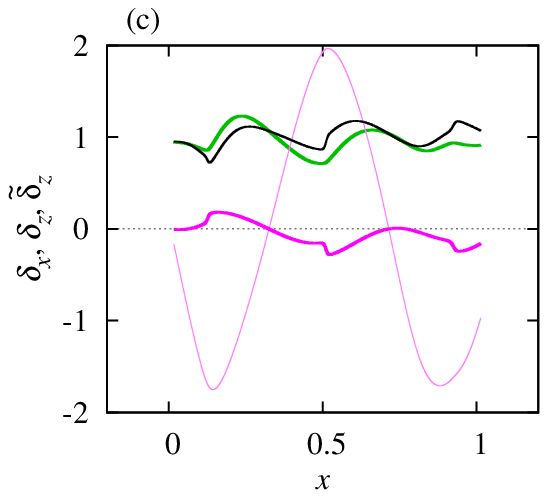}}
\caption{\label{fig:ampl0.06_vsettl0.6_disc-16.2_selisc0}Same as Fig.
\ref{fig:ampl0.06_vsettl0.6_disc-2.7_selisc0} for a deeper accumulation depth, $\zacc = 16.2$
(marked by a horizontal dashed line in Fig.
\ref{fig:pos_ampl0.06_vsettl0.6_t27}).}
\end{figure}
If we let the line fall more and prescribe $z = -\zacc = -16.2$
(seen in Fig. \ref{fig:pos_ampl0.06_vsettl0.6_t27}), the
``undulation'' of the total factor $\mathcal{F}$, of course,
becomes stronger, see Fig.
\ref{fig:densities_factor_total_foldsummed_with_histogram_ampl0.06_vsettl0.6_disc-16.2_selisc0}.
It may be surprising that stretching and projection both have
even much stronger effect, shown in Fig.
\ref{fig:densities_factors_ampl0.06_vsettl0.6_disc-16.2_selisc0},
but they are approximately anticorrelated, so that they more or
less cancel out each other (note that $\mathcal{S}$ and
$\mathcal{P}$ are multiplied in \eqref{eq:fact_sh_t_u}, and are
shown on a logarithmic scale in
\ref{fig:densities_factors_ampl0.06_vsettl0.6_disc-16.2_selisc0}).
Qualitatively, this can be understood as a result of the
relative vertical displacements being much larger than the
horizontal ones, which is clearly observable in Fig.
\ref{fig:pos_ampl0.06_vsettl0.6_t27}: both the stretching and
the tilting of the line result mainly from the vertical
deformation, and, when different parts of the line are
accumulated on the same level, with a nearly vertical velocity
($v_x/v_z$ is still small), this deformation is practically
removed, resulting in an approximately homogeneous horizontal
line after projection.

An open question is why horizontal displacements are much
smaller than vertical ones. Note that the amplitude of the
shear flow is the same in the vertical and the horizontal
directions. The phenomenon is certainly due to the symmetry
breaking introduced by the settling term in the velocity
\eqref{eq:eqmotion_noninertial}. We shall return to the
possible importance of this phenomenon in Section
\ref{subsec:pardep_A}.

Note in Fig.
\ref{fig:densities_factors_ampl0.06_vsettl0.6_disc-16.2_selisc0}
that stretching almost always dilutes the original density,
while projection almost always densifies it. While this already
follows from the geometry of the line in Fig.
\ref{fig:pos_ampl0.06_vsettl0.6_t27}, we can easily provide
with a more general explanation: On the one hand, anyhow we
initialize our material line segment, it will gradually align
with the stretching direction (as opposed to shrinking). As for
the projection, on the other hand, the simple horizontal
projection of an (in our case, curved) line is always shorter
than the original line. This can only be altered by a strong
horizontal velocity component, but $v_x/v_z$ is small here.

Figure
\ref{fig:densities_terms_ampl0.06_vsettl0.6_disc-16.2_selisc0}
confirms our visual observation (in Fig.
\ref{fig:pos_ampl0.06_vsettl0.6_t27}) of the strong vertical
and relatively weak horizontal deformation, and emphasizes the
importance of the smallness of $v_x/v_z$ in avoiding strong
modifications of the density.

\begin{figure}[h!]
	\subfloat{\label{fig:densities_factor_total_foldsummed_with_histogram_ampl0.06_vsettl0.6_disc-16.2_selisc5}\includegraphics{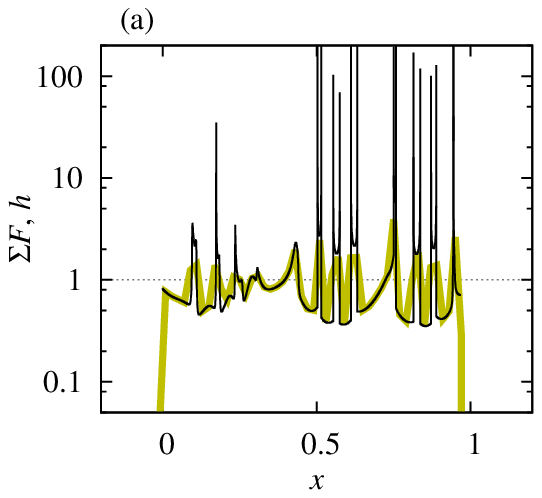}}
	\subfloat{\label{fig:densities_factors_ampl0.06_vsettl0.6_disc-16.2_selisc5}\includegraphics{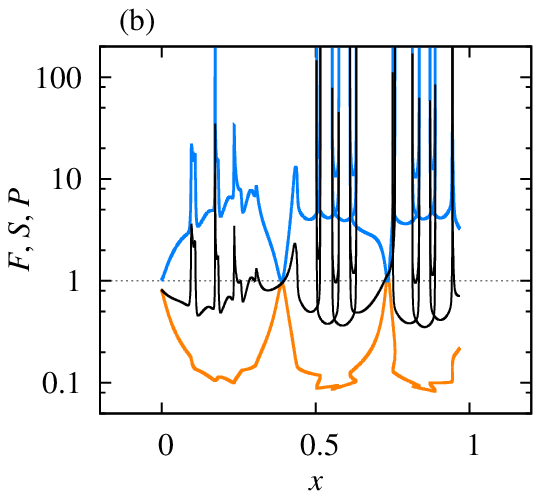}}
	\subfloat{\label{fig:densities_terms_ampl0.06_vsettl0.6_disc-16.2_selisc5}\includegraphics{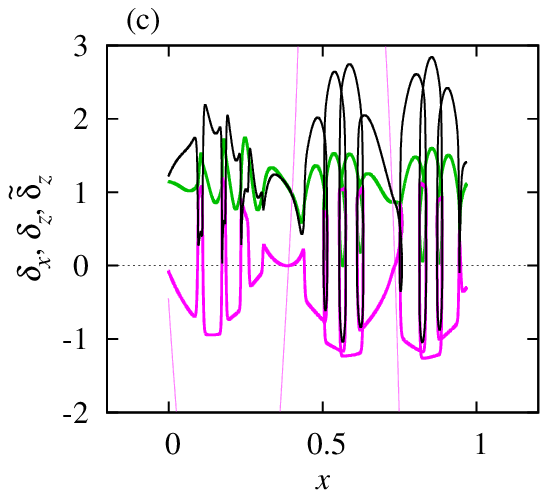}}
\caption{\label{fig:ampl0.06_vsettl0.6_disc-16.2_selisc5}Same as Fig.
\ref{fig:ampl0.06_vsettl0.6_disc-2.7_selisc0} for $\zacc = 97.2$ (marked by a horizontal dashed line in Fig.
\ref{fig:pos_ampl0.06_vsettl0.6_t161}), and panel (a) showing the total factor $\mathcal{F}$ summed
over the different branches of the material line segment.}
\end{figure}
In Fig. \ref{fig:ampl0.06_vsettl0.6_disc-16.2_selisc5},
accumulation is prescribed at an even deeper depth, $z = -\zacc
= -97.2$ (seen in Fig. \ref{fig:pos_ampl0.06_vsettl0.6_t161}).
As Fig. \ref{fig:pos_ampl0.06_vsettl0.6_t161} shows, the line
segment has undergone foldings by the time it reaches this
accumulation level. At the folding points (note that they occur
in pairs), where the tangent of the line coincides with the
local velocity (see Eq. \eqref{eq:caustic_withn}), caustics
appear: the projection factor $\mathcal{P}$ tends to infinity
(blue line in Fig.
\ref{fig:densities_factors_ampl0.06_vsettl0.6_disc-16.2_selisc5}),
and this is carried over also to the total factor $\mathcal{F}$
(black line in Fig.
\ref{fig:densities_factors_ampl0.06_vsettl0.6_disc-16.2_selisc5}).
To obtain the total density forming along the accumulation
level, the total factors $\mathcal{F}$ corresponding to each of
the branches of the material line have to be summed up, see
Fig.
\ref{fig:densities_factor_total_foldsummed_with_histogram_ampl0.06_vsettl0.6_disc-16.2_selisc5}.
Fig.
\ref{fig:densities_factor_total_foldsummed_with_histogram_ampl0.06_vsettl0.6_disc-16.2_selisc5}
also shows that our histogram $h$ is not able to resolve the
caustics and fine structures.

The novelty in Fig.
\ref{fig:densities_terms_ampl0.06_vsettl0.6_disc-16.2_selisc5}
is that even the weighted parametric derivative
$\tilde{\delta}_z$ grows to considerable magnitudes, this is
how it becomes possible that the sum
$\delta_x+\tilde{\delta}_z$ crosses zero, where the caustics
are found (see Eq. \eqref{eq:paracaustic}). The unweighted
parametric derivative $\delta_z$ is so large that it typically
does not fit to the scale of the plot, which concentrates on
the other quantities. Note also that $v_x$ changes sign near
the caustics, which is not surprising: this is how folds can
appear.

\begin{figure}[h!]
	\subfloat{\label{fig:pos_ampl0.25_vsettl0.6_t3}\includegraphics{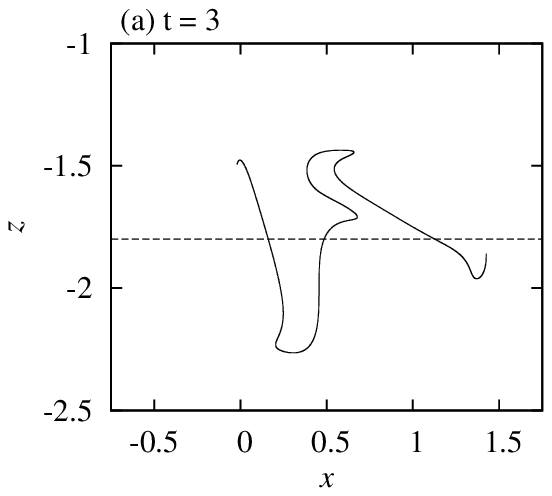}}
	\subfloat{\label{fig:pos_ampl0.25_vsettl0.6_t6}\includegraphics{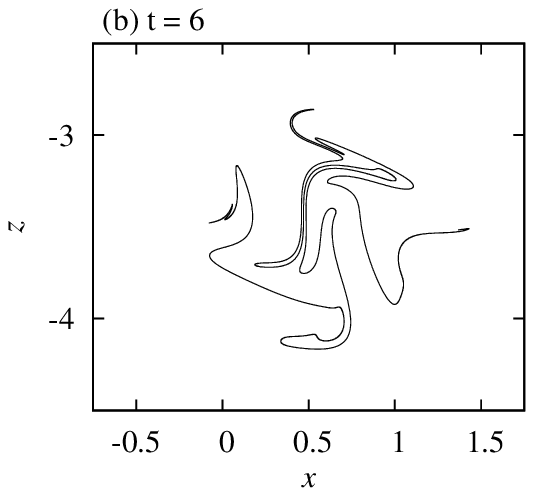}}
	\subfloat{\label{fig:pos_ampl0.25_vsettl0.6_t11}\includegraphics{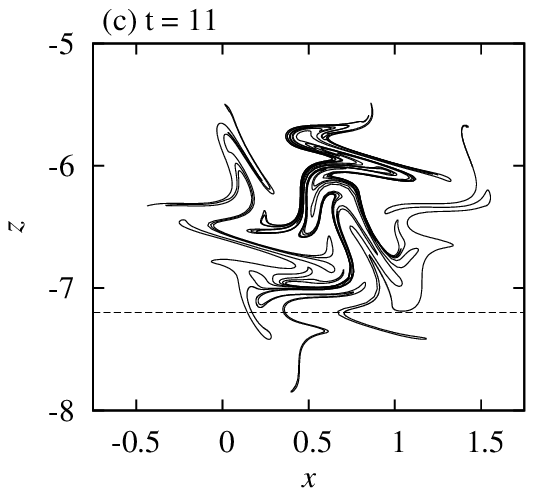}}
\caption{\label{fig:pos_ampl0.25_vsettl0.6}Same as Fig. \ref{fig:pos_ampl0.06_vsettl0.6}
for $A = 0.25$. The dashed lines in panels (a) and (c) mark the accumulation levels taken for Figs.
\ref{fig:ampl0.25_vsettl0.6_disc-1.8_selisc0} and \ref{fig:ampl0.25_vsettl0.6_disc-1.8_selisc3}.}
\end{figure}
Now we change our parameter setting to obtain a completely
chaotic case, when the phase portrait shows homogeneous mixing.
The time evolution of the geometry of the line of particles is
shown in Fig. \ref{fig:pos_ampl0.25_vsettl0.6}. Strong
stretching occurs at the very beginning, which is accompanied
soon by several folds (Fig.
\ref{fig:pos_ampl0.25_vsettl0.6_t3}). This is the situation
that resembles the most to our motivating example in Fig.
\ref{fig:benguela}. Later on, rather complicated structures
develop (Fig. \ref{fig:pos_ampl0.25_vsettl0.6_t6}). Finally,
the line follows finer and finer structures of the typical
fractal filamentation of chaos (Fig.
\ref{fig:pos_ampl0.25_vsettl0.6_t11}). Unlike in Fig.
\ref{fig:pos_ampl0.06_vsettl0.6}, anisotropy in the relative
displacement of neighboring particles is not obvious.

\begin{figure}[h!]
	\subfloat{\label{fig:densities_factor_total_foldsummed_with_histogram_ampl0.25_vsettl0.6_disc-1.8_selisc0}\includegraphics{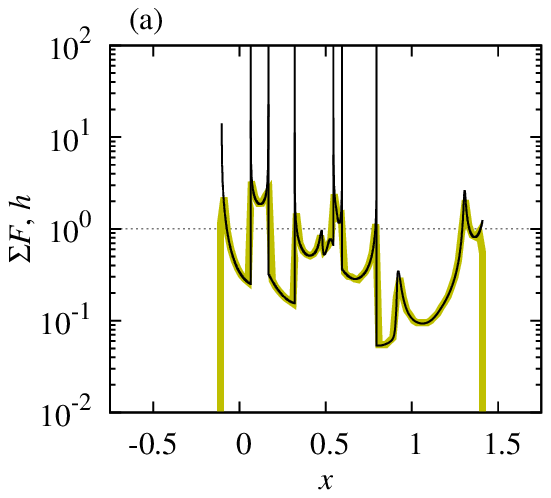}}
	\subfloat{\label{fig:densities_factors_ampl0.25_vsettl0.6_disc-1.8_selisc0}\includegraphics{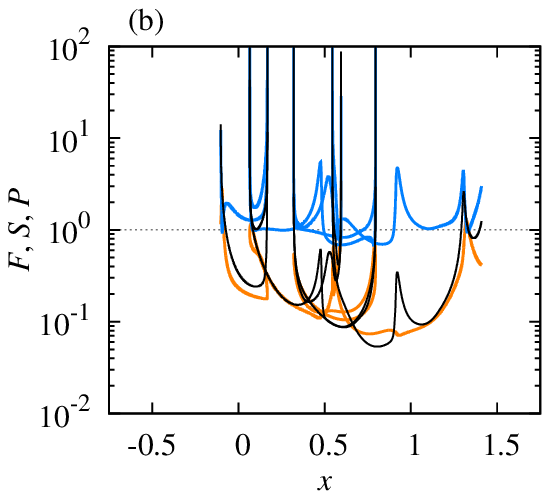}}
	\subfloat{\label{fig:densities_terms_ampl0.25_vsettl0.6_disc-1.8_selisc0}\includegraphics{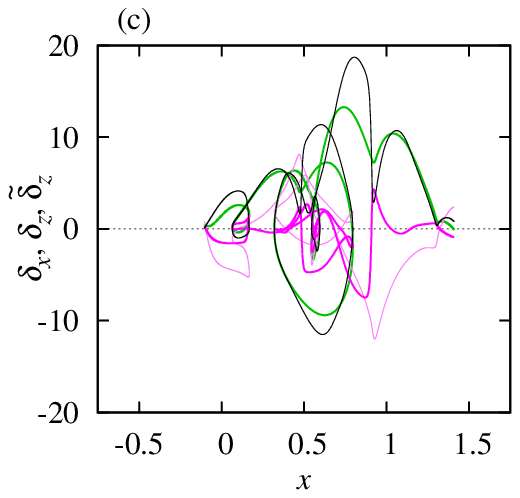}}
\caption{\label{fig:ampl0.25_vsettl0.6_disc-1.8_selisc0}Same as Fig.
\ref{fig:ampl0.06_vsettl0.6_disc-16.2_selisc5} for $A = 0.25$ and accumulation level $\zacc = 1.8$
(shown as a horizontal line in \ref{fig:pos_ampl0.25_vsettl0.6_t3}).}
\end{figure}
For an accumulation level that is reached during the early
stages of the development of the structures (see Fig.
\ref{fig:pos_ampl0.25_vsettl0.6_t3}), the summed total factor
$\sum\mathcal{F}$, as well as the histogram $h$, exhibits
considerable inhomogeneities, see Fig.
\ref{fig:densities_factor_total_foldsummed_with_histogram_ampl0.25_vsettl0.6_disc-1.8_selisc0}.
(In this case, they are only the peaks of the caustics that are
not resolved by $h$.) Therefore, we can say that chaos first
\emph{inhomogeneizes} the initially uniform distribution.

In Fig.
\ref{fig:densities_factors_ampl0.25_vsettl0.6_disc-1.8_selisc0},
we can observe that stretching dilutes, and projection
densifies, in accordance with our general argumentation that we
gave when discussing Fig.
\ref{fig:densities_factors_ampl0.06_vsettl0.6_disc-16.2_selisc0}.

In Fig.
\ref{fig:densities_terms_ampl0.25_vsettl0.6_disc-1.8_selisc0},
we can see that $\delta_x$ and $\delta_z$ have similar
magnitude, lacking the anisotropy observed in Figs.
\ref{fig:ampl0.06_vsettl0.6_disc-2.7_selisc0}-\ref{fig:ampl0.06_vsettl0.6_disc-16.2_selisc5}.
Furthermore, $\tilde{\delta}_z$ is not much smaller than
$\delta_z$, due to the relatively strong amplitude $A$ of the
flow compared to $W$.

\begin{figure}[h!]
	\subfloat{\label{fig:densities_factor_total_foldsummed_with_histogram_ampl0.25_vsettl0.6_disc-1.8_selisc3}\includegraphics{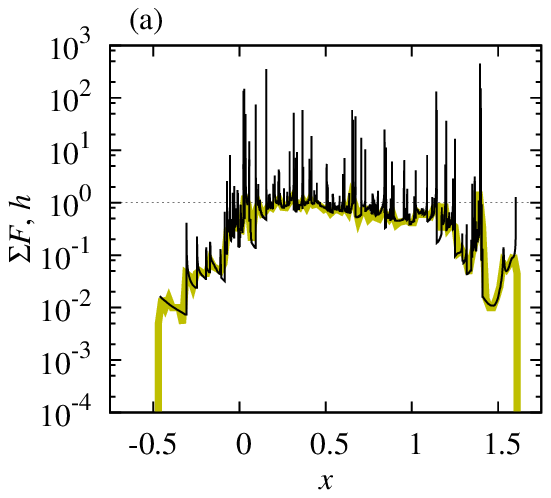}}
	\subfloat{\label{fig:densities_factors_ampl0.25_vsettl0.6_disc-1.8_selisc3}\includegraphics{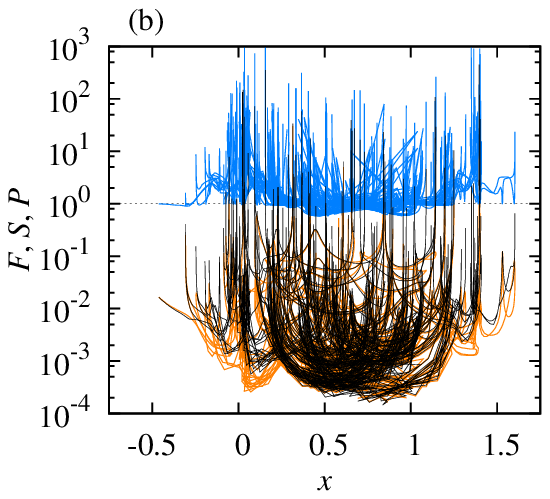}}
	\subfloat{\label{fig:densities_terms_ampl0.25_vsettl0.6_disc-1.8_selisc3}\includegraphics{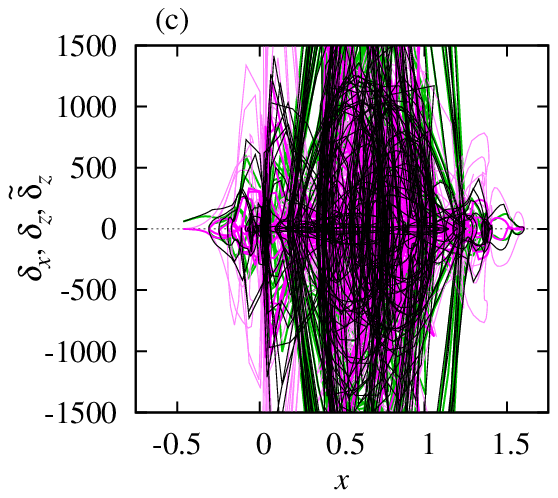}}
\caption{\label{fig:ampl0.25_vsettl0.6_disc-1.8_selisc3}Same as Fig.
\ref{fig:ampl0.25_vsettl0.6_disc-1.8_selisc0} for $A = 0.25$ and $\zacc = 7.2$ (shown as a horizontal line in
\ref{fig:pos_ampl0.25_vsettl0.6_t11}).}
\end{figure}
For an accumulation level placed at a depth where chaotic
filamentation is rather developed (see Fig.
\ref{fig:pos_ampl0.25_vsettl0.6_t11}), the summed total factor
$\sum\mathcal{F}$, shown in Fig.
\ref{fig:densities_factor_total_foldsummed_with_histogram_ampl0.25_vsettl0.6_disc-1.8_selisc3},
is composed of extremely many contributions from individual
branches, and thus exhibits extremely many corresponding
caustics. However, apart from the caustics and some fine-scale
structures, the resulting shape is quite simple: it is a single
bump. The histogram $h$, not being able, of course, to resolve
the caustics, only ``detects'' this bump. This coarse-grained
structure is resembling somewhat to the Gaussian that is
expected to appear for asymptotically long times, and clearly
indicates that chaos now \emph{homogeneizes} earlier
inhomogeneities (cf. Fig.
\ref{fig:ampl0.25_vsettl0.6_disc-1.8_selisc0}), unless the
density is investigated on a fine scale. Due to conservation of
mass and the not very enhanced horizontal extension, the bulk
of the bump is close to $1$ in Fig.
\ref{fig:densities_factor_total_foldsummed_with_histogram_ampl0.25_vsettl0.6_disc-1.8_selisc3}.

This is not so, however, before summing up the contributions
from the different branches, see the black line,
$\mathcal{F}$, in Fig.
\ref{fig:densities_factors_ampl0.25_vsettl0.6_disc-1.8_selisc3}.
It is clear that stretching (represented by $\mathcal{S}$,
orange line), which dilutes, ``wins'' as opposed to projection
(represented by $\mathcal{P}$, blue line), which densifies.
Chaos naturally involves very strong stretching. Projection,
however, originates from local geometric properties: it is
determined solely by the local orientation of the material line
and the local velocity of the fluid where and when a particle
of the line arrives at the accumulation depth. The local
orientation of the line is, after strong enough mixing,
practically random (with a possibly nonuniform distribution,
determined by the flow). After reaching this randomness, the
projection factor $\mathcal{P}$ will not grow any more, i.e.,
its magnitude saturates. At the same time, stretching can
always grow. Note, however, that the projection factor
$\mathcal{P}$ should also increase at the beginning.

In this last, chaotic case, the parametric derivatives of the
horizontal and vertical positions behave in a completely
irregular way, see Fig.
\ref{fig:densities_terms_ampl0.25_vsettl0.6_disc-1.8_selisc3}.
The only conclusion that we can draw from Fig.
\ref{fig:densities_terms_ampl0.25_vsettl0.6_disc-1.8_selisc3},
based on some breaks in some lines, is that the resolution of
the filamentation is reaching its limit with the current number
of the particles.

\section{Systematic study of parameter dependence}
\label{sec:pardep}

It is an interesting question how inhomogeneities and the
underlying effects, as presented in the previous Section,
respond to changes in the parameters. For characterizing the
basic properties of the quantities analyzed in Section
\ref{subsec:ex_ill}, we evaluate their average and standard
deviation along the accumulation level. The former gives the
net effect (dilution or densification), while the latter
characterizes the strength of inhomogeneities. We emphasize
that evaluation along the accumulation level means that we
evaluate the statistics with respect to horizontal length, but
we do not sum up over possible different branches of the
material line that are present at the same point within the
accumulation level (except for the summed total factor
$\sum\mathcal{F}$ and the normalized histogram $h$, the
definition of which implies summation). In particular, the
average of a quantity $\phi$, where $\phi$ is either
$\mathcal{S}$, $\mathcal{P}$, $\mathcal{F}$, $\delta_x$,
$\delta_z$ or $\tilde{\delta}_z$, is obtained as
\BE
\langle \phi \rangle = \sum_{i=1}^{n} \frac{1}{x_i-x_{i-1}} \int_{x_{i-1}}^{x_i} \phi \,\rmd x ,
\label{eq:average}
\EE
where $x$ is the horizontal coordinate along the accumulation
level, $x_1$ and $x_n$ are the positions where the beginning
and the endpoint of the material line reach the accumulation
level, respectively, $x_i$ for $i \in \{2,\ldots,n-1\}$ are the
positions of the caustics where the line undergoes a fold (note
that $x_{i+1} < x_i$ if $x_i > x_{i-1}$ and vice versa, which
implies that $n$, and also the number of the caustics, $n-2$,
are even). The formula for the standard deviation is similar.
Beyond averages and standard deviations, we shall also consider
the number of the caustics.

We mention here that the standard deviation is not well-defined
if caustics are present. Since caustics, as mentioned, are
$1/\sqrt{x}$-type singularities in the density, the integral
over their square, $1/x$, does not remain finite. Indeed, we
numerically found that the standard deviation calculated from a
normalized histogram $h$ grows approximately as $-\log\,\Delta
x$ with the bin size $\Delta x$ of the histogram whenever the
number of the caustics is greater than zero (should it be $2$
or several thousands), which is a characteristic of numerical
integrals over $1/x$. Therefore, for such parameter values, we
plot the standard deviation of the normalized histogram $h$,
separately for a smaller and for a larger bin size, and do not
show the numerically obtained standard deviation for
$\sum\mathcal{F}$, $\mathcal{F}$ or $\mathcal{P}$. When
calculating $h$ for different parameter values, we keep the
number $N$ of the bins constant (instead of the bin size), and
indicate $N$ in the lower index of $h$ as $h_N$.

We concentrate on the dependence on the following parameters:
the settling velocity $W$, the accumulation depth $\zacc$, and
the amplitude of the fluctuating part of the velocity field,
which corresponds to $A$ in Eq. \eqref{eq:shearflow_detail}.
An important combination of these quantities is $\tau =
\zacc/W$, which is roughly proportional to the time needed for
the material line to reach the accumulation level.
The difference of $\tau$ from the actual settling time, when
averaged over the particles, is due to the inhomogeneous
spatial distribution of the particles, as explained in Section
\ref{subsec:ex_model}. Note, furthermore, that different parts
of the material line reach the accumulation level at different
times. This results in a smearout of the settling time along
the line, which is also reflected in the properties that we
investigate. As the vertical extension of the material line
grows in time, or with increasing depth, the importance of this
phenomenon also increases. At the beginning, the spread of the
settling time is smaller than the characteristic time scale of
the flow (i.e., unity), but later on it grows above this
characteristic time. In spite of all this, the time $\tau$
gives a good guidance for the interpretation of what can be
observed.

\subsection{Dependence on the settling velocity $W$}
\label{subsec:pardep_W}

\begin{figure}[h!]
	\subfloat{\label{fig:comp_average_factors_ampl0.07_iscpervsettl-12.5}\includegraphics{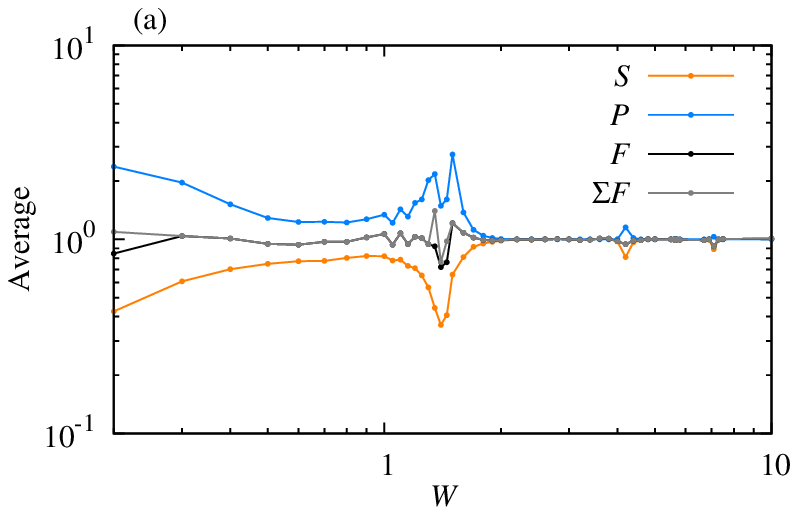}}
	\subfloat{\label{fig:comp_stddeviation_factors_ampl0.07_iscpervsettl-12.5}\includegraphics{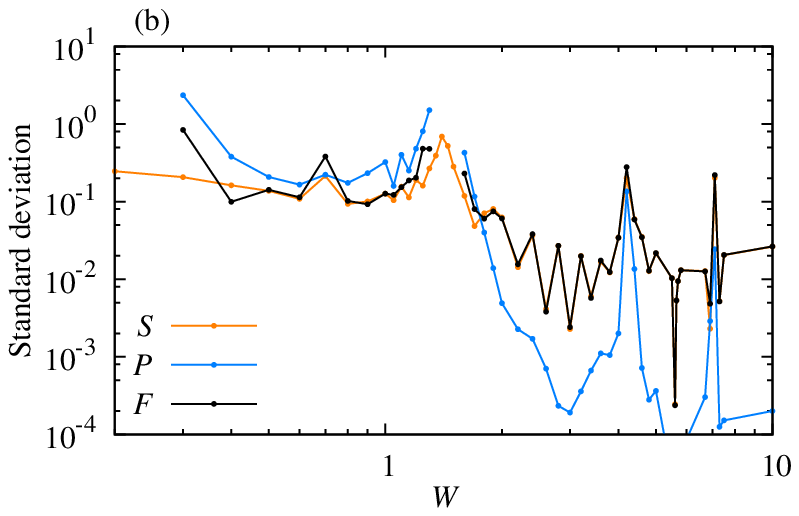}}\\
	\subfloat{\label{fig:comp_average_terms_ampl0.07_iscpervsettl-12.5}\includegraphics{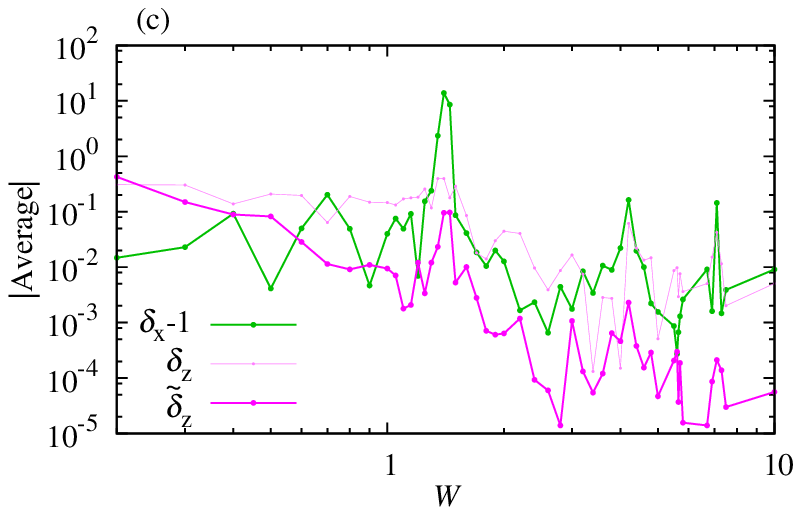}}
	\subfloat{\label{fig:comp_stddeviation_terms_ampl0.07_iscpervsettl-12.5}\includegraphics{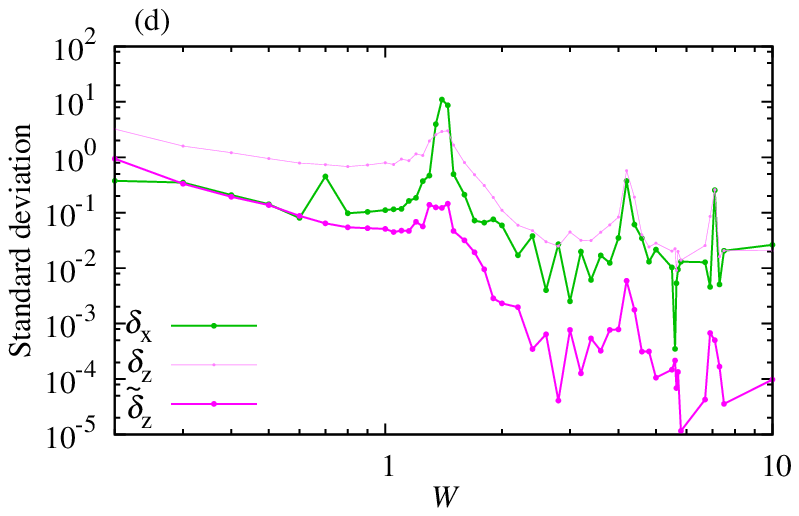}}
\caption{\label{fig:comp_ampl0.07_iscpervsettl-12.5}(a) The average and (b) the standard deviation, as a function of the settling velocity $W$, of the stretching factor $\mathcal{S}$, of the projection factor $\mathcal{P}$, and of the total factor $\mathcal{F}$, the latter also summed up over the different branches of the line in panel (a) (for the chosen values of $\tau$ and $A$, the result mostly coincides with the non-summed-up factor). (c)-(d) The same for the parametric derivative of the horizontal position $\delta_x$ (minus $1$ for comparability), the parametric derivative of the vertical position $\delta_z$, and the weighted parametric derivative of the vertical position $\tilde{\delta}_z$. $\tau = \zacc/W = 12.5$ is kept constant. The amplitude of the shear flow is $A = 0.07$.}
\end{figure}
Keeping $\zacc$ (and $A$) constant for increasing $W$ leads to
a decrease in $\tau$ and a corresponding weakening in both the
average and the standard deviation of all effects under
investigation, since they have less time available to act on
the material line. However, when keeping $\tau$ constant for
increasing $W$, we can still experience a reduction in all
effects represented by $\mathcal{S}$, $\mathcal{P}$,
$\delta_x$, $\delta_z$ and $\tilde{\delta}_z$, hence also in
the total factor $\mathcal{F}$, as the example in Fig.
\ref{fig:comp_ampl0.07_iscpervsettl-12.5} illustrates. This is
due to the fact that particles falling faster experience the
inhomogeneities of the shear flow at a higher frequency, as a
result of which these inhomogeneities average out (similarly as
shown in \cite{Bezuglyy2006} and then applied in
\cite{Gustavsson2014} in a damped noisy setting). We have found
this phenomenon to be present independently of the amplitude
$A$, including whether the flow is observed to be chaotic or
not.

An additional feature in Fig.
\ref{fig:comp_ampl0.07_iscpervsettl-12.5} is the presence of
resonances at $W = \sqrt{2}$ and its odd multiples, affecting
some small neighborhood around these values. $W = \sqrt{2}$
corresponds to a special case when the vertical displacement
that would arise from the settling velocity $W$ alone during
one time period of the shear flow (taken to be unity in
\eqref{eq:shearflow_detail}) coincides with the spatial period
$\sqrt{2}$ of the flow in the $z$ coordinate. But this is a
phenomenon very specific to the choice of our kinematic flow,
and would not exist in a generic, spatially or temporally
nonperiodic flow.

For the smallest value, $W = 0.2$, and near $W = \sqrt{2}$,
caustics and more than one branch is present for the setting
of Fig. \ref{fig:comp_ampl0.07_iscpervsettl-12.5}. As a
consequence, the average of the total factor $\mathcal{F}$ is
not the same as that of its summed up version $\sum\mathcal{F}$
(Fig.
\ref{fig:comp_average_factors_ampl0.07_iscpervsettl-12.5}), and
their standard deviation, as well as that of the projection
factor $\mathcal{P}$, is not defined (in Fig.
\ref{fig:comp_stddeviation_factors_ampl0.07_iscpervsettl-12.5}).

An interesting observation in Fig. \ref{fig:comp_stddeviation_factors_ampl0.07_iscpervsettl-12.5} is that the effect of projection becomes completely negligible compared to that of stretching (for the standard deviation at least) for increasing $W$. Without being able to provide an explanation, we cannot judge to what extent this property is universal, but it becomes clear that
one effect can be more important than the other in some situations.

\subsection{Dependence on the depth $\zacc$}
\label{subsec:pardep_zacc}

The dependence on $\zacc$, when keeping $W$ and $A$ constant,
is composed of two ``signals'' for each quantity. One
corresponds to the spatial and the temporal periodicity of the
flow, causing quasiperiodic oscillations in the strengths of
the investigated effects on fine scales. While quasiperiodic
oscillations would not be present in a generic flow, the
presence of fluctuations is natural. The other ``signal'', more
relevant for us, is a much smoother trend observable on coarser
scales. We shall concentrate on the coarse-grained behavior.
Generally, the individual effects become stronger with
increasing $\zacc$, which results from the longer time
available for them to act (this time is roughly proportional to
$\tau$). At the same time, there are several nontrivialities,
and we shall highlight the main points on the example of a
chaotic case. The regular case is qualitatively similar with
different functional forms, and a full description of both
cases is found in Appendix \ref{sec:pardep_zacc_detailed}.

\begin{figure}[h!]
	\subfloat{\label{fig:comp_average_factors_ampl0.25_vsettl0.6}\includegraphics{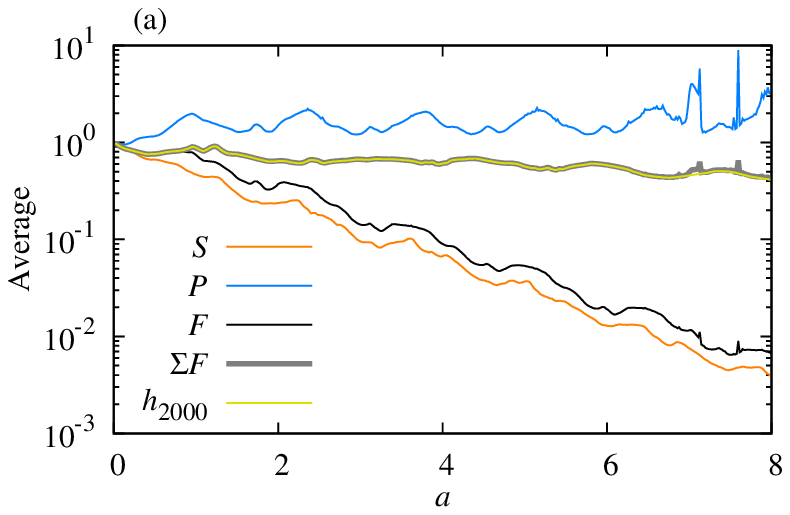}}
	\subfloat{\label{fig:comp_stddeviation_factors_ampl0.25_vsettl0.6}\includegraphics{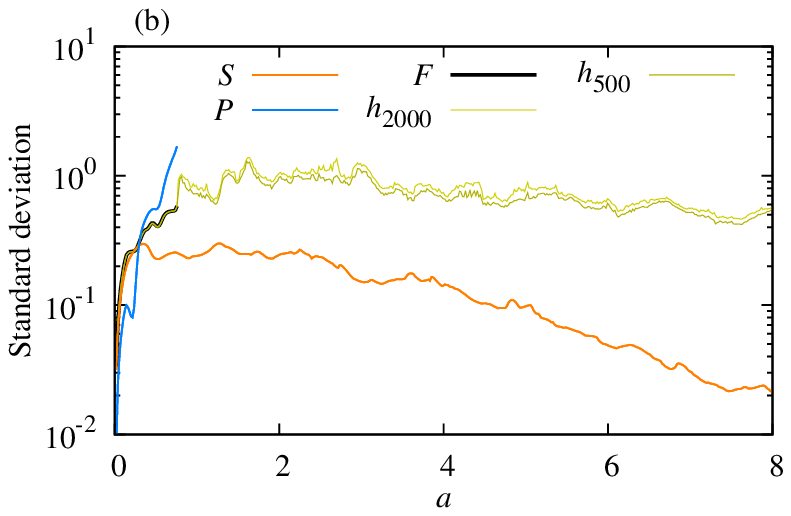}}
\caption{\label{fig:comp_ampl0.25_vsettl0.6}(a) The average and (b) the standard deviation, as a function of the accumulation depth $\zacc$, of the indicated quantities. $W = 0.6$ is kept constant, as well as the amplitude $A = 0.25$ of the shear flow.}
\end{figure}
The chaotic example is presented in Fig.
\ref{fig:comp_ampl0.25_vsettl0.6} (we note that individual
spikes in the plots for $\zacc > 6$ are numerical artefacts due
to the presence of caustics). Fig.
\ref{fig:comp_average_factors_ampl0.25_vsettl0.6} indicates
that both $\langle \mathcal{S} \rangle$ and $\langle
\mathcal{F} \rangle$ decrease exponentially as a function of
the depth $\zacc$. For the stretching, this can be regarded as
a direct consequence of chaos, taking into account that depth
is roughly proportional to settling time. The average $\langle
\mathcal{P} \rangle$ of the projection factor increases only
moderately (possibly related to saturational effects), and this
is why the stretching behavior determines the total factor,
resulting in an exponential dependence. Still, if we sum up the
total factor over line branches, its average $\langle
\sum\mathcal{F} \rangle$ remains approximately constant (see
the gray line in Fig.
\ref{fig:comp_average_factors_ampl0.25_vsettl0.6}) because of
mass conservation. That is, in spite of the strong reshapement
of the material line, there is no net densification or much net
dilution. Only a slight dilution is observable, for larger
$\zacc$, when the flow is able to advect parts of the material
line horizontally farther away from the unit-sized horizontal
section in which it was initialized. Calculating the average
$\langle h_N \rangle$ of the normalized histogram $h_N$ yields
the same result as $\langle \sum\mathcal{F} \rangle$ (for a
wide range of bin numbers $N$, of which only one is shown),
except for the absence of the artificial individual spikes.

As for the inhomogeneities, Fig.
\ref{fig:comp_stddeviation_factors_ampl0.25_vsettl0.6} shows
that the total factor $\mathcal{F}$ exhibits an increasing
inhomogeneity with increasing depth $\zacc$, for small $\zacc$,
with a slight slowing down in the rate before the first
caustics appear and it is not meaningful to continue the graph.
This behavior seems to result from those of the stretching
factor $\mathcal{S}$ and the projection factor $\mathcal{P}$,
with equal importance. In particular, the standard deviation of
the projection factor $\mathcal{P}$ is also increasing as long
as it exists, due to the access to more and more different
degrees of tiltness. Since different parts of the (phase) space
are stretched differently, an even sharper increase is
observable at the very beginning for the stretching factor
$\mathcal{S}$, too. This increase, however, levels off very
soon, and then turns to a pronounced decrease. This might be
related to the decreasing magnitude of the factor $\mathcal{S}$
itself in average, but also to the stretching of any given part
of the line being influenced by more and more parts of the
(phase) space, more and more homogeneously. Maybe this mixing
is why the standard deviation of the normalized histogram $h_N$
also starts decreasing for deeper depths $\zacc$, for any
particular choice $N$ of the coarse-graining (though taking on
slightly different values depending on the number $N$ of the
bins). At the same time, the coarse-grained quantifier $h_N$
follows, of course, the behavior of the total factor
$\mathcal{F}$ very accurately for small depths $\zacc$, i.e.,
it detects the inhomogeneization, and does not depend on the
choice of $N$.

To summarize, both in chaotic and regular settings (see
Appendix \ref{sec:pardep_zacc_detailed}), the emergence of the
inhomogeneities mostly takes place at the beginning of the
settling process (observable for small accumulation depths
$\zacc$). As soon as caustics appear, the standard deviation
diverges, but any given coarse-graining exhibits
homogeneization on the long term (for large $\zacc$).

\begin{figure}[h!]
	\subfloat{\label{fig:comp_zero_count_ampl0.07_vsettl0.6}\includegraphics{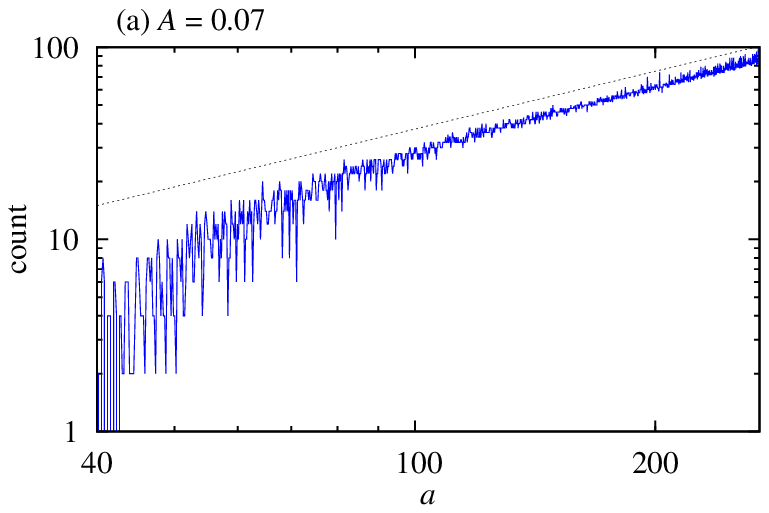}}
	\subfloat{\label{fig:comp_zero_count_ampl0.25_vsettl0.6}\includegraphics{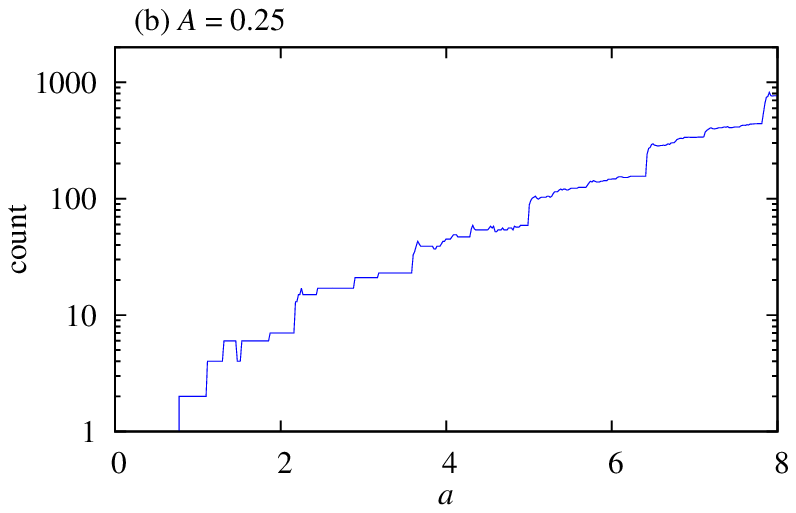}}
\caption{\label{fig:comp_zero_count_vsettl0.6}The number of the caustics, as a function of the accumulation depth $\zacc$. For comparison, a linear law is shown in panel (a). $W = 0.6$, and the amplitudes $A$ are as indicated.}
\end{figure}

It is important to point out that the number of the caustics,
besides the usual fluctuations (which are present in any
statistics of any quantity), increases without bounds: it
increases linearly and exponentially as a function of the depth
$\zacc$ in the regular and the chaotic case, respectively, see
Figs. \ref{fig:comp_zero_count_ampl0.07_vsettl0.6} and
\ref{fig:comp_zero_count_ampl0.25_vsettl0.6}). This indicates
that inhomogeneization always continues on infinitely small
spatial scales, due to the perpetual mixing of the (phase)
space.

\subsection{Dependence on the amplitude $A$, and a balance between stretching and projection}
\label{subsec:pardep_A}

Even in the presence of strong oscillations as a function of
$\zacc$, and of resonances as a function of $W$, the dependence
on the amplitude $A$ of the shear flow can be investigated
without being influenced by the spatial and temporal
periodicities. We concentrate here on the factors
$\mathcal{S}$, $\mathcal{P}$ and $\mathcal{F}$.

\begin{figure}[h!]
	\subfloat{\label{fig:comp_average_factors_vsettl0.6_isc-7.50}\includegraphics{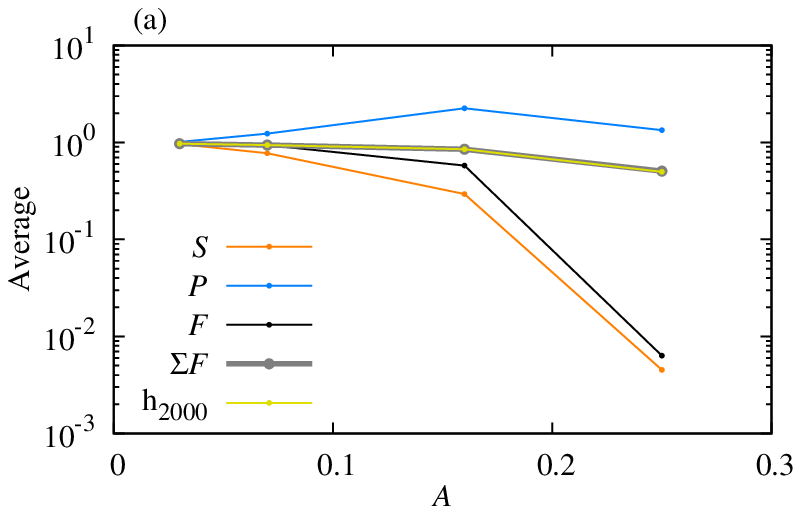}}
	\subfloat{\label{fig:comp_stddeviation_factors_vsettl0.6_isc-7.50}\includegraphics{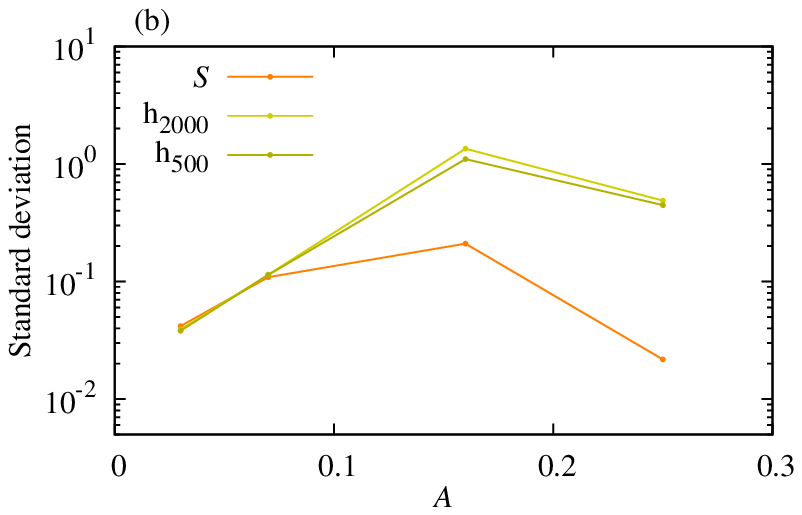}}
\caption{\label{fig:comp_vsettl0.6_isc-7.50}(a) The average and (b) the standard deviation, as a function of the amplitude $A$ of the shear flow, of the indicated quantites. $W = 0.6$ is kept constant, as well as $\tau = 12.5$.}
\end{figure}
As Fig. \ref{fig:comp_average_factors_vsettl0.6_isc-7.50}
illustrates, the average stretching factor $\langle \mathcal{S}
\rangle$ exhibits a stronger response to increasing amplitude
than the average projection factor $\langle \mathcal{P}
\rangle$, resulting in a net dilution in the average total
factor $\langle \mathcal{F} \rangle$. The projection factor
$\langle \mathcal{P} \rangle$ actually saturates and even turns
to a little (unexplained) weakening. Mass conservation keeps
the total factor, when summed up over the different branches,
approximately constant on average, with a slight dilution for
larger $A$, like in Fig.
\ref{fig:comp_average_factors_ampl0.25_vsettl0.6} for larger
$\zacc$. In total, the dependences in Fig.
\ref{fig:comp_average_factors_vsettl0.6_isc-7.50} are
remarkably similar to those in Fig.
\ref{fig:comp_average_factors_ampl0.25_vsettl0.6} (and Fig.
\ref{fig:comp_average_factors_ampl0.07_vsettl0.6} as well).

Due to the presence of caustics, only the stretching factor
$\mathcal{S}$ and the coarse-grained normalized histograms
$h_N$ are presented in Fig.
\ref{fig:comp_stddeviation_factors_vsettl0.6_isc-7.50}, which
shows the standard deviations, and there is an inevitable
dependence on $N$. Nevertheless, the matching with the
corresponding figures showing the dependence on the depth
(Figs. \ref{fig:comp_stddeviation_factors_ampl0.25_vsettl0.6}
and Fig.
\ref{fig:comp_stddeviation_factors_ampl0.07_vsettl0.6}) is also
remarkably good. We conclude that increasing the amplitude $A$
has a similar effect as increasing the depth $\zacc$, which is
a consequence of a stronger rearrangement of the material in
both cases.

Note in Fig. \ref{fig:comp_average_factors_vsettl0.6_isc-7.50} that $\langle \mathcal{F} \rangle$ seems to tend to $1$ for decreasing $A$ more quickly than $\langle \mathcal{S} \rangle$ and $\langle \mathcal{P} \rangle$ themselves, and, in particular, it practically never goes \emph{above} $1$. That is, for decreasing $A$ it might not be permitted to happen that projection would start to dominate stretching, even though this would be the inverse of what happens for increasing $A$. The prohibition of projection ``winning'' is also indicated by the experience that stretching and projection tend to be anticorrelated for decreasing $A$ (not shown, but cf. Fig. \ref{fig:densities_factors_ampl0.06_vsettl0.6_disc-2.7_selisc0} and the related discussion). All this would mean that stretching and projection would become exactly balanced (cancelling out each other) in the limit of small $A$. The existence of this balancing limit would also imply that \emph{stretching and projection effects cannot occur without each other} even for arbitrarily small perturbations of a uniform flow.

The existence of this balancing limit between stretching and
projection, with a vanishing effect on the final density, might
be a plausible assumption based on the observation that the
horizontal displacements are much smaller than the vertical
ones for the rather small amplitude $A$ of Figs.
\ref{fig:densities_terms_ampl0.06_vsettl0.6_disc-2.7_selisc0}-\ref{fig:densities_terms_ampl0.06_vsettl0.6_disc-16.2_selisc5}.
Furthermore, taking into account the similar dependence on the
amplitude $A$ and the depth $\zacc$, and the smaller magnitude
of $\langle \mathcal{F} \rangle$ in Figs.
\ref{fig:comp_average_factors_ampl0.25_vsettl0.6} and
\ref{fig:comp_average_factors_ampl0.07_vsettl0.6} than $\langle
\mathcal{S} \rangle$ or $\langle \mathcal{P} \rangle$, a
similar balancing might also occur for small depths $\zacc$.
However, note in Fig.
\ref{fig:densities_factors_ampl0.06_vsettl0.6_disc-2.7_selisc0}
that stretching and projection do not cancel out each other
\emph{locally} in space, and Fig.
\ref{fig:comp_stddeviation_factors_ampl0.07_iscpervsettl-12.5}
also suggests that stretching can be more important than
projection in certain setups with \emph{weak} mixing. As the
reason for the anisotropy in the magnitudes of the relative
displacements (e.g. in Figs.
\ref{fig:densities_terms_ampl0.06_vsettl0.6_disc-2.7_selisc0}-\ref{fig:densities_terms_ampl0.06_vsettl0.6_disc-16.2_selisc5})
is unexplained, we are not able to draw solid conclusions about
the existence of a balancing limit between stretching and
projection.


\section{Discussion and conclusion}
\label{sec:concl}

In this paper, we have described the mechanisms, stretching and
projection, that give rise to inhomogeneities in the density of
a layer of noninertial particles when the particles, after
falling in a $d$-dimensional fluid flow (such that the particle
velocity field is incompressible), are accumulated on a
particular level. We have explored in an example flow how
different characteristics of the accumulated density depend on
generic parameters.

In order to check if our numerical observations are generic, we
carried out the main analyses in a different example flow, a
spatially periodic sheared vortex flow with temporal modulation
\cite{Feudel2005,Lindner2017}. Considering a completely chaotic
case in the presence of temporally periodic modulation, we
found all of our qualitative conclusions to hold. What is even
more important, breaking temporal periodicity did not introduce
any relevant alteration.

The emergence of inhomogeneities from a homogeneous
distribution, pointed out in this context first in
\cite{Monroy2017}, might be surprising in volume-preserving
flows. In the special setting when the initial conditions are
distributed in a $(d-1)$-dimensional subset of the
$d$-dimensional domain of the fluid flow, the
$(d-1)$-dimensional density defined along the evolving subset
is not conserved. However, this fact cannot be regarded as the
basic source of inhomogeneity. If we have a full-dimensional
set of initial conditions, distributed over a continuous range
of levels, we can apply the results of this paper to the
horizontal layers, and then integrate over the initial vertical
coordinate to obtain the final, $(d-1)$-dimensional density
after the accumulation --- and inhomogeneities arising within
the individual layers are expected to be carried over to this
final density. As a conclusion, we can say that the finite
support of an (otherwise homogeneous) initial distribution is
at the origin of the observed inhomogeneities at the
accumulation level (without any coarse-graining, cf. Section \ref{sec:intro}), but the mechanism by which they develop
involves the stretching and the projection processes described
above.

\begin{figure}[h!]
\includegraphics{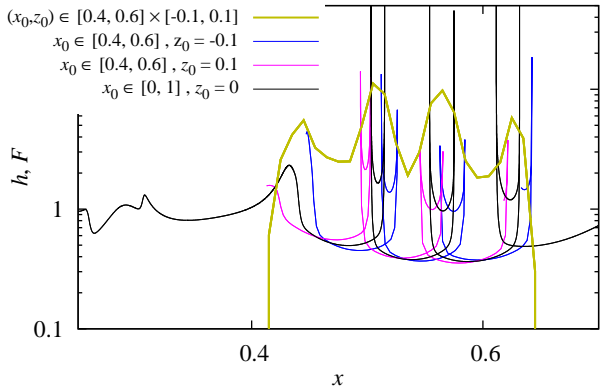}
\caption{\label{fig:histogram_densities_smallbox}The normalized histogram $h$ resulting from the small square $(x_0,z_0) \in [0.4,0.6]\times[-0.1,0.1]$ of initial conditions (green), the total factor $\mathcal{F}$ corresponding to the lowest and the highest rows of the small square ($z_0 = -0.1$ and $0.1$, blue and magenta, respectively), and the total factor $\mathcal{F}$ corresponding to an initial line segment of unit length at $z_0 = 0$ (the black line of Fig. \ref{fig:densities_factors_ampl0.06_vsettl0.6_disc-16.2_selisc5}, black here, too), as a function of the position along the accumulation level. The bin size of the histogram is $0.01$. $A = 0.06$, $W = 0.6$, $\zacc = 97.2$.}
\end{figure}

To illustrate the above considerations, we first show
that a $2$-dimensional set of initial conditions in the
$2$-dimensional shear flow problem (see Section
\ref{subsec:ex_model}) also leads to inhomogeneities in the
accumulated density. In particular, we distribute $10\,201$
initial conditions on a uniform grid in a small square,
$(x_0,z_0) \in [0.4,0.6]\times[-0.1,0.1]$. We numerically
approximate the resulting accumulated density by calculating a
normalized histogram $h$, which is shown in Fig.
\ref{fig:histogram_densities_smallbox} and is clearly
inhomogeneous. For comparison, we also include the total
factors $\mathcal{F}$ (not summed up over the different
branches) that come from the lowest and the highest rows of the
initial square ($z_0 = -0.1$ and $0.1$, respectively), as well
as that corresponding to the total factor in Fig.
\ref{fig:densities_factors_ampl0.06_vsettl0.6_disc-16.2_selisc5},
which is obtained with the same parameters but from a
horizontal line segment of unit length at $z_0 = 0$. The
factors from the lowest and the highest rows of the initial
square closely embrace the factor from Fig.
\ref{fig:densities_factors_ampl0.06_vsettl0.6_disc-16.2_selisc5},
exhibiting similar features (e.g. the caustics). The final
density corresponding to the full square, as a function of the
position, shows a similar shape to those of the individual
factors, but the fine-scale structures are smeared out. In
particular, all caustics disappear. Although this final density
has a character different from those of the total factors of
the individual lines, we can still say that the mechanisms
leading to the final shape are strongly related to those
(stretching and projection) that work for the individual lines.

\begin{figure}[h!]
\includegraphics{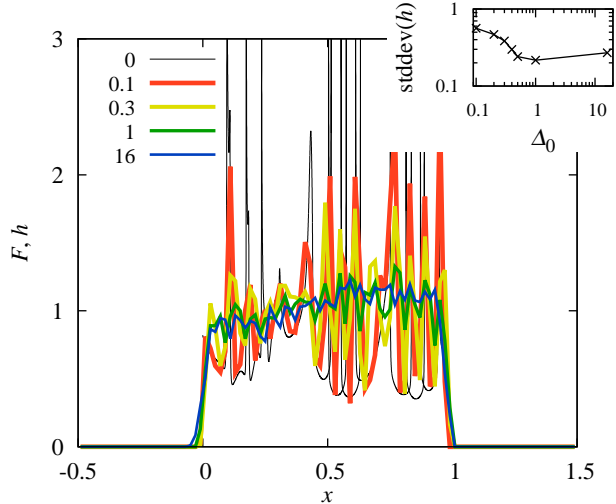}
\caption{\label{fig:histogram_densities_box}The normalized histogram $h$, as a function of the position along the accumulation level, resulting from rectangles of initial conditions with $(x_0,z_0) \in [0,1]\times[0,\Delta_0]$, where $\Delta_0$ is the thickness of the rectangle. The bin size of the histogram is $0.02$. Inset: the standard deviation of $h$ along the accumulation level, as a function of the thickness $\Delta_0$. $A = 0.06$, $W = 0.6$, $\zacc = 97.2$.}
\end{figure}
In Fig. \ref{fig:histogram_densities_box} a sequence of
densities is obtained from increasingly thicker layers of
initial conditions (each having a unit horizontal width, $x_0
\in [0,1]$) for the same parameter setting as in Fig.
\ref{fig:histogram_densities_smallbox}. When the initial
thickness $\Delta_0$ is $0$, we recover the total factor in
Fig.
\ref{fig:densities_factors_ampl0.06_vsettl0.6_disc-16.2_selisc5}
with strong inhomogeneities, including caustics. When
increasing the thickness, the caustics immediately disappear
(i.e., no caustics can exist for any nonzero thickness), but
all other inhomogeneities are smoothed out gradually. We obtain
a rather homogeneous final horizontal density $\sigh$ for
$\Delta_0 = 1$, which does not change much for a much larger
thickness of $\Delta_0 = 16$. The standard deviation of the
final horizontal density, given in the inset of Fig.
\ref{fig:histogram_densities_box}, decreases as a function of
the initial thickness $\Delta_0$ for small thickness values,
while it levels off for larger values, presumably as a result
of the slow smearing out of the initially step-like
distribution. We have thus confirmed that a finite initial
support, smaller than (or, at most, comparable to) the
characteristic length scale of the flow (being unity in Fig.
\ref{fig:histogram_densities_box}), is needed for
inhomogeneities to emerge from an initially homogeneous
distribution within its support. Note that a reduced
dimensionality or a finite support of the subset in which the
initial conditions are distributed represents a strong
inhomogeneity itself, but stretching and projection, determined
by the geometry of the flow, give rise to additional
inhomogeneities.


More generally, we can say that some kind of inhomogeneity is
required in the initial distribution, but the advection in the
flow introduces modifications to this distribution, and these
modifications are characteristic to the geometry of the flow.
For initial distributions with a full-dimensional
($d$-dimensional) support, stretching and projection are still
the two mechanisms that modify the distribution. For generic
shapes of the initial support, we expect the properties to be
the most closely related to stretching and projection of those
$(d-1)$-dimensional (hyper-) surfaces that are aligned with the
unstable foliation of the phase space from the beginning. For
initial supports with a small extension in a particular
direction, stretching and projection of a $(d-1)$-dimensional
(hyper-) surface perpendicular to this direction is expected to
provide with a good approximation.

The relevance of stretching and projection also extends to the
case when $v_z > 0$ is allowed, which has not been investigated
here. For this case, our results can be generalized by taking
the first intersection of the investigated trajectories with
the accumulation surface, and formally extending the
accumulation to infinitely long times. A special property of
such a setup is the typical presence of a chaotic saddle
\cite{Lai2011} in the domain of the flow, the unstable manifold
of which may leave an important imprint on the particular shape
of the distribution that is observed on the accumulation
surface.

However, note that the unstable manifold gets importance for
time scales of increasing duration, and its properties become
actually observable for asymptotically long times. As discussed
in Section \ref{sec:intro}, relevant time scales in our
motivating example are much shorter than asymptotically long
ones, and our investigation in Sections
\ref{sec:ex}-\ref{sec:pardep} revealed interesting
phenomenology before reaching asymptotic behavior.

The study of dynamical systems has traditionally concentrated
on asymptotically long time scales \cite{Ott1993}, even in open
systems \cite{Lai2011}. The motivation for studying this regime
probably has its roots in equilibrium statistical physics, in
which any macroscopic time scale is infinitely long compared to
the characteristic time scales of the individual components of
the system. A common argument, furthermore, is that long-term
behavior dominates observations of the system as opposed to
short-term transients. Our study underlines the practical
relevance of non-asymptotic behavior, which can exhibit
remarkably rich phenomenology \cite{Ottino1989,Pierrehumbert1993}.

A natural next step is the application of our results to a
realistic oceanic setting to study the sedimentation of
biogenic particles. From a theoretical point of view, the
corresponding novelties are the anisotropy of the velocity
field (cf. Section \ref{subsec:eqmotion}) and its nonperiodic
dependence on time. For a complete description, biogeochemical reactions and (dis)aggregation processes of the particles may need to be taken into account. Once all
relevant ingredients are incorporated, the results should be
useful to interpret instrumentally observed data, e.g. from
sediment traps, as well.

\begin{acknowledgments}
We are thankful for Zolt\'an Vandrus for checking the analytical derivations. We acknowledge financial support from the Spanish grants LAOP
CTM2015-66407-P (AEI/FEDER, EU) and ESOTECOS
FIS2015-63628-C2-1-R (AEI/FEDER, EU), and from the Hungarian
grant NKFI-124256 (NKFIH).
\end{acknowledgments}

\appendix

\section{Justification of the particle equation of motion, Eq. \eqref{eq:eqmotion_noninertial}}
\label{sec:ad:ParticleEquations}

Starting from the Maxey--Riley--Gatignol equations describing the
motion of small particles in a flow, and for small enough
values of the Stokes number $\mathrm{St}$, it can be
shown\cite{Haller2008} that, after a short transient time (of
order $\mathrm{St}$), the particle dynamics approaches a slow manifold on
which the particle position $\bfX$ is described, to first order
in $\mathrm{St}$, by
\begin{equation}
\dot{\bfX} = \bfv_\mathrm{fluid}(\bfX,t) + \mathrm{St} (\beta-1) \left( \frac{\mathrm{D}\mathbf{v}_\mathrm{fluid}}{\mathrm{D}t} + \frac{1}{\mathrm{Fr}^2} \hat{\mathbf{k}} \right) ,
\label{eq:MRG}
\end{equation}
where $\bfv_\mathrm{fluid}$ is the nondimensionalized velocity
field of the fluid flow, $\mathrm{D}/\mathrm{D}t$ denotes the
advective derivative following the fluid velocity, with $\mathrm{D}\mathbf{v}_\mathrm{fluid}/\mathrm{D}t$
describing particle inertia, and $\mathbf{k}$ is the vertical
unit vector pointing upwards (in the $z$ direction).

In case $\mathrm{St}$ is very small, it may justify the complete neglection of
the first-order term in $\mathrm{St}$ in \eqref{eq:MRG}. However, if $\mathrm{Fr}$ is considerably smaller than unity such that $\mathrm{St} \ll \mathrm{St}/\mathrm{Fr}^2 \ll 1$, the gravitational term $\hat{\mathbf{k}}/\mathrm{Fr}^2$ may not be negligible even if the inertial term $\mathrm{D}\mathbf{v}_\mathrm{fluid}/\mathrm{D}t$ is. This situation leads to the well-known\cite{siegel1997,Siegel2008,Qiu2014,Roullier2014,Monroy2017} approximation
\BE
\dot{\bfX} = \bfv(\bfX,t) \equiv \bfv_\mathrm{fluid}(\bfX,t) -
W \hat{\mathbf{k}} \ .
\label{eq:MRG0}
\EE

In cases, however, when $\mathrm{Fr}$ is so small that even $\mathrm{St}/\mathrm{Fr}^2 \ll 1$ does \emph{not} hold, the perturbative derivation of Eq. \eqref{eq:MRG} is invalid. Nevertheless, in such circumstances it is straightforward to amend the derivation by
Haller and Sapsis\cite{Haller2008} by considering the regime
$\mathrm{St}\to 0$ and $\mathrm{Fr}\to 0$ with
the dimensionless terminal settling velocity of a particle in
still fluid, $W = (1-\beta) \mathrm{St}/\mathrm{Fr}^2$ (Eq. \eqref{eq:settlingvelocity}), remaining
constant. In this situation, for small enough values of $\mathrm{St}$, a
normally hyperbolic slow manifold still exists, which is
globally attracting, and in which the motion is described, to
first order in $\mathrm{St}$, by
\begin{equation}
\dot{\bfX} = \bfv_\mathrm{fluid}(\bfX,t) - W \hat{\mathbf{k}} +
\mathrm{St} \left( (\beta-1)\frac{\mathrm{D}\mathbf{v}_\mathrm{fluid}}{\mathrm{D}t} + W \hat{\mathbf{k}}\cdot \nabla \bfv_\mathrm{fluid}\right) \ .
\label{eq:MRG2}
\end{equation}

For very small $\mathrm{St}$, when it is appropriate to neglect the first-order term in $\mathrm{St}$ in Eq. \eqref{eq:MRG2}, we recover Eq. \eqref{eq:MRG0}, which is also Eq. \eqref{eq:eqmotion_noninertial} of the main text. We thus conclude that whenever $\mathrm{St} \ll 1$ and $\mathrm{St} \ll W \sim \mathrm{St}/\mathrm{Fr}^2$, irrespectively of whether $W \sim \mathrm{St}/\mathrm{Fr}^2 \ll 1$ or not, Eq. \eqref{eq:eqmotion_noninertial} is a good approximation. From a physical point of view, $\mathrm{St} \ll W$, or, equivalently, $1/\mathrm{Fr}^2 \gg 1$ means that gravity is strong.

As discussed in Section \ref{subsec:eqmotion}, $\mathrm{St}$ is typically
very small for the type of particles we are interested in within a broad range of
scales of oceanic flows, but $\mathrm{Fr}$ is very small as well under the same circumstances, so that $\mathrm{St} \ll 1$ and $\mathrm{St} \ll W \sim \mathrm{St}/\mathrm{Fr}^2$ are both satisfied. Therefore, according to our conclusion above, Eq. \eqref{eq:eqmotion_noninertial} provides with the relevant approximation, whether or not $W \sim \mathrm{St}/\mathrm{Fr}^2$ is small compared to unity (although it is typically not in our motivating situation, as mentioned in the main text).

\section{The derivation of Eqs. \eqref{eq:sh_t_bfx0}-\eqref{eq:J_t_bfx0}}
\label{sec:ad:F_t_bfx0}

For this derivation, we introduce the notation
$\rmd^{d-1}\mathbf{S}$ for integrations over $d-1$-dimensional
(hyper-) surfaces when no parameterization is specified.
Integrations over $d$-dimensional volumes are denoted by
$\rmd^d\bfX$.

The surface density $\sigh$ at a given point $\bfx$, with
$z=-\zacc$, of the accumulation level accumulated up to some
time instant $t$ from the time of initialization $t_0$, can be
computed from the mass that has crossed the accumulation level
at $\bfx$:
\begin{equation}
\sigh(\bfx,t;t_0) = \lim_{|S_B|\to 0}\frac{1}{|S_B|} \int^t_{t_0} \int_{S_B} \rho(t',\bfX)\bfv(t',\bfX) \cdot \rmd^{d-1}\mathbf{S} \rmd t' ,
\label{eq:sh}
\end{equation}
where $S_B$ is a surface in the neighborhood of $\bfx$ within
the accumulation level (defined by $z=-\zacc$), and
$\rho(t,\bfX)$ is the full-dimensional density.

Let us consider all trajectories, initialized at $z=z_0$, that
arrive at the accumulation level within $S_B$. We denote by $S$
the (hyper-) surface that encloses all these trajectories from
their initialization to $z=-\zacc$. Since the density is
assumed to be the constant zero above $z=z_0$, and trajectories
do not cross the side of the surface $S$ by definition, we can
extend the domain of integration from $S_B$ to $S$ in
\eqref{eq:sh}:
\begin{align}
\sigh(\bfx,t;t_0) &= \lim_{|S_B|\to 0}\frac{1}{|S_B|}\int^t_{t_0} \int_{S} \rho(t',\bfX)\bfv(t',\bfX) \cdot \rmd^{d-1}\mathbf{S} \rmd t' \nonumber \\
&= \lim_{|S_B|\to 0}\frac{1}{|S_B|} \int^t_{t_0} \int_{V} \frac{\partial}{\partial \bfX'} \cdot ( \rho(t',\bfX')\bfv(t',\bfX') ) \rmd^d\bfX' \rmd t' ,
\label{eq:sh_V}
\end{align}
where we have used the divergence theorem, and $V$ is the
$d$-volume enclosed by $S$. Using the mass conservation
expressed by the continuity equation
\begin{equation}
\frac{\partial \rho(t,\bfX)}{\partial t} = - \frac{\partial}{\partial \bfX} \left( \rho(t,\bfX)\bfv(t,\bfX) \right) ,
\label{eq:cont}
\end{equation}
we obtain
\begin{equation}
\sigh(\bfx,t;t_0) = - \lim_{|S_B|\to 0}\frac{1}{|S_B|} \int^t_{t_0} \int_{V} \frac{\partial \rho(t',\bfX')}{\partial t'} \rmd^d\bfX' \rmd t' ,
\label{eq:sh_V_cont}
\end{equation}

For a fixed volume, we can take the partial derivative outside
the spatial integral as a total derivative:
\begin{align}
\sigh(\bfx,t;t_0) &= - \lim_{|S_B|\to 0}\frac{1}{|S_B|} \int^t_{t_0} \frac{\rmd}{\rmd t'} \int_{V} \rho(t',\bfX') \rmd^d\bfX' \rmd t' \nonumber \\
&= \lim_{|S_B|\to 0}\frac{1}{|S_B|} \left( \int_{V} \rho(t_0,\bfX') \rmd^d\bfX' - \int_{V} \rho(t,\bfX') \rmd^d\bfX' \right) .
\label{eq:sh_V_intt}
\end{align}
For sufficiently large $t$, i.e., after the arrival of all
trajectories within $V$ to the accumulation level, the second
term in \eqref{eq:sh_V_intt} vanishes (the trajectories have
already crossed the accumulation level). In the limit $|S_B|\to
0$ we only have one trajectory within $V$, which arrives to the
accumulation level at the point $\bfx$, and the preimage of
which at $t=t_0$ we denote by $\bfx_0$ at $z=z_0$. In this
limit, the ``sufficiently large'' $t$ is $t \geq
t(f_z=-\zacc,\bfx_0)$, and $\sigh(\bfx,t \geq
t(f_z=-\zacc,\bfx_0);t_0)$ is what was denoted in Section
\ref{subsec:formulae_general} as
$\sigh(t(f_z=-\zacc,\bfx_0),\bfx_0)$. As for the first term in
\eqref{eq:sh_V_intt}, it is the total initial mass within $V$,
which can be, again in the limit $|S_B|\to 0$, written as the
product of the initial surface density $\sigma(t=t_0;\bfx_0)$
at $\bfx_0$ and the area $|S_U|$ of the infinitesimal surface
$S_U$ at $z=z_0$ that corresponds to the trajectories arriving
at $z=-\zacc$ within $S_B$. For some given $\bfx_0$, we can
thus write:
\begin{equation}
\sigh(t(f_z=-\zacc,\bfx_0),\bfx_0) = \lim_{|S_B|\to 0}\frac{1}{|S_B|} \sigma(t=t_0;\bfx_0) |S_U| .
\label{eq:pre_sh_t_bfx0}
\end{equation}
Since $\sigma(t=t_0;\bfx_0)$ itself is unaffected by the limit,
\eqref{eq:pre_sh_t_bfx0} is the same as \eqref{eq:sh_t_bfx0}
with
\begin{equation}
\mathcal{F}(t(f_z=-\zacc,\bfx_0),\bfx_0) = \lim_{|S_B|\to 0}\frac{|S_U|}{|S_B|} ,
\label{eq:pre_F_t_bfx0}
\end{equation}
that is, the total factor is just the ratio of the areas of the
initial and the final infinitesimal surfaces, one image of the
other, neighboring horizontally (with $z=z_0$ and $z=-\zacc$)
the starting and the endpoint of the trajectory, respectively.

We are now looking for the mathematical relation linking these
areas. This relation is provided by the transformation of the
coordinates from those of the initial surface to those of the
final one:
\begin{align}
|S_B| &= \int_{S_B} \rmd^{d-1}\mathbf{S} = \int_{S_B} \rmd^{d-1}\bfx \nonumber \\
&= \int_{S_U} \det\left( \left.\frac{\partial (f_1(t(f_z=-\zacc,\bfx_0),\bfx_0),\ldots,f_{d-1}(t(f_z=-\zacc,\bfx_0),\bfx_0))}{\partial \bfx_0}\right|_{f_z} \right) \rmd^{d-1}\bfx_0 .
\label{eq:S_B_transformed}
\end{align}
Substituting this and $|S_U| = \int_{S_U} \rmd^{d-1}\mathbf{S} = \int_{S_U} \rmd^{d-1}\bfx_0$ into \eqref{eq:pre_F_t_bfx0}, we recover \eqref{eq:F_t_bfx0}-\eqref{eq:J_t_bfx0} in the indicated limit.

\section{The derivation of Eq. \eqref{eq:constz_to_constt}}
\label{sec:ad:constz_to_constt}

Let us compare two forms of the total differential of the
horizontal position of the endpoint of a trajectory:
\begin{align}
\rmd f_i(t,\bfx_0) &= \sum_{j=1}^{d-1} \left.\frac{\partial f_i(t,\bfx_0)}{\partial x_{0j}}\right|_t \rmd x_{0j} + \left.\frac{\partial f_i(t,\bfx_0)}{\partial t}\right|_{\bfx_0} \rmd t , \label{eq:diff_x_t} \\
\rmd f_i(t(f_z,\bfx_0),\bfx_0) &= \sum_{j=1}^{d-1} \left.\frac{\partial f_i(t(f_z,\bfx_0),\bfx_0)}{\partial x_{0j}}\right|_{f_z} \rmd x_{0j} + \left.\frac{\partial f_i(t(f_z,\bfx_0),\bfx_0)}{\partial f_z}\right|_{\bfx_0} \rmd f_z \label{eq:diff_x_z}
\end{align}
for $i \in \{1,\ldots,d-1\}$. For \eqref{eq:diff_x_t}, the
independent variables are $t$ and $\bfx_0$, while for
\eqref{eq:diff_x_z}, they are $f_z$ and $\bfx_0$ (cf. Section
\ref{subsec:formulae_general}). Similarly to
\eqref{eq:diff_x_t}, we can write the total differential of
$f_z$ itself as follows:
\begin{equation}
\rmd f_z(t,\bfx_0) = \sum_{j=1}^{d-1} \left.\frac{\partial f_z(t,\bfx_0)}{\partial x_{0j}}\right|_t \rmd x_{0j} + \left.\frac{\partial f_z(t,\bfx_0)}{\partial t}\right|_{\bfx_0} \rmd t .
\label{eq:diff_z_t}
\end{equation}
Substituting \eqref{eq:diff_z_t} into \eqref{eq:diff_x_z} and
comparing the result with \eqref{eq:diff_x_t}, which must be
equal to \eqref{eq:diff_x_z}, gives
\begin{align}
\left.\frac{\partial f_i(t,\bfx_0)}{\partial x_{0j}}\right|_t &= \left.\frac{\partial f_i(t(f_z,\bfx_0),\bfx_0)}{\partial x_{0j}}\right|_{f_z} + \left.\frac{\partial f_i(t(f_z,\bfx_0),\bfx_0)}{\partial f_z}\right|_{\bfx_0} \left.\frac{\partial f_z(t,\bfx_0)}{\partial x_{0j}}\right|_t , \label{eq:x_dx0} \\
\left.\frac{\partial f_i(t,\bfx_0)}{\partial t}\right|_{\bfx_0} &= \left.\frac{\partial f_i(t(f_z,\bfx_0),\bfx_0)}{\partial f_z}\right|_{\bfx_0} \left.\frac{\partial f_z(t,\bfx_0)}{\partial t}\right|_{\bfx_0} . \label{eq:x_dt}
\end{align}
for all $i,j \in \{1,\ldots,d-1\}$. Taking into account in
\eqref{eq:x_dt} that, along a given trajectory characterized by
$\bfx_0$,
\begin{equation}
\left.\frac{\partial f_i(t,\bfx_0)}{\partial t}\right|_{\bfx_0} = v_i(t,\bff(t,\bfx_0))
\label{eq:v_zh}
\end{equation}
for $i \in \{1,\ldots,d\}$, substituting $\left.\frac{\partial f_i(t(f_z,\bfx_0),\bfx_0)}{\partial f_z}\right|_{\bfx_0}$ from \eqref{eq:x_dt} into \eqref{eq:x_dx0} yields \eqref{eq:constz_to_constt}.

\section{The derivation of Eq. \eqref{eq:F_t_bfx0_alt}}
\label{sec:ad:F_t_bfx0_alt}

According to a Jacobi-type formula \cite{Harville2008},
\begin{equation}
|J^{-1}|\frac{\partial|J|}{\partial\alpha} = \mathrm{Tr}\,\left( \frac{\partial J}{\partial\alpha} J^{-1} \right) ,
\label{eq:Jacobiformula}
\end{equation}
where $\alpha$ can be an arbitrary variable. We shall choose
$\alpha = f_z$, and solve the differential equation
\eqref{eq:Jacobiformula} for $|J|$ for a fixed $\bfx_0$ (i.e.,
``along a trajectory''). In order to do this, we shall
transform the right-hand side of \eqref{eq:Jacobiformula} to
express it in terms of the velocity field.

We first introduce a new quantity, the derivative of the final
position $\bff$ with respect to the vertical coordinate $f_z$:
\begin{align}
\tv_i(t(f_z,\bfx_0),\bfx_0) &= \left.\frac{\partial f_i(t(f_z,\bfx_0),\bfx_0)}{\partial f_z}\right|_{\bfx_0} \nonumber \\
&= \left.\frac{\partial f_i(t,\bfx_0)}{\partial t}\right|_{\bfx_0,t=t(f_z,\bfx_0)} \left.\frac{\partial t(f_z,\bfx_0)}{\partial f_z}\right|_{\bfx_0} \nonumber \\
&= \frac{v_i(t(f_z,\bfx_0),\bff(t(f_z,\bfx_0),\bfx_0))}{v_z(t(f_z,\bfx_0),\bff(t(f_z,\bfx_0),\bfx_0))} ,
\label{eq:tbfvh}
\end{align}
for $i \in \{1,\ldots,d\}$ --- that is, our new quantity
$\tbfv$ is simply a rescaled version of the original velocity
$\bfv$.

Now let us perform a change in the variables: in a sufficiently
small neighborhood of a given trajectory (characterized by
$\bfx_0$), let us regard the coordinates $\bff =
(f_1,\ldots,f_d)$ of the endpoint of the trajectory as
independent variables, instead of $\bfx_0$ and $f_z$ --- then
$\bfx_0$ becomes a function of $\bff$, $\bfx_0 = \bfx_0(\bff)$.
This change is possible only in a local sense, since the
function $\bff = \bff(t(f_z,\bfx_0),\bfx_0)$ is not invertible,
but we can usually find a small neighborhood around a given
trajectory where it is. The exceptions are trajectories for
which $\det(J) = 0$. For such trajectories, the factor
$\mathcal{F}$ tends to infinity (the positions $\bff$ of the
endpoints of these trajectories correspond to those of the
caustics, cf. Section \ref{subsec:formulae_st_and_pr}, and
\eqref{eq:paracaustic} and \eqref{eq:caustic_withn} in
particular), so that the computation of $\mathcal{F}$ becomes
irrelevant.

In terms of the new independent variables, let us take the
horizontal part of the divergence of the rescaled velocity
$\tbfv$, while keeping $f_z$ constant (note that this implies
the variation of $\bfx_0$, cf. the discussion below):
\begin{align}
\sum_{i=1}^{d-1} \left. \frac{\partial}{\partial f_i} \tv_i(t(f_z,\bfx_0(\bff)),\bfx_0(\bff)) \right|_{f_z} &= \sum_{i,j=1}^{d-1} \left( \left.\frac{\partial \tv_i(t(f_z,\bfx_0),\bfx_0)}{\partial x_{0j}}\right|_{f_z,\bfx_0=\bfx_0(\bff)} \left.\frac{\partial x_{0j}(\bff)}{\partial f_i}\right|_{f_z} \right) \nonumber \\
&= \sum_{i,j=1}^{d-1} \left( \left.\frac{\partial \left( \left.\frac{\partial f_i(t(f_z,\bfx_0),\bfx_0)}{\partial f_z}\right|_{\bfx_0,\bfx_0=\bfx_0(\bff)} \right)}{\partial x_{0j}}\right|_{f_z,\bfx_0=\bfx_0(\bff)} \left.\frac{\partial x_{0j}(\bff)}{\partial f_i}\right|_{f_z} \right) \nonumber \\
&= \sum_{i,j=1}^{d-1} \left( \left.\frac{\partial \left( \left.\frac{\partial f_i(t(f_z,\bfx_0),\bfx_0)}{\partial x_{0j}}\right|_{f_z,\bfx_0=\bfx_0(\bff)} \right)}{\partial f_z}\right|_{\bfx_0,\bfx_0=\bfx_0(\bff)} \left.\frac{\partial x_{0j}(\bff)}{\partial f_i}\right|_{f_z} \right) ,
\label{eq:divergence}
\end{align}
where we applied the chain rule in the first line, substituted
\eqref{eq:tbfvh} for the second line, and, for the third line,
took advantage of the fact that the partial derivative is taken
at a constant $f_z$. If it were taken at a constant time $t$,
changing the order of the derivations would not be possible.
What we obtain in the third line of \eqref{eq:divergence} is
exactly the right-hand side of \eqref{eq:Jacobiformula}. After
changing the independent variables back to $\bfx_0$ and $f_z$,
and substituting \eqref{eq:divergence} into
\eqref{eq:Jacobiformula}, we can solve the differential
equation in terms of $f_z$, keeping $\bfx_0$ constant, which
corresponds to following one particular trajectory. With the
initial condition that $|J|$ is the identity matrix for
$f_z=z_0$, we obtain \eqref{eq:F_t_bfx0_alt}, for which we
introduce a new notation. In particular, due to the different
role of $f_z$ compared to the horizontal components of the
endpoint $\bff$ of the trajectory, we find it useful to
introduce the vector $\bffp = (f_1,\ldots,f_{d-1})$, composed
of the horizontal coordinates of the trajectory. At a constant
$f_z$ (or at a constant $t$, as in Appendix
\ref{sec:ad:F_t_bfx0_alt_constt}), it is $\bffp$ that
identifies the particular trajectory, this is why the variation
of $\bffp$ at a constant $f_z$ (or $t$) implies the variation
of $\bfx_0$.

\section{Transforming Eq. \eqref{eq:F_t_bfx0_alt}}
\label{sec:further_transform}

In this Appendix, we collect the results for how Eq.
\eqref{eq:F_t_bfx0_alt} can be transformed and evaluated
numerically, and discuss the derivations in further appendices.

As shown in Appendix \ref{sec:ad:F_t_bfx0_alt_constt}, Eq.
\eqref{eq:F_t_bfx0_alt} can be written with derivatives taken
at a constant time $t$:
\begin{align}
\mathcal{F}(t(f_z=-\zacc,\bfx_0),\bfx_0) =& \exp \left( -\int_{t_0}^{t(f_z=-\zacc,\bfx_0)} \sum_{i=1}^{d-1} \left[ \left.\frac{\partial v_i(t,\bff(t,\bfx_0(t,\bffp)))}{\partial f_i}\right|_{t,\bffp = \bffp(t,\bfx_0)} \nonumber \right.\right. \\
&- \left.\left. \left.\frac{\partial v_z(t,\bff(t,\bfx_0(t,\bffp)))}{\partial f_k}\right|_{t,\bffp = \bffp(t,\bfx_0)} \frac{v_k(t,\bff(t,\bfx_0))}{v_z(t,\bff(t,\bfx_0))} \nonumber \right.\right. \\
&+ \left.\left. \frac{\left.\frac{\partial f_z(t,\bfx_0(t,\bffp))}{\partial f_i}\right|_{t,\bffp = \bffp(t,\bfx_0)} \left.\frac{\partial}{\partial t} \left( \frac{v_i(t,\bff(t,\bfx_0))}{v_z(t,\bff(t,\bfx_0))} \right)\right|_{f_1,\ldots,f_{d-1}}}{\sum_{k=1}^{d-1} \left.\frac{\partial f_z(t,\bfx_0(t,\bffp))}{\partial f_i}\right|_{t,\bffp = \bffp(t,\bfx_0)} \frac{v_i(t,\bff(t,\bfx_0))}{v_z(t,\bff(t,\bfx_0))} - 1} \right] \right) ,
\label{eq:F_t_bfx0_alt_constt}
\end{align}
where the derivatives taken with respect to the coordinates
$f_i$, $i \in \{1,\ldots,d-1\}$, at a constant $t$ again
correspond to varying $\bfx_0$, which implies that these
derivatives are taken along the surface to which the initial
sheet of particles evolves up to time $t$. The distinguishing
property of this formula is that all of its components can be
evaluated locally, except for $\left.\frac{\partial
f_z(t,\bfx_0(t,\bffp))}{\partial f_i}\right|_{t,\bffp =
\bffp(t,\bfx_0)}$. These latter quantities describe the
tiltness of the surface, and they can be obtained by solving
the following differential equation (see Appendix
\ref{sec:ad:tiltness} for the derivation):
\begin{align}
\frac{\rmd}{\rmd t}\left( \left.\frac{\partial f_z(t,\bfx_0(t,\bffp))}{\partial f_i}\right|_t \right) =& \left.\frac{\partial v_z(t,\bff(t,\bffp)}{\partial x_i}\right|_t + \left.\frac{\partial f_z(t,\bfx_0(t,\bffp))}{\partial f_i}\right|_t \left.\frac{\partial v_z(t,\bff(t,\bffp)}{\partial z}\right|_t \nonumber \\
-& \sum^{d-1}_{j=1} \left( \left.\frac{\partial v_j(t,\bff(t,\bffp))}{\partial x_i}\right|_t + \left.\frac{\partial f_z(t,\bfx_0(t,\bffp))}{\partial f_i}\right|_t \left.\frac{\partial v_j(t,\bff(t,\bffp))}{\partial z}\right|_t \right) \left.\frac{\partial f_z(t,\bfx_0(t,\bffp))}{\partial f_j}\right|_t ,
\label{eq:tiltness}
\end{align}
with the initial condition $\left.\frac{\partial f_z(t,\bfx_0(t,\bffp))}{\partial f_i}\right|_{t = t_0,\bffp = \bffp(t = t_0,\bfx_0)} = 0$. The solution can numerically be evaluated along any single trajectory.

It is important to note that typically, $\left.\frac{\partial
f_z(t,\bfx_0(t,\bffp))}{\partial f_i}\right|_{t,\bffp =
\bffp(t,\bfx_0)} \simeq 0$ in realistic oceanic flows, and
neglecting the second and the third terms in
\eqref{eq:tiltness} yields
\BE
\left.\frac{\partial f_z(t,\bfx_0(t,\bffp))}{\partial f_i}\right|_{t,\bffp = \bffp(t,\bfx_0)} \simeq \int_{t_0}^t \left.\frac{\partial v_z(t',\bff(t',\bffp = \bffp(t',\bfx_0))}{\partial x_i}\right|_{t'} \rmd t' .
\label{eq:tiltness_approx}
\EE

\section{The derivation of Eq. \eqref{eq:F_t_bfx0_alt_constt}}
\label{sec:ad:F_t_bfx0_alt_constt}

First, the integral in \eqref{eq:F_t_bfx0_alt} can be
transformed to a temporal one as
\begin{align}
&\mathcal{F}(t(f_z=-\zacc,\bfx_0),\bfx_0) \nonumber \\
&= \exp \Bigg( -\int_{t_0}^{t(f_z=-\zacc,\bfx_0)} \sum_{i=1}^{d-1} \left. \frac{\partial}{\partial f_i} \left( \frac{\hat{v}_i(f_z,\bffp)}{\hat{v}_z(f_z,\bffp)} \right) \right|_{f_z,f_z=f_z(t,\bfx_0),\bffp=\bffp(t,\bfx_0))} v_z(t,\bff(t,\bfx_0)) \rmd t \Bigg) .
\label{eq:F_t_bfx0_alt2}
\end{align}

In order to replace the derivatives taken at constant depth
$f_z$ by derivatives that are taken at constant time $t$, we
perform a calculation similar to that expressed by
\eqref{eq:diff_x_t}-\eqref{eq:x_dt} in Appendix
\ref{sec:ad:constz_to_constt} for $\tbfv$ instead of $\bff$. We
first recognize that we can regard $\bffp =
(f_1,\ldots,f_{d-1})$ (the horizontal coordinates of a
trajectory) and $t$ as independent variables, too, instead of
$\bff = (f_1,\ldots,f_{d})$ (i.e., $\bffp$ and $f_z$) as in
Appendix \ref{sec:ad:F_t_bfx0_alt}. We are interested here in
the direct dependence of $\tbfv$ on these new independent
variables. To emphasize this, we introduce the notations
$\htbfv(t,\bffp)$ and $\htbfv(f_z,\bffp)$ in this Appendix
instead of the more complicated notation
$\tbfv(t,\bfx_0(t,\bffp))$ and
$\tbfv(t(f_z,\bfx_0(f_z,\bffp)),\bfx_0(f_z,\bffp))$, and we
compare the total differential of $\htbfv$ expressed in terms
of $t$ and $\bffp$ and that expressed in terms of $f_z$ and
$\bffp$:
\begin{align}
\rmd \htv_i(t,\bffp) &= \sum_{j=1}^{d-1} \left.\frac{\partial \htv_i(t,\bffp)}{\partial f_j}\right|_t \rmd f_j + \left.\frac{\partial \htv_i(t,\bffp)}{\partial t}\right|_{\bffp} \rmd t , \label{eq:diff_tv_t} \\
\rmd \htv_i(f_z,\bffp) &= \sum_{j=1}^{d-1} \left.\frac{\partial \htv_i(f_z,\bffp)}{\partial f_j}\right|_{f_z} \rmd f_j + \left.\frac{\partial \htv_i(f_z,\bffp)}{\partial f_z}\right|_{\bffp} \rmd f_z \label{eq:diff_tv_z}
\end{align}
for $i \in \{1,\ldots,d-1\}$ (note that $\htv_z = 1$ is constant).

Before we can compare \eqref{eq:diff_tv_t} and
\eqref{eq:diff_tv_z}, we have to take into account that $f_z$
itself can be regarded as a function of our independent
variables $\bffp$ and $t$, hence
\begin{equation}
\rmd f_z(t,\bffp) = \sum_{j=1}^{d-1} \left.\frac{\partial f_z(t,\bffp)}{\partial f_j}\right|_t \rmd f_j + \left.\frac{\partial f_z(t,\bffp)}{\partial t}\right|_{\bffp} \rmd t .
\label{eq:diff_z_t_bffp}
\end{equation}
Substituting this into \eqref{eq:diff_tv_t} and comparing the
result with \eqref{eq:diff_tv_z}, we obtain
\begin{align}
\left.\frac{\partial \htv_i(t,\bffp)}{\partial f_j}\right|_t &= \left.\frac{\partial \htv_i(f_z,\bffp)}{\partial f_j}\right|_{f_z} + \left.\frac{\partial \htv_i(f_z,\bffp)}{\partial f_z}\right|_{\bffp} \left.\frac{\partial f_z(t,\bffp)}{\partial f_j}\right|_t , \label{eq:tbfvh_dx} \\
\left.\frac{\partial \htv_i(t,\bffp)}{\partial t}\right|_{\bffp} &= \left.\frac{\partial \htv_i(f_z,\bffp)}{\partial f_z}\right|_{\bffp} \left.\frac{\partial f_z(t,\bffp)}{\partial t}\right|_{\bffp} \label{eq:tbfvh_dt}
\end{align}
for $i,j \in \{1,\ldots,d-1\}$.

Still similarly to what is done in Section
\ref{sec:ad:constz_to_constt}, we can substitute
$\left.\frac{\partial \htv_i(f_z,\bffp)}{\partial
f_z}\right|_{\bffp}$ from \eqref{eq:tbfvh_dt} into
\eqref{eq:tbfvh_dx}. Additionally, $\left.\frac{\partial
f_z(t,\bffp)}{\partial t}\right|_{\bffp}$ in
\eqref{eq:tbfvh_dt} can be expressed based on
\eqref{eq:diff_z_t_bffp}. Namely, \eqref{eq:diff_z_t_bffp} can
be written as
\begin{equation}
\frac{\rmd f_z(t,\bffp)}{\rmd t} = \sum_{j=1}^{d-1} \left.\frac{\partial f_z(t,\bffp)}{\partial f_j}\right|_t \frac{\rmd f_j}{\rmd t} + \left.\frac{\partial f_z(t,\bffp)}{\partial t}\right|_{\bffp} ,
\label{eq:diff_z_t_alt}
\end{equation}
in which we can identify the velocities according to
\eqref{eq:v_zh}, but with total derivatives, if we exclude
varying $\bfx_0$: this choice means that we are following
particular trajectories, and the horizontal components $f_j$,
$j \in \{1,\ldots,d-1\}$, become functions of the time $t$. In
this way, we obtain
\begin{equation}
\left.\frac{\partial f_z(t,\bffp)}{\partial t}\right|_{\bffp} = v_z(t,\bff(t,\bffp)) - \sum_{j=1}^{d-1} \left.\frac{\partial f_z(t,\bffp)}{\partial f_j}\right|_t v_j(t,\bff(t,\bffp)) .
\label{eq:diff_z_t_rearranged}
\end{equation}
Taking this into account in \eqref{eq:tbfvh_dt} when
substituting $\left.\frac{\partial \htv_i(f_z,\bffp)}{\partial
f_z}\right|_{\bffp}$ from \eqref{eq:tbfvh_dt} into
\eqref{eq:tbfvh_dx}, we obtain
\begin{equation}
\left.\frac{\partial \htv_i(f_z,\bffp)}{\partial f_j}\right|_{f_z} = \left.\frac{\partial \htv_i(t,\bffp)}{\partial f_j}\right|_t - \frac{\left.\frac{\partial \htv_i(t,\bffp)}{\partial t}\right|_{\bffp}}{v_z(t,\bff(t,\bffp)) - \sum_{k=1}^{d-1} \left.\frac{\partial f_z(t,\bffp)}{\partial f_k}\right|_t v_k(t,\bff(t,\bffp))} \left.\frac{\partial f_z(t,\bffp)}{\partial f_j}\right|_t
\label{eq:constz_to_constt_2}
\end{equation}
for $i,j \in \{1,\ldots,d-1\}$. Note that the derivatives taken
at constant time $t$ with respect to $f_j$, $j \in
\{1,\ldots,d-1\}$, correspond to varying the selected
trajectory, as explained in Appendix \ref{sec:ad:F_t_bfx0_alt},
so that these derivatives are \emph{not} taken at a constant
vertical coordinate: instead, they are taken along the surface
to which the initial sheet of particles evolves up to time $t$.

Expanding the definition \eqref{eq:tbfvh},
\eqref{eq:constz_to_constt_2} can be applied to express
$\sum_{i=1}^{d-1} \left. \frac{\partial}{\partial f_i} \left(
\frac{\hat{v}_i(f_z,\bffp)}{\hat{v}_z(f_z,\bffp)} \right)
\right|_{f_z,f_z=f_z(t,\bfx_0),\bffp=\bffp(t,\bfx_0))}$ in
\eqref{eq:F_t_bfx0_alt2}. After expanding the first term,
simplification, and changing the variables from $t$ and $\bffp$
to $t$ and $\bfx_0$, we recover \eqref{eq:F_t_bfx0_alt_constt}.

\section{The derivation of Eq. \eqref{eq:tiltness}}
\label{sec:ad:tiltness}

In this derivation, we regard the quantities $f_z$ and
$\left.\frac{\partial f_z(t,\bfx_0(t,\bffp))}{\partial
f_i}\right|_{t}$ to be functions of $t$ and $\bffp$ (we express
this by writing $f_z(t,\bffp)$ and $\left.\frac{\partial
f_z(t,\bffp)}{\partial f_i}\right|_{t}$), and, as explained in
relation with \eqref{eq:diff_z_t_rearranged}, we regard $\bffp$
itself to be a function of $t$, which means that we choose to
follow particular trajectories.

After applying the definition of the total derivative to
$\left.\frac{\partial f_z(t,\bffp)}{\partial f_i}\right|_{t}$
(in terms of $t$ and $\bffp$), we can make a rearrangement to
obtain
\begin{align}
\frac{\rmd}{\rmd t}\left( \left.\frac{\partial f_z(t,\bffp)}{\partial f_i}\right|_t \right) =& \frac{\partial}{\partial f_i}\left.\left( \left.\frac{\partial f_z(t,\bffp)}{\partial t}\right|_{\bffp} + \sum^{d-1}_{j=1}v_j(t,\bff(t,\bffp))\left.\frac{\partial f_z(t,\bffp)}{\partial f_j}\right|_t\right)\right|_t \nonumber \\
&- \sum^{d-1}_{j=1} \left.\frac{\partial v_j(t,\bff(t,\bffp)}{\partial f_i}\right|_t \left.\frac{\partial f_z(t,\bffp)}{\partial f_j}\right|_t .
\label{eq:tiltness_dt}
\end{align}
From \eqref{eq:diff_z_t_rearranged}, we can substitute the
vertical velocity component $v_z$ to obtain
\begin{equation}
\frac{\rmd}{\rmd t}\left( \left.\frac{\partial f_z(t,\bffp)}{\partial f_i}\right|_t \right) = \left.\frac{\partial v_z(t,\bff(t,\bffp)}{\partial f_i}\right|_t - \sum^{d-1}_{j=1} \left.\frac{\partial v_j(t,\bff(t,\bffp))}{\partial f_i}\right|_t \left.\frac{\partial f_z(t,\bffp)}{\partial f_j}\right|_t .
\label{eq:tiltness_dt_alt}
\end{equation}

In \eqref{eq:tiltness_dt_alt} we are still facing the problem
that the partial derivatives, taken with respect to the
horizontal coordinates of the trajectory, are taken along the
material sheet, not with respect to the Cartesian coordinates
of the domain of the fluid flow. The transformation between the
two types of coordinates is given by the relation
\begin{equation}
\left.\frac{\partial }{\partial f_k}\right|_t = \left.\frac{\partial }{\partial x_k}\right|_t + \left.\frac{\partial f_z(t,\bffp)}{\partial f_k}\right|_t \left.\frac{\partial }{\partial z}\right|_t ,
\label{eq:traj_to_Cartesian}
\end{equation}
where $(x_1,\ldots,x_{d-1},z)$ denote the desired Cartesian
coordinates. By substituting \eqref{eq:traj_to_Cartesian} into
\eqref{eq:tiltness_dt_alt}, we recover \eqref{eq:tiltness}.

\section{The derivation of Eqs. \eqref{eq:s_t_u}-\eqref{eq:S_t_u}}
\label{sec:ad:S_t_u}

First we formalize densities on surfaces embedded into volumes.
Let us consider a $d$ dimensional space, into which a
$(d-1)$-dimensional surface is embedded. Let us take a
parameterization $\bff$ of this surface by $\bfu$, where $\bff$
is a $d$ dimensional vector, and $\bfu$ is a $d-1$ dimensional
vector. In case mass is (or particles are) only distributed on
the surface, then the $d$ dimensional density $\rho(\bfX)$ at
the position $\bfX$ of the $d$ dimensional space can be written
as
\BE
\rho(\bfX) = \int_{D_\bff} \sigma(\bfu) \delta(\bfX-\bff(\bfu)) \rmd^{d-1}\bfu ,
\label{eq:rho}
\EE
where $D_\bff$ is the domain of $\bff(\bfu)$, and
$\sigma(\bfu)$ characterizes the distribution of mass (or
particles) within the surface. In particular, it gives the
surface density with respect to the coordinates that
parameterize the surface. In physical problems one is usually
interested in the surface density that is taken with respect to
length (area, etc. for higher dimensions), which is obtained by
choosing the parameter(s) to be the arc length (and its
generalizations for higher dimensions, in the sense that
integrating 1 with respect to the parameters gives the
[generalized] area of the [generalized] surface).

In the special case $d=2$ the parameter vector $\bfu$
simplifies to a scalar $u$, so that
\BE
\rho(\bfX) = \int_0^l \sigma(u) \delta(\bfX-\bff(u)) \rmd u \ ,
\label{eq:rho2d}
\EE
where $l$ is the length of the line segment parameterized by $u$.

A line segment of initial conditions (a material line of
particles) at the time of initialization $t = t_0$ in a $d=2$
dimensional flow, parameterized by its arc length $u$, shall be
denoted as
\BE
\bff(t=t_0;u) = \bff_0(u) \ .
\EE
Any later image (at time $t$) of any point of this line segment
is obtained via the time evolution of the flow, and the line
segment can thus still be parameterized by $u$:
\BE
\bff(t;u) = \bfP(\bff(t=t_0;u),t_0;t) = \bfP(\bff_0(u),t_0;t) \ ,
\label{eq:fev_int}
\EE
where $\bfP$ is the flow map, that is,
$\bfX=\bfP(\bfX_0,t_0;t)$ gives the position $\bfX$ at time $t$
of the fluid element that was at $\bfX_0$ at time $t_0$. It
follows that
\BE
\dot{\bff}(t;u) = \bfv(\bff(t;u),t) \ ,
\label{eq:fev_der}
\EE
where $\bfv(\bfX,t)$ is the velocity field at the position $\bfX$ at time $t$.

The initial ``surface'' density (with respect to arc length) is
given by $\sigma(t=t_0;u)$, by which
\BE
\rho(\bfX,t=t_0) = \int_0^{l} \sigma(t=t_0;u) \delta(\bfX-\bff(t=t_0;u)) \rmd u \ .
\EE
The question is how this ``surface'' density transforms with time evolution.

Certainly,
\begin{align}
\rho(\bfX,t) &= \int_0^{l} \sigma(t=t_0;u) \delta(\bfP(\bfX,t;t_0)-\bff(t=t_0;u)) \rmd u \nonumber \\
&= \int_0^{l} \sigma(t=t_0;u) \delta(\bfX-\bff(t;u)) \rmd u \ ,
\label{eq:rho_t_1}
\end{align}
because of (\ref{eq:fev_int}) (note the reverse time evolution
here), and since the determinant of the Jacobian of
$\bfP(\bfX,t;t_0)$ is $1$ in volume-preserving flows. The
problem is that $u$ is \emph{not an arc length} of the image of
the material line, i.e., of $\bff(t;u)$. We have to transform
the integration such that it is taken with respect to the arc
length $s$ of the image $\bff(t;u)$.

This means that we are looking for a function $u =
\varphi_t(s)$ such that
\BE
\int_0^{s} \left| \frac{\rmd \bff(t;\varphi_t(s'))}{\rmd s'} \right| \rmd s' = s
\EE
for all $s$ within the full length of the image of the material
line. Let us transform the integration by the change $u =
\varphi_t(s)$ itself as
\BE
\int_0^{u=\varphi_t(s)} \left| \frac{\rmd \bff(t;u')}{\rmd u'} \right| \rmd u' = s \ .
\EE
Now let us take the derivative of both sides with respect to $s$:
\BE
\left| \frac{\rmd \bff(t;u)}{\rmd u} \right|_{u=\varphi_t(s)} \frac{\rmd \varphi_t(s)}{\rmd s} = 1 \ ,
\EE
that is,
\BE
\frac{\rmd \varphi_t(s)}{\rmd s} = \left| \frac{\rmd \bff(t;u)}{\rmd u} \right|_{u=\varphi_t(s)}^{-1} \ .
\label{eq:duds}
\EE
This, along with the condition $\varphi_t(s=0)=0$, defines
$\varphi_t(s)$ uniquely.

Now let us perform the change of the integration variable in (\ref{eq:rho_t_1}) as
\BE
\rho(\bfX,t) = \int_0^{s=\varphi_t^{-1}(l)} \sigma(t=t_0;u=\varphi_t(s)) \delta(\bfX-\bff(t;u=\varphi_t(s))) \frac{\rmd \varphi_t(s)}{\rmd s} \rmd s \ ,
\label{eq:rho_t_2}
\EE
where $\frac{\rmd \varphi_t(s)}{\rmd s}$ is given by
(\ref{eq:duds}). In fact, it is enough to know $\frac{\rmd
\varphi_t(s)}{\rmd s}$ (and not $\varphi_t(s)$ itself), since
from (\ref{eq:rho_t_2}) we can read off the surface density of
the image of the material line (at time $t$) with respect to
the arc length of this image:
\begin{align}
\sigma(t;u=\varphi_t(s)) &= \sigma(t=t_0;u=\varphi_t(s)) \frac{\rmd \varphi_t(s)}{\rmd s} \nonumber \\
&= \sigma(t=t_0;u=\varphi_t(s)) \left| \frac{\rmd \bff(t;u)}{\rmd u} \right|_{u=\varphi_t(s)}^{-1} .
\label{eq:s_arcl}
\end{align}
Keeping $u$ as a parameter of $\sigma$ is reasonable if we
follow the time evolution of $\sigma$ in a Lagrangian sense:
the ``surface'' density $\sigma$ associated with a given
particle is characterized by a particular value of $u$,
describing the initial position of our particle along the
initial line segment. In terms of $u$ instead of the final
position of the particle, \eqref{eq:s_arcl} reads as
\BE
\sigma(t;u) = \sigma(t=t_0;u) \left| \frac{\rmd \bff(t;u)}{\rmd u} \right|^{-1} ,
\label{eq:s_arcl_2}
\EE
which is \eqref{eq:s_t_u}-\eqref{eq:S_t_u}.

\section{Explanation for Eq. \eqref{eq:P_t_u}}
\label{sec:ad:P_t_u}

\begin{figure}[h!]
\includegraphics{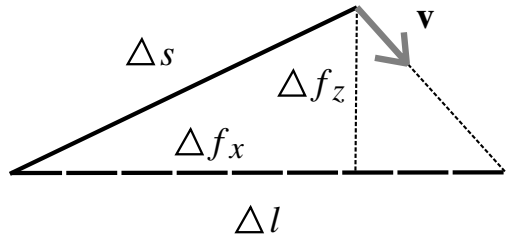}
\caption{\label{fig:projection}The geometry of the projection of an infinitesimal line segment taking into account kinematic effects. See text for details.}
\end{figure}
One configuration of the simple geometry relating the pre- and
the post-projection length of an infinitesimal segment of the
material line (at time $t$ around a position that is
characterized by $u$) is illustrated in Fig.
\ref{fig:projection}. The thick solid line in black is the
pre-projection state of the infinitesimal line segment (of
length $\Delta s$, where $s$ is the arc length along the
material line and increases from the left to the right in Fig.
\ref{fig:projection}), while the post-projection length is
marked by a thick dashed line. The orientation of the former is
determined by the dynamics and is given by the two components,
$\Delta f_x$ and $\Delta f_z$, of the infinitesimal line
segment corresponding to $\Delta s$, while the latter is
horizontal. The post-projection length $\Delta l$ is determined
by the direction of the velocity $\bfv$ as indicated by a thin
dashed line in the figure (the velocity is assumed to be
uniform in the infinitesimal domain considered here), and we
are interested in the ratio of $\Delta l$ and $\Delta s$:
\begin{equation}
\mathcal{P} = \lim_{\Delta s \to 0} \frac{\Delta s}{\Delta l} .
\label{eq:P_expl}
\end{equation}

As a vertical thin dashed line indicates in Fig.
\ref{fig:projection}, the post-projection line segment can be
divided to two sections, and $\Delta l$ can be calculated as
the sum of the lengths of these two sections. The length of the
left-hand-side section is simply $\Delta f_x$, while that of
the right-hand-side one can be obtained as $\Delta f_z$
multiplied by the tangent of the angle enclosed by the two thin
dashed lines. The latter is the opposite of the ratio of the
two components of the velocity, so that we have $- \Delta f_z
v_x/v_z$ for the length of the right-hand-side segment.
Therefore, we have
\begin{align}
\mathcal{P} &= \lim_{\Delta s \to 0} \frac{\Delta s}{\Delta f_x - \Delta f_z \frac{v_x}{v_z}} \nonumber \\
&= \lim_{\Delta s \to 0} \frac{1}{\frac{\Delta f_x}{\Delta s} - \frac{\Delta f_z}{\Delta s} \frac{v_x}{v_z}} \nonumber \\
&= \left( \frac{\rmd f_x}{\rmd s} - \frac{\rmd f_z}{\rmd s} \frac{v_x}{v_z} \right)^{-1}
\label{eq:pre_P_expl_result}
\end{align}
for the configuration presented if Fig. \ref{fig:projection}.

Other configurations (when $s$ increases from the right to the
left or when the velocity has an essentially different
direction) can be treated in a similar way, and the general
result for the projection factor $\mathcal{P}$ differs from
\eqref{eq:pre_P_expl_result} only in an absolute value:
\begin{equation}
\mathcal{P} = \left| \frac{\rmd f_x}{\rmd s} - \frac{\rmd f_z}{\rmd s} \frac{v_x}{v_z} \right|^{-1} .
\label{eq:P_expl_result}
\end{equation}
With the notation indicating that we are investigating an
infinitesimal line segment of the material line at a time
instant $t$ and that is characterized by the initial position
$u$, we recognize \eqref{eq:P_t_u}.

\section{Stretching and projection for 3D flows}
\label{sec:st_and_pr_3D}

For $d=3$, let $\bff(t=t_0;\bfu)$ be the parametric form of a
planar sheet of initial conditions at time $t = t_0$,
parameterized by the vector $\bfu = (u_1, u_2)$ that is an
appropriate generalization of arc length (in the sense that
$\int\int_\mathbf{min}^\mathbf{max} 1 \,\rmd^2 \bfu$ gives the
initial area of the sheet). Let $\sigma(t=t_0;\bfu)$ be the
initial density within the sheet at $\bfu$. The density
$\sigma_\mathbf{x}(t;\bfu)$ at a point of the accumulation
level whose initial preimage was characterized by $\bfu$ is
given by
\BE
\sigh(t;\bfu) = \sigma(t=t_0;\bfu) \mathcal{F}(t;\bfu) = \sigma(t=t_0;\bfu) \mathcal{S}(t;\bfu) \mathcal{P}(t;\bfu) ,
\label{eq:fact_sh_t_vecu}
\EE
where $\mathcal{F}(t;u)$ is the total factor multiplying the
initial density,
\BE
\mathcal{S}(t;\bfu) = \left| \frac{\partial
\bff(t;\bfu)}{\partial u_1} \times \frac{\partial
\bff(t;\bfu)}{\partial u_2} \right|^{-1},  \label{eq:S_t_vecu}
\EE
with $\times$ denoting the vector cross product, is the factor
representing stretching ($\bff(t;\bfu)$ stands for the image of
the sheet at $t$), and
\BE
\mathcal{P}(t;\bfu) = \left| \left( \frac{\partial \bff(t;\bfu)}{\partial s_1} - \frac{\bfv(\bff(t;\bfu),t)}{v_z(\bff(t;\bfu),t)} \frac{\partial f_z(t;\bfu)}{\partial s_1} \right) \times \left( \frac{\partial \bff(t;\bfu)}{\partial s_2} - \frac{\bfv(\bff(t;\bfu),t)}{v_z(\bff(t;\bfu),t)} \frac{\partial f_z(t;\bfu)}{\partial s_2} \right) \right|^{-1}
\label{eq:P_t_vecu}
\EE
is the factor representing projection ($\bfs = (s_1, s_2)$ is
the appropriate generalization of the arc length for the
\emph{image} of the sheet).

More direct forms are
\begin{align}
\mathcal{F}(t;\bfu) &= \left| \left( \frac{\partial \bff(t;\bfu)}{\partial u_1} - \frac{\bfv(\bff(t;\bfu),t)}{v_z(\bff(t;\bfu),t)} \frac{\partial f_z(t;\bfu)}{\partial u_1} \right) \times \left( \frac{\partial \bff(t;\bfu)}{\partial u_2} - \frac{\bfv(\bff(t;\bfu),t)}{v_z(\bff(t;\bfu),t)} \frac{\partial f_z(t;\bfu)}{\partial u_2} \right) \right|^{-1} \\
&= \left| \left( \frac{\partial f_x(t;\bfu)}{\partial u_1} - \frac{v_x(\bff(t;\bfu),t)}{v_z(\bff(t;\bfu),t)} \frac{\partial f_z(t;\bfu)}{\partial u_1} \right) \cdot \left( \frac{\partial f_y(t;\bfu)}{\partial u_2} - \frac{v_y(\bff(t;\bfu),t)}{v_z(\bff(t;\bfu),t)} \frac{\partial f_z(t;\bfu)}{\partial u_2} \right) \right. \nonumber \\
& \left. - \left( \frac{\partial f_y(t;\bfu)}{\partial u_1} - \frac{v_y(\bff(t;\bfu),t)}{v_z(\bff(t;\bfu),t)} \frac{\partial f_z(t;\bfu)}{\partial u_1} \right) \cdot \left( \frac{\partial f_x(t;\bfu)}{\partial u_2} - \frac{v_x(\bff(t;\bfu),t)}{v_z(\bff(t;\bfu),t)} \frac{\partial f_z(t;\bfu)}{\partial u_2} \right) \right|^{-1} .
\label{eq:simple_F_t_vecu}
\end{align}
The second equation is due to the fact that the vectorial
product in the first line has only one nonzero component.
Unlike for $d=2$, the parametric derivatives of the position do
not appear here in a simple combination. Nevertheless,
\eqref{eq:simple_F_t_vecu} is equivalent to
\eqref{eq:F_t_bfx0}-\eqref{eq:constz_to_constt} for $d=3$.
Formulae for $d > 3$ can be constructed similarly.

Appendix \ref{sec:n_t_vecu} transforms the results to a matrix
formulation, and links them to the local normal vector of the
sheet at the time and place of its arrival to the accumulation
level.

\section{The projection factor in 3D flows expressed in terms of the normal vector}
\label{sec:n_t_vecu}

Note that the right-hand side of \eqref{eq:simple_F_t_vecu} can
be written as
\BE
\mathcal{F}(t;\bfu) = \left| \det\left(J(t;\bfu)\right) \right|^{-1} ,
\label{eq:F_t_vecu_withdet}
\EE
where
\BE
J_{ij}(t;\bfu) = \frac{\partial f_i(t;\bfu)}{\partial u_j} - \frac{v_i(\bff(t;\bfu),t)}{v_z(\bff(t;\bfu),t)} \frac{\partial f_z(t;\bfu)}{\partial u_j}
\label{eq:Jij}
\EE
is a 2 by 2 matrix with $i,j \in \{x,y\}$. For $\det J$, the
matrix determinant lemma \cite{Harville2008} can be applied:
\BE
\det\left(J(t;\bfu)\right) = \det\left( \frac{\partial \bffp(t;\bfu)}{\partial \bfu} \right) \left( 1 - \sum_{i,j \in \{x,y\}} \frac{\partial f_z(t;\bfu)}{\partial u_i} \left.\frac{\partial u_i(t;\bffp)}{\partial f_j}\right|_{\bffp=\bffp(t;\bfu)} \frac{v_j(\bff(t;\bfu),t)}{v_z(\bff(t;\bfu),t)} \right) ,
\label{eq:detJ}
\EE
where $\bffp(t;\bfu)$ denotes the vector formed from the first
two components of $\bff(t;\bfu)$, and $\bfu(t;\bffp)$ is the
inverse of $\bffp(t;\bfu)$. The second factor in
\eqref{eq:detJ} can be simplified using the chain rule (note
the sum for $i$), and one obtains
\BE
\det\left(J(t;\bfu)\right) = \det\left( \frac{\partial \bffp(t;\bfu)}{\partial \bfu} \right) \left( 1 - \sum_{j \in \{x,y\}} \left.\frac{\partial f_z(t;\bfu(t;\bffp))}{\partial f_j}\right|_{\bffp=\bffp(t;\bfu)} \frac{v_j(\bff(t;\bfu),t)}{v_z(\bff(t;\bfu),t)} \right) .
\label{eq:detJ_simple}
\EE

Now let us introduce the normal vector $\bfn$ of the surface $\bff$:
\begin{align}
\bfn(t;\bfu) &= \left( \frac{\partial \bff(t;\bfu)}{\partial u_1} \times \frac{\partial \bff(t;\bfu)}{\partial u_2} \right) \left| \frac{\partial \bff(t;\bfu)}{\partial u_1} \times \frac{\partial \bff(t;\bfu)}{\partial u_2} \right|^{-1} \nonumber \\
&= \left( \frac{\partial \bff(t;\bfu)}{\partial u_1} \times \frac{\partial \bff(t;\bfu)}{\partial u_2} \right) \mathcal{S}(t;\bfu) ,
\label{eq:n_t_vecu}
\end{align}
where the second line is obtained by substituting
\eqref{eq:S_t_vecu}. Note that
\BE
n_z(t;\bfu) = \det\left( \frac{\partial \bffp(t;\bfu)}{\partial \bfu} \right) \mathcal{S}(t;\bfu) ,
\label{eq:nz_t_vecu}
\EE
which can be substituted in \eqref{eq:detJ_simple}.
Furthermore, it can be shown that
\BE
n_i(t;\bfu) = n_z(t;\bfu) \left.\frac{\partial f_z(t;\bfu(t;\bffp))}{\partial f_i}\right|_{\bffp=\bffp(t;\bfu)}
\label{eq:nxy_t_vecu}
\EE
for $i \in \{x,y\}$, so that \eqref{eq:detJ_simple} can be
written as
\begin{align}
\det\left(J(t;\bfu)\right) &= \mathcal{S}(t;\bfu)^{-1} \left( n_z(t;\bfu) + n_x(t;\bfu) \frac{v_x(\bff(t;\bfu),t)}{v_z(\bff(t;\bfu),t)} + n_y(t;\bfu) \frac{v_y(\bff(t;\bfu),t)}{v_z(\bff(t;\bfu),t)} \right) \nonumber \\
&= \mathcal{S}(t;\bfu)^{-1} \frac{\bfn(t;\bfu) \cdot \bfv(\bff(t;\bfu),t)}{v_z(\bff(t;\bfu),t)} .
\label{eq:detJ_moresimple}
\end{align}

According to \eqref{eq:detJ_moresimple} and
\eqref{eq:F_t_vecu_withdet}, we finally have
\BE
\mathcal{F}(t;\bfu) = \mathcal{S}(t;\bfu) \left|\frac{v_z(\bff(t;\bfu),t)}{\bfn(t;\bfu) \cdot \bfv(\bff(t;\bfu),t)} \right| ,
\label{eq:F_t_vecu_withn}
\EE
from which
\BE
\mathcal{P}(t;\bfu) = \left| \frac{v_z(\bff(t;\bfu),t)}{\bfn(t;\bfu) \cdot \bfv(\bff(t;\bfu),t)} \right|
\label{eq:P_t_vecu_withn}
\EE
also follows.

One sees from \eqref{eq:F_t_vecu_withn} or
\eqref{eq:P_t_vecu_withn} that the caustics (divergences in the
denominator) are located where the local normal vector of the
sheet is perpendicular to the local velocity, similarly to the
$d=2$ case.

\section{Additional details about the dependence on the depth $\zacc$
of several quantities in the double-shear flow}
\label{sec:pardep_zacc_detailed}

\begin{figure}[h!]
	\subfloat{\label{fig:comp_average_factors_ampl0.07_vsettl0.6}\includegraphics{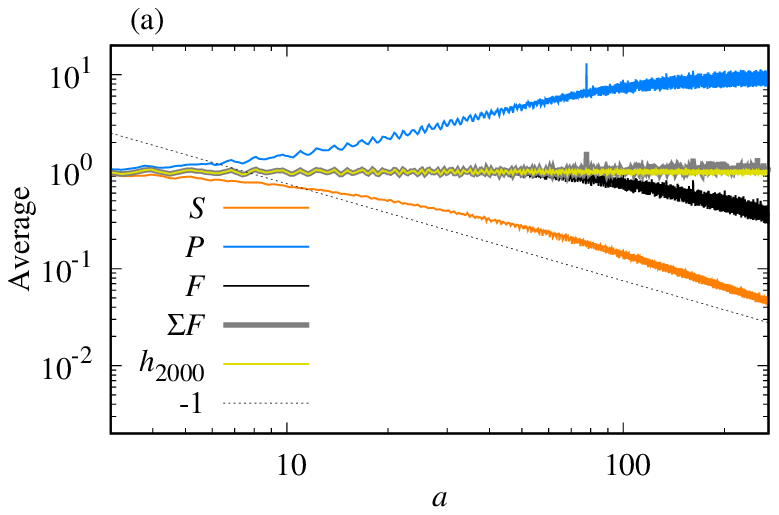}}
	\subfloat{\label{fig:comp_stddeviation_factors_ampl0.07_vsettl0.6}\includegraphics{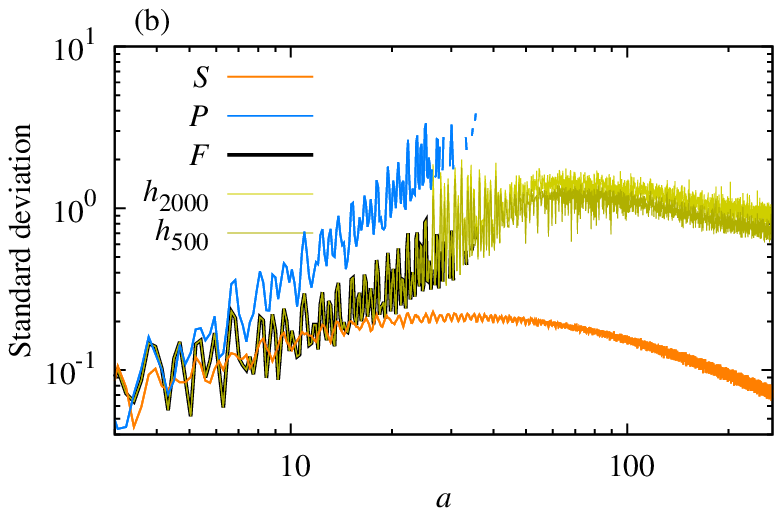}}\\
	\subfloat{\label{fig:comp_average_terms_ampl0.07_vsettl0.6}\includegraphics{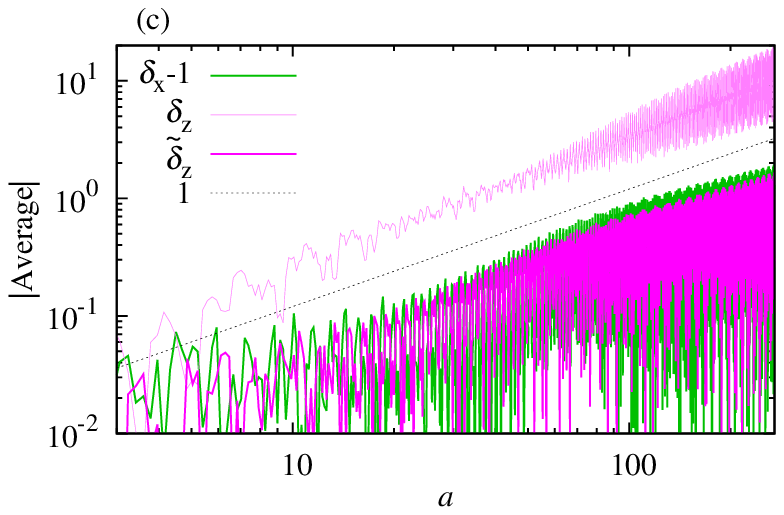}}
	\subfloat{\label{fig:comp_stddeviation_terms_ampl0.07_vsettl0.6}\includegraphics{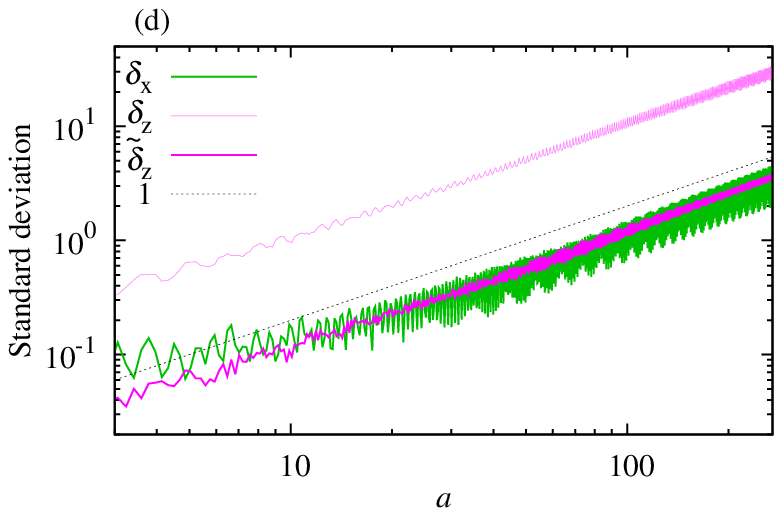}}
\caption{\label{fig:comp_ampl0.07_vsettl0.6}(a) The average and
(b) the standard deviation, as a function of the accumulation depth $\zacc$,
of the same quantities as in Fig. \ref{fig:comp_ampl0.25_vsettl0.6}.
(c) The average and (d) the standard deviation of the parametric derivative of the horizontal
position $\delta_x$ (minus $1$ for comparability), the parametric derivative of the vertical
position $\delta_z$, and the weighted parametric derivative of the vertical position
$\tilde{\delta}_z$. For comparison, the dashed lines mark power laws with the exponents
indicated in the legend. $W = 0.6$ and $A = 0.07$.}
\end{figure}
Figure \ref{fig:comp_ampl0.07_vsettl0.6} exhibits an example
when chaos is not observable. The average of each factor
introduced in Section \ref{subsec:formulae_st_and_pr}, shown in
Fig. \ref{fig:comp_average_factors_ampl0.07_vsettl0.6}, starts
with small slopes for small depths $\zacc$. The average of the
stretching factor, $\langle \mathcal{S} \rangle$, bends down
for increasing $\zacc$. In particular, for intermediate values
of $\zacc$, it might seem to follow a $1/\zacc$ dependence.
This would agree with the expectation that the reciprocal of
the stretching factor, corresponding to the length of the line,
should increase proportionally to time in this non-chaotic
situation --- but the line bends down even more for increasing
$\zacc$. The average projection factor, $\langle \mathcal{P}
\rangle$, increases with depth, which is a natural consequence
of the undulation of the material line. Approximately $\langle
\mathcal{P} \rangle \approx 1/\langle \mathcal{S} \rangle$
until $\zacc \approx 20$, and the increase of $\langle
\mathcal{P} \rangle$ gets slower for larger values of $\zacc$,
which might be related to the bounded nature of the average
effect of the projection. As a consequence of this behavior,
the average total factor $\langle \mathcal{F} \rangle$ (still
in Fig. \ref{fig:comp_average_factors_ampl0.07_vsettl0.6}) is
practically $1$ up to $a \approx 20$, and exhibits a slight
decrease above (even though $\mathcal{F} = \mathcal{S}
\mathcal{P}$ only pointwise, and $\langle \mathcal{F} \rangle
\neq \langle \mathcal{S} \rangle \langle \mathcal{P} \rangle$).
We are facing thus a net dilution for increasing depth, but
without a known simple functional form. However, this dilution
is local along the line: when summing up over the different
branches of the line, the average $\langle \sum\mathcal{F}
\rangle$ of the summed total factor does not follow the
decrease in $\langle \mathcal{F} \rangle$. Instead, it remains
remarkably close to a constant. The approximate conservation of
$\langle \sum\mathcal{F} \rangle$ is not a surprise, since mass
is conserved, and the horizontal support of the distribution of
the particles changes very little in our example. The average
$\langle h_N \rangle$ of the normalized histogram $h_N$, for $N
= 2000$ and also in a wide range in $N$ around this choice, is
practically the same as $\langle \sum\mathcal{F} \rangle$
(without the spike near $\zacc = 80$, which indicates that
probably this little spike is a numerical artifact).

The parametric derivatives, $\langle \delta_x \rangle$ and $\langle \delta_z \rangle$ (their absolute values are shown in Fig. \ref{fig:comp_average_terms_ampl0.07_vsettl0.6}), obey a clear linear law as a function of $\zacc$ (at least in terms of the envelope in the case of $\langle \delta_x \rangle$, which otherwise oscillates so strongly that sign changes occur in each ``period'' of the oscillation). This might be so because these quantities directly concern the final positions observed at the accumulation depth. Weighting $\delta_z$ by the local (time-dependent!) velocity does not ruin the linear functional relation, see $\langle \tilde{\delta}_z \rangle$ in Fig. \ref{fig:comp_average_terms_ampl0.07_vsettl0.6}. The effect of this weighting is the reduction of the magnitude to approximately match that of $\langle \delta_x \rangle$ (without weighting, $\langle \delta_z \rangle$ is much larger, cf. Figs. \ref{fig:densities_terms_ampl0.06_vsettl0.6_disc-2.7_selisc0}-\ref{fig:densities_terms_ampl0.06_vsettl0.6_disc-16.2_selisc5} and the related discussion), and the enhancement of the oscillations. The linear nature agrees with the regular dynamics, and with the proportional increase of the settling time for increasing depth $\zacc$.

Turning to the characterization of the inhomogeneities, which
we do by investigating the standard deviation of the relevant
quantities, we can conclude from Fig.
\ref{fig:comp_stddeviation_factors_ampl0.07_vsettl0.6} that the
total factor $\mathcal{F}$ (black line) exhibits an increasing
inhomogeneity with increasing depth $\zacc$ for approximately
$\zacc < 30$, while there are no caustics, i.e., while it is
meaningful to plot the line. The similarity in the functional
form to that of that corresponding to the projection factor
$\mathcal{P}$ (blue line) suggests the strong influence of this
latter factor. The stretching factor is also getting more and
more inhomogeneous at the beginning, but the increase in its
standard deviation slows down, and turns to a homogeneization,
which might be related to the decreasing magnitude of the
factor $\mathcal{S}$ itself in average (as discussed in
relation with Fig.
\ref{fig:comp_average_factors_ampl0.07_vsettl0.6}). As long as
the standard deviation of the total factor $\mathcal{F}$ is
meaningful, the standard deviation of the normalized histogram
$h_N$ matches that of $\mathcal{F}$ very closely both for small
($N = 2000$) and large ($N = 500$) bin size, i.e., clearly
indicates stronger and stronger inhomogeneities. However, later
on, when the standard deviation of $h_N$ becomes $N$-dependent
(at approximately $\zacc = 50$), inhomogeneization does not
continue any more, but it turns to a homogeneization for any
$N$, similarly to what is seen in the chaotic case (Fig.
\ref{fig:comp_stddeviation_factors_ampl0.25_vsettl0.6}).

The standard deviations of the parametric derivatives, as shown
in Fig. \ref{fig:comp_stddeviation_terms_ampl0.07_vsettl0.6},
increase in a simple, ballistic manner, in agreement with the
regular features of the flow.

\begin{figure}[h!]
	\subfloat{\label{fig:comp_average_terms_ampl0.25_vsettl0.6}\includegraphics{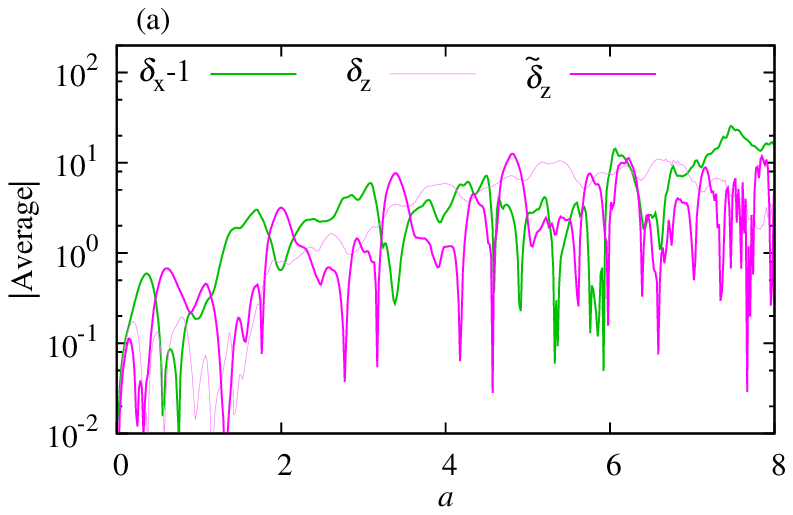}}
	\subfloat{\label{fig:comp_stddeviation_terms_ampl0.25_vsettl0.6}\includegraphics{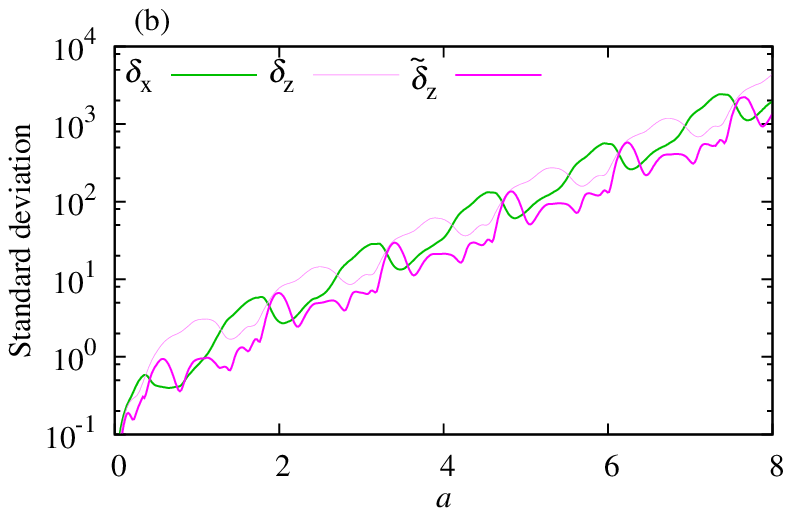}}
\caption{\label{fig:comp_terms_ampl0.25_vsettl0.6}Same as Figs. \ref{fig:comp_average_terms_ampl0.07_vsettl0.6}-\ref{fig:comp_stddeviation_terms_ampl0.07_vsettl0.6} for $A = 0.25$.}
\end{figure}
In the chaotic case, illustrated by Figs.
\ref{fig:comp_ampl0.25_vsettl0.6} and
\ref{fig:comp_terms_ampl0.25_vsettl0.6}, the tendencies are
always the same as in the regular case, the main difference
lies in the functional forms of the dependences. The dependence
of the averages and the standard deviations of the particular
factors on $\zacc$ are discussed in Section
\ref{subsec:pardep_zacc} of the main text.

Fig. \ref{fig:comp_average_terms_ampl0.25_vsettl0.6} indicates
that the (unweighted and weighted) parametric derivatives
exhibit a pronounced strengthening with $\zacc$ in average, but
the particular functional form is not clear. Note that, in
contrast with the regular case of Fig.
\ref{fig:comp_average_terms_ampl0.07_vsettl0.6}, the unweighted
parametric derivative $\delta_z$ of the vertical location is
not larger in magnitude than the other two quantities in the
plot (cf. Figs.
\ref{fig:densities_terms_ampl0.25_vsettl0.6_disc-1.8_selisc0}-\ref{fig:densities_terms_ampl0.25_vsettl0.6_disc-1.8_selisc3}).
As for the standard deviations of the (unweighted and weighted)
parametric derivatives, Fig.
\ref{fig:comp_stddeviation_terms_ampl0.25_vsettl0.6} shows that
they do not increase with $\zacc$ in a diffusional way (which
could be expected in the presence of chaos), but follow an
exponential law.

Summarizing the dependence on the accumulation depth $\zacc$,
we always experience a net local dilution with increasing
depth, which is ``neutralized'' after summing up the different
branches of the material line due to mass conservation. The
inhomogeneities emerge at the beginning of the settling process
(observable for small accumulation depths), and, if
coarse-grained, they exhibit a decay on the long term (for
large accumulation depths).

%


\begin{thebibliography}{34}%
\makeatletter
\providecommand \@ifxundefined [1]{%
 \@ifx{#1\undefined}
}%
\providecommand \@ifnum [1]{%
 \ifnum #1\expandafter \@firstoftwo
 \else \expandafter \@secondoftwo
 \fi
}%
\providecommand \@ifx [1]{%
 \ifx #1\expandafter \@firstoftwo
 \else \expandafter \@secondoftwo
 \fi
}%
\providecommand \natexlab [1]{#1}%
\providecommand \enquote  [1]{``#1''}%
\providecommand \bibnamefont  [1]{#1}%
\providecommand \bibfnamefont [1]{#1}%
\providecommand \citenamefont [1]{#1}%
\providecommand \href@noop [0]{\@secondoftwo}%
\providecommand \href [0]{\begingroup \@sanitize@url \@href}%
\providecommand \@href[1]{\@@startlink{#1}\@@href}%
\providecommand \@@href[1]{\endgroup#1\@@endlink}%
\providecommand \@sanitize@url [0]{\catcode `\\12\catcode `\$12\catcode
  `\&12\catcode `\#12\catcode `\^12\catcode `\_12\catcode `\%12\relax}%
\providecommand \@@startlink[1]{}%
\providecommand \@@endlink[0]{}%
\providecommand \url  [0]{\begingroup\@sanitize@url \@url }%
\providecommand \@url [1]{\endgroup\@href {#1}{\urlprefix }}%
\providecommand \urlprefix  [0]{URL }%
\providecommand \Eprint [0]{\href }%
\providecommand \doibase [0]{http://dx.doi.org/}%
\providecommand \selectlanguage [0]{\@gobble}%
\providecommand \bibinfo  [0]{\@secondoftwo}%
\providecommand \bibfield  [0]{\@secondoftwo}%
\providecommand \translation [1]{[#1]}%
\providecommand \BibitemOpen [0]{}%
\providecommand \bibitemStop [0]{}%
\providecommand \bibitemNoStop [0]{.\EOS\space}%
\providecommand \EOS [0]{\spacefactor3000\relax}%
\providecommand \BibitemShut  [1]{\csname bibitem#1\endcsname}%
\let\auto@bib@innerbib\@empty
\bibitem [{\citenamefont {Michaelides}(1997)}]{Michaelides1997}%
  \BibitemOpen
  \bibfield  {author} {\bibinfo {author} {\bibfnamefont {E.~E.}\ \bibnamefont
  {Michaelides}},\ }\bibfield  {title} {\enquote {\bibinfo {title}
  {{Hydrodynamic Force and Heat/Mass Transfer From Particles, Bubbles, and
  Drops}},}\ }\href {\doibase 10.1115/1.1537258} {\bibfield  {journal}
  {\bibinfo  {journal} {Journal of Fluids Engineering}\ }\textbf {\bibinfo
  {volume} {125}},\ \bibinfo {pages} {209--238} (\bibinfo {year}
  {1997})}\BibitemShut {NoStop}%
\bibitem [{\citenamefont {Falkovich}\ and\ \citenamefont
  {Fouxon}(2002)}]{Falkovich2002}%
  \BibitemOpen
  \bibfield  {author} {\bibinfo {author} {\bibfnamefont {G.}~\bibnamefont
  {Falkovich}}\ and\ \bibinfo {author} {\bibfnamefont {M.~G.}\ \bibnamefont
  {Fouxon}, \bibfnamefont {I~Stepanov}},\ }\bibfield  {title} {\enquote
  {\bibinfo {title} {Acceleration of rain initiation by cloud turbulence},}\
  }\href@noop {} {\bibfield  {journal} {\bibinfo  {journal} {Nature}\ }\textbf
  {\bibinfo {volume} {419}},\ \bibinfo {pages} {151--154} (\bibinfo {year}
  {2002})}\BibitemShut {NoStop}%
\bibitem [{\citenamefont {Sabine}\ \emph {et~al.}(2004)\citenamefont {Sabine},
  \citenamefont {Feely}, \citenamefont {Gruber}, \citenamefont {Key},
  \citenamefont {Lee}, \citenamefont {Bullister}, \citenamefont {Wanninkhof},
  \citenamefont {Wong}, \citenamefont {Wallace}, \citenamefont {Tilbrook},
  \citenamefont {Millero}, \citenamefont {Peng}, \citenamefont {Kozyr},
  \citenamefont {Ono},\ and\ \citenamefont {Rios}}]{Sabine2004}%
  \BibitemOpen
  \bibfield  {author} {\bibinfo {author} {\bibfnamefont {C.}~\bibnamefont
  {Sabine}}, \bibinfo {author} {\bibfnamefont {R.}~\bibnamefont {Feely}},
  \bibinfo {author} {\bibfnamefont {N.}~\bibnamefont {Gruber}}, \bibinfo
  {author} {\bibfnamefont {R.}~\bibnamefont {Key}}, \bibinfo {author}
  {\bibfnamefont {K.}~\bibnamefont {Lee}}, \bibinfo {author} {\bibfnamefont
  {J.}~\bibnamefont {Bullister}}, \bibinfo {author} {\bibfnamefont
  {R.}~\bibnamefont {Wanninkhof}}, \bibinfo {author} {\bibfnamefont
  {C.}~\bibnamefont {Wong}}, \bibinfo {author} {\bibfnamefont {D.}~\bibnamefont
  {Wallace}}, \bibinfo {author} {\bibfnamefont {B.}~\bibnamefont {Tilbrook}},
  \bibinfo {author} {\bibfnamefont {F.}~\bibnamefont {Millero}}, \bibinfo
  {author} {\bibfnamefont {T.}~\bibnamefont {Peng}}, \bibinfo {author}
  {\bibfnamefont {A.}~\bibnamefont {Kozyr}}, \bibinfo {author} {\bibfnamefont
  {T.}~\bibnamefont {Ono}}, \ and\ \bibinfo {author} {\bibfnamefont
  {A.}~\bibnamefont {Rios}},\ }\bibfield  {title} {\enquote {\bibinfo {title}
  {The oceanic sink for anthropogenic {C}{O}2},}\ }\href@noop {} {\bibfield
  {journal} {\bibinfo  {journal} {Science}\ }\textbf {\bibinfo {volume}
  {305}},\ \bibinfo {pages} {367--371} (\bibinfo {year} {2004})}\BibitemShut
  {NoStop}%
\bibitem [{\citenamefont {Logan}\ and\ \citenamefont
  {Wilkinson}(1990)}]{Logan1990}%
  \BibitemOpen
  \bibfield  {author} {\bibinfo {author} {\bibfnamefont {B.~E.}\ \bibnamefont
  {Logan}}\ and\ \bibinfo {author} {\bibfnamefont {D.~B.}\ \bibnamefont
  {Wilkinson}},\ }\bibfield  {title} {\enquote {\bibinfo {title} {Fractal
  geometry of marine snow and other biological aggregates},}\ }\href {\doibase
  10.4319/lo.1990.35.1.0130} {\bibfield  {journal} {\bibinfo  {journal}
  {Limnology and Oceanography}\ }\textbf {\bibinfo {volume} {35}} (\bibinfo
  {year} {1990}),\ 10.4319/lo.1990.35.1.0130}\BibitemShut {NoStop}%
\bibitem [{\citenamefont {Buesseler}\ \emph {et~al.}(2007)\citenamefont
  {Buesseler}, \citenamefont {Antia}, \citenamefont {Chen}, \citenamefont
  {Fowler}, \citenamefont {Gardner}, \citenamefont {Gustafsson}, \citenamefont
  {Harada}, \citenamefont {Michaels}, \citenamefont {Rutgers van~der Loeff},
  \citenamefont {Sarin}, \citenamefont {Steinberg},\ and\ \citenamefont
  {Trull}}]{Buesseler2007}%
  \BibitemOpen
  \bibfield  {author} {\bibinfo {author} {\bibfnamefont {K.~O.}\ \bibnamefont
  {Buesseler}}, \bibinfo {author} {\bibfnamefont {A.~N.}\ \bibnamefont
  {Antia}}, \bibinfo {author} {\bibfnamefont {M.}~\bibnamefont {Chen}},
  \bibinfo {author} {\bibfnamefont {S.~W.}\ \bibnamefont {Fowler}}, \bibinfo
  {author} {\bibfnamefont {W.~D.}\ \bibnamefont {Gardner}}, \bibinfo {author}
  {\bibfnamefont {O.}~\bibnamefont {Gustafsson}}, \bibinfo {author}
  {\bibfnamefont {K.}~\bibnamefont {Harada}}, \bibinfo {author} {\bibfnamefont
  {A.~F.}\ \bibnamefont {Michaels}}, \bibinfo {author} {\bibfnamefont
  {M.}~\bibnamefont {Rutgers van~der Loeff}}, \bibinfo {author} {\bibfnamefont
  {M.}~\bibnamefont {Sarin}}, \bibinfo {author} {\bibfnamefont {D.~K.}\
  \bibnamefont {Steinberg}}, \ and\ \bibinfo {author} {\bibfnamefont
  {T.}~\bibnamefont {Trull}},\ }\bibfield  {title} {\enquote {\bibinfo {title}
  {An assessment of the use of sediment traps for estimating upper ocean
  particle fluxes},}\ }\href@noop {} {\bibfield  {journal} {\bibinfo  {journal}
  {Journal of Marine Research}\ }\textbf {\bibinfo {volume} {65}},\ \bibinfo
  {pages} {345--416} (\bibinfo {year} {2007})}\BibitemShut {NoStop}%
\bibitem [{\citenamefont {Mitchell}\ \emph {et~al.}(2008)\citenamefont
  {Mitchell}, \citenamefont {H.}, \citenamefont {Seuront}, \citenamefont
  {Wolk},\ and\ \citenamefont {Li}}]{Mitchell2008}%
  \BibitemOpen
  \bibfield  {author} {\bibinfo {author} {\bibfnamefont {J.}~\bibnamefont
  {Mitchell}}, \bibinfo {author} {\bibfnamefont {Y.}~\bibnamefont {H.}},
  \bibinfo {author} {\bibfnamefont {L.}~\bibnamefont {Seuront}}, \bibinfo
  {author} {\bibfnamefont {F.}~\bibnamefont {Wolk}}, \ and\ \bibinfo {author}
  {\bibfnamefont {H.}~\bibnamefont {Li}},\ }\bibfield  {title} {\enquote
  {\bibinfo {title} {Phytoplankton patch patterns: Seascape anatomy in a
  turbulent ocean},}\ }\href@noop {} {\bibfield  {journal} {\bibinfo  {journal}
  {Journal of Marine Systems}\ }\textbf {\bibinfo {volume} {69}},\ \bibinfo
  {pages} {247–253} (\bibinfo {year} {2008})}\BibitemShut {NoStop}%
\bibitem [{\citenamefont {Siegel}\ and\ \citenamefont
  {Deuser}(1997)}]{siegel1997}%
  \BibitemOpen
  \bibfield  {author} {\bibinfo {author} {\bibfnamefont {D.~A.}\ \bibnamefont
  {Siegel}}\ and\ \bibinfo {author} {\bibfnamefont {W.~G.}\ \bibnamefont
  {Deuser}},\ }\bibfield  {title} {\enquote {\bibinfo {title} {{Trajectories of
  sinking particles in the Sargasso Sea: modeling of statistical funnels above
  deep-ocean sediment traps}},}\ }\href@noop {} {\bibfield  {journal} {\bibinfo
   {journal} {Deep-Sea Research Part I-Oceanographic Research Papers}\ }\textbf
  {\bibinfo {volume} {44}},\ \bibinfo {pages} {1519 -- 1541} (\bibinfo {year}
  {1997})}\BibitemShut {NoStop}%
\bibitem [{\citenamefont {Schlitzer}, \citenamefont {Usbeck},\ and\
  \citenamefont {Fischer}(2003)}]{Schlitzer2003}%
  \BibitemOpen
  \bibfield  {author} {\bibinfo {author} {\bibfnamefont {R.}~\bibnamefont
  {Schlitzer}}, \bibinfo {author} {\bibfnamefont {R.}~\bibnamefont {Usbeck}}, \
  and\ \bibinfo {author} {\bibfnamefont {G.}~\bibnamefont {Fischer}},\
  }\enquote {\bibinfo {title} {Inverse modeling of particulate organic carbon
  fluxes in the south atlantic},}\ in\ \href@noop {} {\emph {\bibinfo
  {booktitle} {The South Atlantic in the Late Quaternary: Reconstruction of
  Material Budgets and Current Systems}}},\ \bibinfo {editor} {edited by\
  \bibinfo {editor} {\bibfnamefont {G.}~\bibnamefont {Wefer}}, \bibinfo
  {editor} {\bibfnamefont {S.}~\bibnamefont {Mulitza}}, \ and\ \bibinfo
  {editor} {\bibfnamefont {V.}~\bibnamefont {Ratmeyer}}}\ (\bibinfo
  {publisher} {Springer-Verlag Berlin Heidelberg New York Tokyo},\ \bibinfo
  {year} {2003})\ pp.\ \bibinfo {pages} {1--19}\BibitemShut {NoStop}%
\bibitem [{\citenamefont {Siegel}, \citenamefont {Fields},\ and\ \citenamefont
  {Buesseler}(2008)}]{Siegel2008}%
  \BibitemOpen
  \bibfield  {author} {\bibinfo {author} {\bibfnamefont {D.}~\bibnamefont
  {Siegel}}, \bibinfo {author} {\bibfnamefont {E.}~\bibnamefont {Fields}}, \
  and\ \bibinfo {author} {\bibfnamefont {K.~O.}\ \bibnamefont {Buesseler}},\
  }\bibfield  {title} {\enquote {\bibinfo {title} {A bottom-up view of the
  biological pump: Modeling source funnels above ocean sediment traps},}\
  }\href@noop {} {\bibfield  {journal} {\bibinfo  {journal} {Deep-Sea Research
  I}\ }\textbf {\bibinfo {volume} {55}},\ \bibinfo {pages} {108--127} (\bibinfo
  {year} {2008})}\BibitemShut {NoStop}%
\bibitem [{\citenamefont {Qiu}, \citenamefont {Doglioli},\ and\ \citenamefont
  {Carlotti}(2014)}]{Qiu2014}%
  \BibitemOpen
  \bibfield  {author} {\bibinfo {author} {\bibfnamefont {Z.}~\bibnamefont
  {Qiu}}, \bibinfo {author} {\bibfnamefont {A.}~\bibnamefont {Doglioli}}, \
  and\ \bibinfo {author} {\bibfnamefont {F.}~\bibnamefont {Carlotti}},\
  }\bibfield  {title} {\enquote {\bibinfo {title} {Using a {L}agrangian model
  to estimate source regions of particles in sediment traps},}\ }\href
  {\doibase 10.1007/s11430-014-4880-x} {\bibfield  {journal} {\bibinfo
  {journal} {Science China: Earth Sciences}\ }\textbf {\bibinfo {volume}
  {57}},\ \bibinfo {pages} {2447--2456} (\bibinfo {year} {2014})}\BibitemShut
  {NoStop}%
\bibitem [{\citenamefont {Monroy}\ \emph {et~al.}(2017)\citenamefont {Monroy},
  \citenamefont {Hern\'andez-Garc\'{\i}a}, \citenamefont {Rossi},\ and\
  \citenamefont {L\'opez}}]{Monroy2017}%
  \BibitemOpen
  \bibfield  {author} {\bibinfo {author} {\bibfnamefont {P.}~\bibnamefont
  {Monroy}}, \bibinfo {author} {\bibfnamefont {E.}~\bibnamefont
  {Hern\'andez-Garc\'{\i}a}}, \bibinfo {author} {\bibfnamefont
  {V.}~\bibnamefont {Rossi}}, \ and\ \bibinfo {author} {\bibfnamefont
  {C.}~\bibnamefont {L\'opez}},\ }\bibfield  {title} {\enquote {\bibinfo
  {title} {Modeling the dynamical sinking of biogenic particles in oceanic
  flow},}\ }\href {\doibase 10.5194/npg-24-293-2017} {\bibfield  {journal}
  {\bibinfo  {journal} {Nonlinear Processes in Geophysics}\ }\textbf {\bibinfo
  {volume} {24}},\ \bibinfo {pages} {293--305} (\bibinfo {year}
  {2017})}\BibitemShut {NoStop}%
\bibitem [{\citenamefont {Roullier}\ \emph {et~al.}(2014)\citenamefont
  {Roullier}, \citenamefont {Berline}, \citenamefont {Guidi}, \citenamefont
  {Durrieu De~Madron}, \citenamefont {Picheral}, \citenamefont {Sciandra},
  \citenamefont {Pesant},\ and\ \citenamefont {Stemmann}}]{Roullier2014}%
  \BibitemOpen
  \bibfield  {author} {\bibinfo {author} {\bibfnamefont {F.}~\bibnamefont
  {Roullier}}, \bibinfo {author} {\bibfnamefont {L.}~\bibnamefont {Berline}},
  \bibinfo {author} {\bibfnamefont {L.}~\bibnamefont {Guidi}}, \bibinfo
  {author} {\bibfnamefont {X.}~\bibnamefont {Durrieu De~Madron}}, \bibinfo
  {author} {\bibfnamefont {M.}~\bibnamefont {Picheral}}, \bibinfo {author}
  {\bibfnamefont {A.}~\bibnamefont {Sciandra}}, \bibinfo {author}
  {\bibfnamefont {S.}~\bibnamefont {Pesant}}, \ and\ \bibinfo {author}
  {\bibfnamefont {L.}~\bibnamefont {Stemmann}},\ }\bibfield  {title} {\enquote
  {\bibinfo {title} {Particle size distribution and estimated carbon flux
  across the arabian sea oxygen minimum zone},}\ }\href {\doibase
  10.5194/bg-11-4541-2014} {\bibfield  {journal} {\bibinfo  {journal}
  {Biogeosciences}\ }\textbf {\bibinfo {volume} {11}},\ \bibinfo {pages}
  {4541--4557} (\bibinfo {year} {2014})}\BibitemShut {NoStop}%
\bibitem [{\citenamefont {Balkovsky}, \citenamefont {Falkovich},\ and\
  \citenamefont {Fouxon}(2001)}]{Balkovsky2001}%
  \BibitemOpen
  \bibfield  {author} {\bibinfo {author} {\bibfnamefont {E.}~\bibnamefont
  {Balkovsky}}, \bibinfo {author} {\bibfnamefont {G.}~\bibnamefont
  {Falkovich}}, \ and\ \bibinfo {author} {\bibfnamefont {A.}~\bibnamefont
  {Fouxon}},\ }\bibfield  {title} {\enquote {\bibinfo {title} {Intermittent
  distribution of inertial particles in turbulent flows},}\ }\href {\doibase
  10.1103/PhysRevLett.86.2790} {\bibfield  {journal} {\bibinfo  {journal}
  {Phys. Rev. Lett.}\ }\textbf {\bibinfo {volume} {86}},\ \bibinfo {pages}
  {2790--2793} (\bibinfo {year} {2001})}\BibitemShut {NoStop}%
\bibitem [{\citenamefont {Bec}(2003)}]{Bec2003}%
  \BibitemOpen
  \bibfield  {author} {\bibinfo {author} {\bibfnamefont {J.}~\bibnamefont
  {Bec}},\ }\bibfield  {title} {\enquote {\bibinfo {title} {Fractal clustering
  of inertial particles in random flows},}\ }\href {\doibase 10.1063/1.1612500}
  {\bibfield  {journal} {\bibinfo  {journal} {Physics of Fluids}\ }\textbf
  {\bibinfo {volume} {15}},\ \bibinfo {pages} {81--84} (\bibinfo {year}
  {2003})}\BibitemShut {NoStop}%
\bibitem [{\citenamefont {Vilela}\ \emph {et~al.}(2007)\citenamefont {Vilela},
  \citenamefont {T\'el}, \citenamefont {de~Moura},\ and\ \citenamefont
  {Grebogi}}]{Vilela2007}%
  \BibitemOpen
  \bibfield  {author} {\bibinfo {author} {\bibfnamefont {R.~D.}\ \bibnamefont
  {Vilela}}, \bibinfo {author} {\bibfnamefont {T.}~\bibnamefont {T\'el}},
  \bibinfo {author} {\bibfnamefont {A.~P.~S.}\ \bibnamefont {de~Moura}}, \ and\
  \bibinfo {author} {\bibfnamefont {C.}~\bibnamefont {Grebogi}},\ }\bibfield
  {title} {\enquote {\bibinfo {title} {Signatures of fractal clustering of
  aerosols advected under gravity},}\ }\href@noop {} {\bibfield  {journal}
  {\bibinfo  {journal} {Phys. Rev. E}\ }\textbf {\bibinfo {volume} {75}},\
  \bibinfo {pages} {065203(R)} (\bibinfo {year} {2007})}\BibitemShut {NoStop}%
\bibitem [{\citenamefont {Cartwright}\ \emph {et~al.}(2010)\citenamefont
  {Cartwright}, \citenamefont {Feudel}, \citenamefont {K{\'a}rolyi},
  \citenamefont {de~Moura}, \citenamefont {Piro},\ and\ \citenamefont
  {T{\'e}l}}]{Cartwright2010}%
  \BibitemOpen
  \bibfield  {author} {\bibinfo {author} {\bibfnamefont {J.~H.~E.}\
  \bibnamefont {Cartwright}}, \bibinfo {author} {\bibfnamefont
  {U.}~\bibnamefont {Feudel}}, \bibinfo {author} {\bibfnamefont
  {G.}~\bibnamefont {K{\'a}rolyi}}, \bibinfo {author} {\bibfnamefont
  {A.}~\bibnamefont {de~Moura}}, \bibinfo {author} {\bibfnamefont
  {O.}~\bibnamefont {Piro}}, \ and\ \bibinfo {author} {\bibfnamefont
  {T.}~\bibnamefont {T{\'e}l}},\ }\enquote {\bibinfo {title} {Dynamics of
  finite-size particles in chaotic fluid flows},}\ in\ \href {\doibase
  10.1007/978-3-642-04629-2_4} {\emph {\bibinfo {booktitle} {Nonlinear Dynamics
  and Chaos: Advances and Perspectives}}},\ \bibinfo {editor} {edited by\
  \bibinfo {editor} {\bibfnamefont {M.}~\bibnamefont {Thiel}}, \bibinfo
  {editor} {\bibfnamefont {J.}~\bibnamefont {Kurths}}, \bibinfo {editor}
  {\bibfnamefont {M.~C.}\ \bibnamefont {Romano}}, \bibinfo {editor}
  {\bibfnamefont {G.}~\bibnamefont {K{\'a}rolyi}}, \ and\ \bibinfo {editor}
  {\bibfnamefont {A.}~\bibnamefont {Moura}}}\ (\bibinfo  {publisher} {Springer
  Berlin Heidelberg},\ \bibinfo {address} {Berlin, Heidelberg},\ \bibinfo
  {year} {2010})\ pp.\ \bibinfo {pages} {51--87}\BibitemShut {NoStop}%
\bibitem [{\citenamefont {Guseva}, \citenamefont {Feudel},\ and\ \citenamefont
  {T\'el}(2013)}]{Guseva2013}%
  \BibitemOpen
  \bibfield  {author} {\bibinfo {author} {\bibfnamefont {K.}~\bibnamefont
  {Guseva}}, \bibinfo {author} {\bibfnamefont {U.}~\bibnamefont {Feudel}}, \
  and\ \bibinfo {author} {\bibfnamefont {T.}~\bibnamefont {T\'el}},\ }\bibfield
   {title} {\enquote {\bibinfo {title} {Influence of the history force on
  inertial particle advection: Gravitational effects and horizontal
  diffusion},}\ }\href {\doibase 10.1103/PhysRevE.88.042909} {\bibfield
  {journal} {\bibinfo  {journal} {Phys. Rev. E}\ }\textbf {\bibinfo {volume}
  {88}},\ \bibinfo {pages} {042909} (\bibinfo {year} {2013})}\BibitemShut
  {NoStop}%
\bibitem [{\citenamefont {Guseva}\ \emph {et~al.}(2016)\citenamefont {Guseva},
  \citenamefont {Feudel}, \citenamefont {Daitche},\ and\ \citenamefont
  {T\'el}}]{Guseva2016}%
  \BibitemOpen
  \bibfield  {author} {\bibinfo {author} {\bibfnamefont {K.}~\bibnamefont
  {Guseva}}, \bibinfo {author} {\bibfnamefont {U.}~\bibnamefont {Feudel}},
  \bibinfo {author} {\bibfnamefont {A.}~\bibnamefont {Daitche}}, \ and\
  \bibinfo {author} {\bibfnamefont {T.}~\bibnamefont {T\'el}},\ }\bibfield
  {title} {\enquote {\bibinfo {title} {History effects in the sedimentation of
  light aerosols in turbulence: The case of marine snow},}\ }\href {\doibase
  10.1103/PhysRevFluids.1.074203} {\bibfield  {journal} {\bibinfo  {journal}
  {Phys. Rev. Fluids}\ }\textbf {\bibinfo {volume} {1}},\ \bibinfo {pages}
  {074203} (\bibinfo {year} {2016})}\BibitemShut {NoStop}%
\bibitem [{\citenamefont {Ott}(1993)}]{Ott1993}%
  \BibitemOpen
  \bibfield  {author} {\bibinfo {author} {\bibfnamefont {E.}~\bibnamefont
  {Ott}},\ }\href@noop {} {\emph {\bibinfo {title} {Chaos in Dynamical
  Systems}}}\ (\bibinfo  {publisher} {Cambridge University Press, Cambridge,
  UK},\ \bibinfo {year} {1993})\BibitemShut {NoStop}%
\bibitem [{\citenamefont {Ottino}(1989)}]{Ottino1989}%
  \BibitemOpen
  \bibfield  {author} {\bibinfo {author} {\bibfnamefont {J.~M.}\ \bibnamefont
  {Ottino}},\ }\href@noop {} {\emph {\bibinfo {title} {The Kinematics of
  Mixing: Stretching, Chaos and Transport}}}\ (\bibinfo  {publisher} {Cambridge
  University Press, Cambridge},\ \bibinfo {year} {1989})\BibitemShut {NoStop}%
\bibitem [{\citenamefont {Pierrehumbert}\ and\ \citenamefont
  {Yang}(1993)}]{Pierrehumbert1993}%
  \BibitemOpen
  \bibfield  {author} {\bibinfo {author} {\bibfnamefont {R.~T.}\ \bibnamefont
  {Pierrehumbert}}\ and\ \bibinfo {author} {\bibfnamefont {H.}~\bibnamefont
  {Yang}},\ }\bibfield  {title} {\enquote {\bibinfo {title} {Global chaotic
  mixing on isentropic surfaces},}\ }\href@noop {} {\bibfield  {journal}
  {\bibinfo  {journal} {J. Atmos. Sci.}\ }\textbf {\bibinfo {volume} {50}},\
  \bibinfo {pages} {2462--2480} (\bibinfo {year} {1993})}\BibitemShut {NoStop}%
\bibitem [{\citenamefont {Gutknecht}\ \emph {et~al.}(2013)\citenamefont
  {Gutknecht}, \citenamefont {Dadou}, \citenamefont {Le~Vu}, \citenamefont
  {Cambon}, \citenamefont {Sudre}, \citenamefont {Garçon}, \citenamefont
  {Machu}, \citenamefont {Rixen}, \citenamefont {Kock}, \citenamefont {Flohr},
  \citenamefont {Paulmier},\ and\ \citenamefont {Lavik}}]{Gutknecht2013}%
  \BibitemOpen
  \bibfield  {author} {\bibinfo {author} {\bibfnamefont {E.}~\bibnamefont
  {Gutknecht}}, \bibinfo {author} {\bibfnamefont {I.}~\bibnamefont {Dadou}},
  \bibinfo {author} {\bibfnamefont {B.}~\bibnamefont {Le~Vu}}, \bibinfo
  {author} {\bibfnamefont {G.}~\bibnamefont {Cambon}}, \bibinfo {author}
  {\bibfnamefont {J.}~\bibnamefont {Sudre}}, \bibinfo {author} {\bibfnamefont
  {V.}~\bibnamefont {Garçon}}, \bibinfo {author} {\bibfnamefont
  {E.}~\bibnamefont {Machu}}, \bibinfo {author} {\bibfnamefont
  {T.}~\bibnamefont {Rixen}}, \bibinfo {author} {\bibfnamefont
  {A.}~\bibnamefont {Kock}}, \bibinfo {author} {\bibfnamefont {A.}~\bibnamefont
  {Flohr}}, \bibinfo {author} {\bibfnamefont {A.}~\bibnamefont {Paulmier}}, \
  and\ \bibinfo {author} {\bibfnamefont {G.}~\bibnamefont {Lavik}},\ }\bibfield
   {title} {\enquote {\bibinfo {title} {Coupled physical/biogeochemical
  modeling including {O}2-dependent processes in the eastern boundary upwelling
  systems: application in the {B}enguela},}\ }\href {\doibase
  10.5194/bg-10-3559-2013} {\bibfield  {journal} {\bibinfo  {journal}
  {Biogeosciences}\ }\textbf {\bibinfo {volume} {10}},\ \bibinfo {pages}
  {3559--3591} (\bibinfo {year} {2013})}\BibitemShut {NoStop}%
\bibitem [{\citenamefont {Maxey}\ and\ \citenamefont
  {Riley}(1983)}]{Maxey1983}%
  \BibitemOpen
  \bibfield  {author} {\bibinfo {author} {\bibfnamefont {M.~R.}\ \bibnamefont
  {Maxey}}\ and\ \bibinfo {author} {\bibfnamefont {J.~J.}\ \bibnamefont
  {Riley}},\ }\bibfield  {title} {\enquote {\bibinfo {title} {{Equation of
  motion for a small rigid sphere in a nonuniform flow}},}\ }\href {\doibase
  10.1063/1.864230} {\bibfield  {journal} {\bibinfo  {journal} {Physics of
  Fluids}\ }\textbf {\bibinfo {volume} {26}},\ \bibinfo {pages} {883--889}
  (\bibinfo {year} {1983})}\BibitemShut {NoStop}%
\bibitem [{\citenamefont {Haller}\ and\ \citenamefont
  {Sapsis}(2008)}]{Haller2008}%
  \BibitemOpen
  \bibfield  {author} {\bibinfo {author} {\bibfnamefont {G.}~\bibnamefont
  {Haller}}\ and\ \bibinfo {author} {\bibfnamefont {T.}~\bibnamefont
  {Sapsis}},\ }\bibfield  {title} {\enquote {\bibinfo {title} {{Where do
  inertial particles go in fluid flows?}}}\ }\href {\doibase
  10.1016/j.physd.2007.09.027} {\bibfield  {journal} {\bibinfo  {journal}
  {Physica D: Nonlinear Phenomena}\ }\textbf {\bibinfo {volume} {237}},\
  \bibinfo {pages} {573--583} (\bibinfo {year} {2008})}\BibitemShut {NoStop}%
\bibitem [{\citenamefont {Jimenez}(1997)}]{Jimenez1997}%
  \BibitemOpen
  \bibfield  {author} {\bibinfo {author} {\bibfnamefont {J.}~\bibnamefont
  {Jimenez}},\ }\bibfield  {title} {\enquote {\bibinfo {title} {Ocean
  turbulence at milimiter scales},}\ }\href@noop {} {\bibfield  {journal}
  {\bibinfo  {journal} {Scientia Marina}\ }\textbf {\bibinfo {volume} {61}},\
  \bibinfo {pages} {47--56} (\bibinfo {year} {1997})}\BibitemShut {NoStop}%
\bibitem [{\citenamefont {Wilkinson}\ and\ \citenamefont
  {Mehlig}(2005)}]{Wilkinson2005}%
  \BibitemOpen
  \bibfield  {author} {\bibinfo {author} {\bibfnamefont {M.}~\bibnamefont
  {Wilkinson}}\ and\ \bibinfo {author} {\bibfnamefont {B.}~\bibnamefont
  {Mehlig}},\ }\bibfield  {title} {\enquote {\bibinfo {title} {Caustics in
  turbulent aerosols},}\ }\href {http://stacks.iop.org/0295-5075/71/i=2/a=186}
  {\bibfield  {journal} {\bibinfo  {journal} {EPL (Europhysics Letters)}\
  }\textbf {\bibinfo {volume} {71}},\ \bibinfo {pages} {186} (\bibinfo {year}
  {2005})}\BibitemShut {NoStop}%
\bibitem [{\citenamefont {Dr\'otos}\ and\ \citenamefont
  {T\'el}(2011)}]{Drotos2011}%
  \BibitemOpen
  \bibfield  {author} {\bibinfo {author} {\bibfnamefont {G.}~\bibnamefont
  {Dr\'otos}}\ and\ \bibinfo {author} {\bibfnamefont {T.}~\bibnamefont
  {T\'el}},\ }\bibfield  {title} {\enquote {\bibinfo {title} {Chaotic saddles
  in a gravitational field: The case of inertial particles in finite
  domains},}\ }\href@noop {} {\bibfield  {journal} {\bibinfo  {journal} {Phys.
  Rev. E}\ }\textbf {\bibinfo {volume} {83}},\ \bibinfo {pages} {056203}
  (\bibinfo {year} {2011})}\BibitemShut {NoStop}%
\bibitem [{\citenamefont {Pierrehumbert}(1994)}]{Pierrehumbert1994}%
  \BibitemOpen
  \bibfield  {author} {\bibinfo {author} {\bibfnamefont {R.}~\bibnamefont
  {Pierrehumbert}},\ }\bibfield  {title} {\enquote {\bibinfo {title} {Tracer
  microstructure in the large-eddy dominated regime},}\ }\href@noop {}
  {\bibfield  {journal} {\bibinfo  {journal} {Chaos, Solitons, and Fractals}\
  }\textbf {\bibinfo {volume} {4}},\ \bibinfo {pages} {1091--1110} (\bibinfo
  {year} {1994})}\BibitemShut {NoStop}%
\bibitem [{\citenamefont {Feudel}\ \emph {et~al.}(2005)\citenamefont {Feudel},
  \citenamefont {Witt}, \citenamefont {Gellert}, \citenamefont {Kurths},
  \citenamefont {Grebogi},\ and\ \citenamefont {Sanju\'an}}]{Feudel2005}%
  \BibitemOpen
  \bibfield  {author} {\bibinfo {author} {\bibfnamefont {F.}~\bibnamefont
  {Feudel}}, \bibinfo {author} {\bibfnamefont {A.}~\bibnamefont {Witt}},
  \bibinfo {author} {\bibfnamefont {M.}~\bibnamefont {Gellert}}, \bibinfo
  {author} {\bibfnamefont {J.}~\bibnamefont {Kurths}}, \bibinfo {author}
  {\bibfnamefont {C.}~\bibnamefont {Grebogi}}, \ and\ \bibinfo {author}
  {\bibfnamefont {M.}~\bibnamefont {Sanju\'an}},\ }\bibfield  {title} {\enquote
  {\bibinfo {title} {Intersections of stable and unstable manifolds: the
  skeleton of lagrangian chaos},}\ }\href {\doibase
  https://doi.org/10.1016/j.chaos.2004.09.059} {\bibfield  {journal} {\bibinfo
  {journal} {Chaos, Solitons \& Fractals}\ }\textbf {\bibinfo {volume} {24}},\
  \bibinfo {pages} {947--956} (\bibinfo {year} {2005})}\BibitemShut {NoStop}%
\bibitem [{\citenamefont {Lindner}\ and\ \citenamefont
  {Donner}(2017)}]{Lindner2017}%
  \BibitemOpen
  \bibfield  {author} {\bibinfo {author} {\bibfnamefont {M.}~\bibnamefont
  {Lindner}}\ and\ \bibinfo {author} {\bibfnamefont {R.~V.}\ \bibnamefont
  {Donner}},\ }\bibfield  {title} {\enquote {\bibinfo {title} {Spatio-temporal
  organization of dynamics in a two-dimensional periodically driven vortex
  flow: A lagrangian flow network perspective},}\ }\href@noop {} {\bibfield
  {journal} {\bibinfo  {journal} {Chaos 27, 035806}\ }\textbf {\bibinfo
  {volume} {27}},\ \bibinfo {pages} {035806} (\bibinfo {year}
  {2017})}\BibitemShut {NoStop}%
\bibitem [{\citenamefont {Bezuglyy}\ \emph {et~al.}(2006)\citenamefont
  {Bezuglyy}, \citenamefont {Mehlig}, \citenamefont {Wilkinson}, \citenamefont
  {Nakamura},\ and\ \citenamefont {Arvedson}}]{Bezuglyy2006}%
  \BibitemOpen
  \bibfield  {author} {\bibinfo {author} {\bibfnamefont {V.}~\bibnamefont
  {Bezuglyy}}, \bibinfo {author} {\bibfnamefont {B.}~\bibnamefont {Mehlig}},
  \bibinfo {author} {\bibfnamefont {M.}~\bibnamefont {Wilkinson}}, \bibinfo
  {author} {\bibfnamefont {K.}~\bibnamefont {Nakamura}}, \ and\ \bibinfo
  {author} {\bibfnamefont {E.}~\bibnamefont {Arvedson}},\ }\bibfield  {title}
  {\enquote {\bibinfo {title} {Generalized {O}rnstein-{U}hlenbeck processes},}\
  }\href@noop {} {\bibfield  {journal} {\bibinfo  {journal} {Journal of
  Mathematical Physics}\ }\textbf {\bibinfo {volume} {47}},\ \bibinfo {pages}
  {073301} (\bibinfo {year} {2006})}\BibitemShut {NoStop}%
\bibitem [{\citenamefont {Gustavsson}, \citenamefont {Vajedi},\ and\
  \citenamefont {Mehlig}(2014)}]{Gustavsson2014}%
  \BibitemOpen
  \bibfield  {author} {\bibinfo {author} {\bibfnamefont {K.}~\bibnamefont
  {Gustavsson}}, \bibinfo {author} {\bibfnamefont {S.}~\bibnamefont {Vajedi}},
  \ and\ \bibinfo {author} {\bibfnamefont {B.}~\bibnamefont {Mehlig}},\
  }\bibfield  {title} {\enquote {\bibinfo {title} {Clustering of particles
  falling in a turbulent flow},}\ }\href@noop {} {\bibfield  {journal}
  {\bibinfo  {journal} {Phys. Rev. Lett.}\ }\textbf {\bibinfo {volume} {112}},\
  \bibinfo {pages} {214501} (\bibinfo {year} {2014})}\BibitemShut {NoStop}%
\bibitem [{\citenamefont {Lai}\ and\ \citenamefont {T\'el}(2011)}]{Lai2011}%
  \BibitemOpen
  \bibfield  {author} {\bibinfo {author} {\bibfnamefont {Y.-C.}\ \bibnamefont
  {Lai}}\ and\ \bibinfo {author} {\bibfnamefont {T.}~\bibnamefont {T\'el}},\
  }\href@noop {} {\emph {\bibinfo {title} {Transient Chaos}}}\ (\bibinfo
  {publisher} {Springer-Verlag, New York},\ \bibinfo {year} {2011})\BibitemShut
  {NoStop}%
\bibitem [{\citenamefont {Harville}(2008)}]{Harville2008}%
  \BibitemOpen
  \bibfield  {author} {\bibinfo {author} {\bibfnamefont {D.~A.}\ \bibnamefont
  {Harville}},\ }\href@noop {} {\emph {\bibinfo {title} {Matrix algebra from a
  statistician's perspective}}}\ (\bibinfo  {publisher} {Springer-Verlag, New
  York},\ \bibinfo {year} {2008})\BibitemShut {NoStop}%
\end{thebibliography}

\end{document}